%% file: lcb.tex
\input gr-qc
% -----------------------------------------------------------------------------
%
\def\eps{\epsilon}
\def\Oeps(#1){{\cal O}(\eps^{#1})}
\def\One{{\cal O}(\eps)}
\def\m{{\hbox{\tt-}}}
\def\p{{\hbox{\tt+}}}
\title{%
{\Bf Riemann Normal Coordinates, Smooth Lattices}\cr
{\Bf and Numerical Relativity}\cr
\cr
{\rm Leo Brewin}\cr
\cr
{\sl Department of Mathematics}\cr
{\sl Monash University}\cr
{\sl Clayton, Vic. 3168}\cr
{\sl Australia}\cr
{\rm 26-Dec-1996}\cr}

\beginabstract
A new lattice based scheme for numerical relativity will be presented. The
scheme uses the same data as would be used in the Regge calculus (eg. a set of
leg lengths on a simplicial lattice) but it differs significantly in the way
that the field equations are computed. In the new method the standard Einstein
field equations are applied directly to the lattice. This is done by using
locally defined Riemann normal coordinates to interpolate a smooth metric over
local groups of cells of the lattice. Results for the time symmetric initial
data for the Schwarzschild spacetime will be presented. It will be shown that
the scheme yields second order accurate estimates (in the lattice spacing) for
the metric and the curvature. It will also be shown that the Bianchi identities
play an essential role in the construction of the Schwarzschild initial data.

\endabstract

\beginsection{Introduction}
One of the most common techniques employed in numerical general relativity is
that of finite differences. In this approach the coordinate components of the
metric are sampled at a series of well defined points. The connection and
curvatures are then calculated from well known finite difference formulae. Such
formula can be easily derived in a number ways, and in particular, by way of 
a quadratic interpolation of the sampled values. Though this method has produced
excellent results it is reasonable to explore other alternatives.

The alternative to be developed in this paper is to use a lattice of geodesic
segments to provide the samples of the spacetime metric. The curvature will, as
with finite differences, be obtained from a quadratic interpolation of the
sampled metric. This procedure is, however, somewhat more involved than that for
finite differences.

Lattice theories of general relativity such as the Regge calculus \Ref(1) are
not particularly new. What makes our approach new is the way in which the
discrete fields equations are constructed from the lattice variables such as
the leg lengths. In the Regge calculus this is done by way of an action
integral whereas as in our approach the field equations of general relativity
are applied directly to the lattice. We will defer further comparison of the
Regge calculus with our approach until after the main results of this paper have
been presented.

For any explicitly given spacetime it is straightforward, though perhaps
computationally tedious, to construct a lattice of geodesic segments (eg. by
sub-dividing the spacetime into a set of simplices). This lattice will inherit
information about the topology and geometry of the parent spacetime. If the
lattice is chosen correctly then it is possible to record all of the topological
information of the parent spacetime in the lattice. The same is not true for the
metric since the lattice contains only a finite number of geodesic segments. Thus
some information will be lost in going from the parent spacetime to the lattice.
The interesting question is how much information about the parent spacetime can
be recovered from the lattice? In particular, can the lattice be used to estimate
the curvature tensor of the spacetime, and if so, with what accuracy? One can
expect that the answer to this question will entail, just as in the finite
difference case, a locally quadratic interpolation of the sampled data.

For the moment suppose that such an interpolation procedure exists (while putting
aside the issue of accuracy). Then given a set of leg lengths one can compute an
associated set of curvatures. Einstein's equations can then be evaluated and the
leg lengths adjusted accordingly. Clearly this amounts to solving Einstein's
equations for the parent spacetime discretised on a lattice. This is the basis
for using lattices -- they provide an alternative to finite differences as a tool
for numerical relativity.

The procedure for estimating the curvatures from the leg lengths will be based
on Riemann normal coordinates \Ref(2-5) (later referred to as RNC). A simple
algebraic definition of these coordinates is that, for a given point $O$, they
are a local set of coordinates, such that
$$\openup10pt\displaylines{%
\Gamma\strut^{\mu}_{\alpha\beta} = 0 \cr
\Gamma\strut^{\mu}_{\alpha\beta,\nu} +
\Gamma\strut^{\mu}_{\beta\nu,\alpha} +
\Gamma\strut^{\mu}_{\nu\alpha,\beta} = 0 \cr}
$$
at $O$.

An equivalent and perhaps more instructive geometric definition goes as follows.
Consider the point $O$ and a small neighbourhood of $O$. If $P$ is any other
point near to $O$ then there exists a unique geodesic joining $O$ to $P$.
Let $a^\mu$ be the components of the unit vector to this geodesic at $O$ and
let $s$ be the geodesic length from $O$ to $P$. Then the Riemann normal
coordinates of $P$ are defined to be $x^\mu = sa^\mu$. This construction fails
whenever the geodesic joining $O$ to $P$ is not unique (ie. when geodesics
cross). Fortunately the neighbourhood of $O$ can always be chosen to be small
enough so that this problem does not arise. Incidently, this displays the local
nature of Riemann normal coordinates. They cannot be used to cover the whole
manifold. Instead a collection of distinct Riemann normal coordinates must be
constructed.

From this simple definition a number of very useful theorems can be proved
(see the Appendix).
In particular, for the parent spacetime, the metric is of the form
$$
g_{\mu\nu}(x) = g_{\mu\nu} - {1\over3}R_{\mu\alpha\nu\beta}\> x^\alpha x^\beta
                + \Oeps(3)
\eqno\eqndef{RNCmetric}
$$
where $g_{\mu\nu}$ and $R_{\mu\alpha\nu\beta}$ are the components of the
metric and Riemann tensors at the origin $O$ of the RNC and and $\eps$ is a
typical length scale of the neighbourhood in which the RNC is defined.
This equation is nothing more than a Taylor series expansion of the metric
around the origin $O$. Higher order approximations can be generated by continuing
the series. The question of convergence is not very important in our context
because we can always choose the domain of the RNC sufficiently small so that
the truncation error in the above expansion is negligible. Note that the above
expansion can be viewed as a flat space part (the constant term) plus a small
perturbation (the curvature term).

From this metric one can derive formulae for such things as the length of a
geodesic segment and the angle subtended at the vertex of a geodesic triangle.
The length of a geodesic segment with vertices $i$ and $j$ is
given by
$$
L^2_{ij} = g_{\mu\nu}\Delta x^\mu_{ij}\Delta x^\nu_{ij}
          - {1\over3} R_{\mu\alpha\nu\beta}\> x^\mu_i x^\nu_i
                                              x^\alpha_j x^\beta_j
          + \Oeps(5)
\eqno\eqndef{LegsEqtn}
$$
while the angle, $\theta_k$, subtended at the vertex $k$ of the triangle with
vertices $i,j$ and $k$ can be computed from
$$
2L_{ik}L_{jk}\cos\theta_k = L^2_{ik} + L^2_{jk} - L^2_{ij}
   -{1\over3}R_{\mu\alpha\nu\beta}\>\Delta x^\mu_{ik}\Delta x^\nu_{ik}
                                    \Delta x^\alpha_{jk}\Delta x^\beta_{jk}
   + \Oeps(5)
\eqno\eqndef{AnglesEqtn}
$$
These formulae can be easily applied if we are given the various
$g_{\mu\nu},R_{\mu\nu\alpha\beta}$ and $x^\mu_i$ of the parent spacetime.
However, we wish to pose the inverse problem, given just the leg lengths, angles
and the combinatoric information of a lattice, compute (or estimate) the
curvature of the parent spacetime. This can be done by treating the above as a
coupled system of non-linear equations. There is one such system of equations
for each vertex.

We will define a {\it smooth lattice} to be a lattice for which we have solved
the above equations, (\eqnLegsEqtn) and (\eqnAnglesEqtn), which in
turn will be referred to as the {\it smooth lattice equations}. (Though a
better choice might be {\it Riemann lattice}.)

There are a number of attractive features in the smooth lattice approach that
deserve special mention. First, we have made explicit use of the equivalence
principle by setting the connection to zero at the origin of each RNC. This not
only simplifies many calculations
but it also allows us to interpret each RNC as a locally freely falling frame.
This central use of the equivalence principle must surely be to our advantage. The
second point is that the smooth lattice equations are very easy to formulate
and solve. In contrast, the equations in the Regge calculus are very tedious
with numerous inverse trigonometric and hyperbolic functions and are far from
simple to setup and evaluate on a computer. Solving the Regge equations on a
computer could easily be a very time consuming task (much more so than either
for finite differences or for our method). The third point is that as our
method has a solid theoretical basis we are able to use all of the usual tools
of differential geometry and analysis to investigate issues such as convergence
and discretisation errors. Such is not the case with the Regge calculus.

Despite these attractive features one cannot apply the method naively. In any
practical application one may need to address the following issues (amongst
others).

\advance\leftskip+0.5cm

\item{\Mymark} {\it Over which region should each RNC be defined?}
To each vertex there should be assigned a set of simplices from which the
information is to be extracted. The size of that region (ie. the number of
simplices) must reflect the amount of information required. It seems that the
smallest practical region would be that composed of the ball of simplices
attached to the nominated vertex.

\item{\Mymark} {\it Are there any gauge freedoms in choosing the RNC?}
Yes, for example the standard rotations, translations etc. This gauge
freedom can be used to force $g_{\mu\nu}$ and or various $x^\mu_i$
to take on specific values. The remaining quantities must be computed together
with the curvatures from the smooth lattice equations.

\item{\Mymark} {\it Are there as many equations as unknowns?}
Yes, but only in two dimensions and only when the surface consists of
triangles. In higher dimensions the set of equations is usually overdetermined.
This problem can be overcome by either resorting to a least squares solution or
by simply excluding some of the equations. Another option would be to increase
the number of parameters, such as derivatives of the curvature, in the
interpolation of the metric so as to produce a properly determined set of
equations.

\item{\Mymark} {\it Is there a unique solution to the smooth lattice
equations?} In general, no -- the smooth lattice equations are non-linear and
so we can expect more than one solution. The members of a discrete family of
solutions will most probably be related by discrete transformations such as
folding one simplex over another. We would hope that there would be one
exceptional case where all the simplices do not overlap and this would be the
solution we would choose. There is also the possibility of having a continuous
family of solutions. This would suggest that the lattice is improperly defined
in that it lacks sufficient structure (eg. too few leg lengths). This later
case should not be allowed to occur in practice.

\item{\Mymark} {\it Can all of the leg lengths and angles be freely chosen?}
If the curvatures are specified locally, as they are in our system of RNC's,
then there is question as to whether or not there exists a single metric which
has a curvature matching the specified local values. This is nothing more than a
question of integrability -- given a curvature tensor does there exist a
corresponding metric? The answer is yes provided that that curvature satisfies
the Bianchi identities. For our set of RNC's this imposes (surprisingly) a set
of constraints on the choice of the leg lengths and angles.

\advance\leftskip-0.5cm

Some of these issues will be explored in greater detail in later sections.

\beginsection{Schwarzschild initial data}
As a test of the smooth lattice method we should be able to successfully recover
the time symmetric 3-geometry for the Schwarzschild spacetime. This metric can be
written in the form
$$ds^2 = dl^2 + \eta(l)^2\, d\Omega^2$$
where $l$ is the proper distance measured from the throat, $\eta(l)$
is a smooth function of $l$ and $d\Omega^2$ is the standard metric of a unit
2-sphere.

We will present two examples of a smooth lattice discretisations of this metric.
In the first example we will discretise the metric of the unit 2-sphere
while retaining a continuous radial coordinate. We will first use the smooth
lattice method to estimate the curvature of a unit 2-sphere (for which we all
know the exact value, 2). This estimate will then be used in a radial integration
of the Hamiltonian constraint to complete the construction of the 3-metric.

This is a very simple test of the method. A much stronger test will be presented
in our second example where the full 3-metric will be discretised (though we
will use the spherical symmetry to simplify the calculations).

\beginsubsection{Smooth lattice 2-sphere}
The smooth lattice equations will be used in this example solely as a means to
estimate the Riemann curvature scalar of a unit 2-sphere. This is known to have
the value 2 so this may appear to be a somewhat un-necessary application of the
method. However, if the method fails in this simplest of all cases then it
surely will be of little use in any other situation. Once the curvature has
been estimated we will employ it in solving the standard time symmetric
constraint equations for the three metric.

We will estimate $R$ from the known 2-metric of the unit 2-sphere. To do so it
will be necessary to choose a lattice on the 2-sphere, compute the geodesic leg
lengths and finally solve a set of equations (given below) for $R$.

The simplest approximation to a 2-sphere is a regular tetrahedron (see
Figure (\figrfr{FigTetA})). This cannot be expected to provide very accurate
estimates for the curvature. Thus some scheme for refining the lattice will be
required. The scheme used in this example will be to successively sub-divide
each triangle according to the pattern in Figure (\figrfr{FigTetB}).

The metric of the 2-sphere can be written as
$$
ds^2 = d\theta^2 + \sin^2\theta\, d\phi^2
$$
or as the induced metric on the surface $1=x^2+y^2+z^2$ in Euclidean 3-space
with the usual $(x,y,z)$ coordinates. The $(\theta,\phi)$ and $(x,y,z)$
coordinates are related by the usual polar coordinate transformations. The
$(x,y,z)$ coordinates of the four vertices of the original tetrahedron are easily
calculated by appealing to the symmetry of the (regular) tetrahedron. The
coordinates of the vertices of the successive lattices are calculated in a two
step process. First each new vertex is introduced to the centre of each old leg.
This vertex is then displaced along the radial direction out to the unit sphere.
In this way the $(\theta,\phi)$ coordinates of each vertex can be calculated.

The leg lengths for each leg are calculated by solving the geodesic equations
on the 2-sphere as a two point boundary value problem. At the same time we
compute $\int ds$ along this geodesic path. This gives us the leg lengths for
each geodesic segment. We then discard the continuum metric and turn to the
smooth lattice equations to estimate $R$.

\beginsubsection{Smooth lattice equations}
Consider the Riemann normal coordinate frame centred on vertex $O$.
Suppose there are $n$ triangles attached to this vertex and that the vertices,
starting with $O$, are labelled, $0$ to $n$.

We are free to choose our Riemann normal coordinates such that
$g_{\mu\nu}(x_o) = {\rm diag}(1,1),\ (x^\mu_o) = (0,0)$ and
$(x^\mu_1) = (\star,0)$ where $\star$ denotes a number to be computed from the
smooth lattice equations. This exhausts all coordinate freedoms, and thus all
of the remaining $x^\mu_i$ and curvature components must be computed from the
given leg lengths and the smooth lattice equations \eqnrfr{LegsEqtn}.
These equations were applied only to the legs of the triangles attached to
$O$. Note that in 2-dimensions, there is only one independent curvature
component, which we can take to be $R_{1212}$.

Since there are $n$ triangles attached to $O$, there will be $2n$ leg
lengths $L_{ij}$. There are also $n+1$ vertices for which there are $2(n+1)$
coordinates $x^\mu_i$ to compute. However, we have already chosen 3 of the
$2(n+1)$ coordinates. Thus we have to compute $2n-1$ coordinates and one
curvature component from the $2n$ leg lengths $L_{ij}$. Fortunately, we have as
many equations as unknowns. (In fact it is easy to see that this will always be
true in 2-dimensions, provided the surface is fully triangulated.)

The $2n$ equations \eqnrfr{LegsEqtn} were solved for the $x^\mu_i$ and
$R_{1212}$ via a Newton-Raphson method. Starting from flat space, the iterations
converged in about 3-4 iterations (though more iterations were required for the
very coarse approximations of the original tetrahedron).

The estimates so obtained are listed in Table \tblrfr{ApproxR}. Since not
every vertex in each approximation is equivalent to every other vertex (they
have differing local triangulations) the method returns different estimates for
$R$ for each vertex. Hence in the table we have listed the best and worst
estimates for $R$. One can observe that the method converges by a factor of four
with each successive sub-division. As the leg lengths are halved with each
sub-division this implies the error in $R$ varies as ${\cal O}(L^2)$ where $L$
is a typical length scale for leg lengths. That is, the smooth lattice yields
2nd-order accurate estimates for the curvature.

\bgroup

%
% two types of horizontal lines
%
\def\maindivider{\leaders\hrule height 1.5pt depth 0pt\hfil}
\def\medmdivider{\leaders\hrule height 1.0pt depth 0pt\hfil}

\def\Stroke{\vrule width 1.0pt}
\def\MyStroke{\omit\Stroke}
%
% a bit of vertical space
%

%
% table = \hbox { left hand edge, \vbox, right hand edge }
% the \vbox contains the body of the table
%
% \tabskip is the inter-column hglue. setting it in one column
% will only have effect in subsequent columns
%
% \offinterlineskip is used in conjunction with the \vrule entry
% in \halign to force each line in the table to be of a specific
% height and depth
%
% NOTE : At least one line in the \halign must have an entry for ALL
%        of the fields declared in the template (including the final
%        vrule entry, I do this by ending some lines with &\cr not \cr).
% NOTE : The integer parameter in the multispan equals the number of
%        columns in the table plus two.
%
{\offinterlineskip
\vskip 18pt plus 5pt minus 5pt
\centerline{\hfil\vrule width 0pt%
\vtop{\tabskip 0.0cm\halign{%
\vrule height 16pt depth 7pt width 0pt#\tabskip=0.5cm&%
\hfil{#}\hfil&#\vrule width 0pt&%
\hfil{#}\hfil&#\vrule width 0pt&%
\hfil{#}\hfil&#\vrule width 0pt&%
\hfil{#}\hfil&%
\vrule height 16pt depth 7pt width 0pt#\tabskip=0.0cm\cr
%--- end of template
\multispan{9}\maindivider\cr
&\multispan{7}
\hfil \bf Table \tbldef{ApproxR}. Estimates of $\vert R-2\vert$ for a unit
2-sphere\hfil&\cr
\multispan{9}\maindivider\cr
&\bf Sub-division&\MyStroke&Worst estimate&&Best estimate&&Average estimate&\cr
\multispan{9}\medmdivider\cr
&1&\MyStroke&1.19&&1.19&&1.19&\cr
&2&\MyStroke&2.99e-1&&4.91e-2&&1.71e-1&\cr
&3&\MyStroke&6.40e-2&&1.91e-3&&2.55e-2&\cr
&4&\MyStroke&1.54e-2&&5.67e-5&&2.97e-3&\cr
&5&\MyStroke&3.81e-3&&5.18e-6&&2.47e-4&\cr
\multispan{9}\maindivider\cr}}\vrule width 0pt\hfil}}

\egroup

\vskip\parskip

\beginsubsection{The Hamiltonian constraint}\secdef{Two}
Using the above values for $R$ we can now proceed to the construction of the
Schwarzschild 3-metric. Our starting point is to propose a 3-metric in the form
$$
ds^2 = dl^2 + \eta(l)^2\left(d\theta^2 + \sin^2\theta\, d\phi^2\right)
$$
Where $l$ is the radial proper distance measured from the throat. There is
only one non-trivial constraint equation, the Hamiltonian constraint,
$$ 0={\strut}^{(3)}R $$
which must be solved for $\eta(l)$. To begin, write the metric in the
2+1 form
$$\left(g_{\mu\nu}\right) =
\left(\matrix{1&0\cr
              0&\eta^2 h_{\mu\nu}\cr}\right)$$
where $h_{\mu\nu}$ is the metric of the unit 2-sphere. Then it is a
straightforward calculation to show that the Hamiltonian constraint is
$$
{d\over dl}\left({1\over\eta}{d\eta\over dl}\right) = {R\over 4\eta^2}
  - {3\over2}\left({1\over\eta}{d\eta\over dl}\right)^2
\eqno\eqndef{ZeroThreeR}
$$
where $R$ is the scalar curvature of the unit 2-sphere.

The boundary conditions at the throat, where $l=0$, are chosen to be
$$
\eta = 2\hskip2cm{\rm and}\hskip2cm {d\eta\over dl} = 0
\eqno\eqndef{BC}
$$
The first condition is equivalent to setting the ADM mass to $m=1$ (ie.
$\eta=2m$). The second condition is required for $l=0$ to be a minimal surface.

Note that in the above equation $R$ can take on any value not just the
discrete values found in the previous section. As we have already seen that the
smooth lattice method gives accurate and convergent estimates for $R$, we will
in the following discussions allow $R$ to take on any value in the range $2\leq
R\leq 3$.

For each choice of $R$ the initial value problem \eqnrfr{ZeroThreeR} was solved
using a 4-th order Runge Kutta method starting from the throat and integrating
outwards. A step length of $dl=0.2$ was used in each of the following
calculations.

The result of the integration is that we have the 3-metric in the form
$$
ds^2 = dl^2 + \eta^2(l)\, d\Omega^2
\eqno\eqndef{ApproxG}
$$
where $d\Omega^2$ is the metric of the unit 2-sphere. We would like to compare
this with the exact metric corresponding to $R=2$.  One easy way to do this
is to integrate \eqnrfr{ZeroThreeR} again but this time with $R=2$. This will
yield a metric of the form
$$
d{\tilde s}^2 = dl^2 + {\tilde\eta}^2(l)\, d\Omega^2
\eqno\eqndef{ExactG}
$$
The error can then be easily computed as
$$
e(l,R) = -1 + { \eta(l)\over {\tilde\eta}(l) }
\eqno\eqndef{FirstError}
$$
The results are shown in Figures (\figrfr{TwoDPlotA})
and (\figrfr{TwoDPlotB}). The first graph shows that for a fixed value of $R$
the error rises steeply from the throat and then settles to a constant value
independent of $l$. This is easy to understand, it shows that in the
distant almost flat regions of the metric the error in approximating a smooth
2-sphere with a smooth lattice does not depend on the size of the 2-sphere. This
graph also shows that for any fixed $l$ the error vanishes as $R\rightarrow2$ (as
it must). The second plot, Figure (\figrfr{TwoDPlotB}), is just a series of
cross-sections of the first plot, Figure (\figrfr{TwoDPlotA}), at specific
values of $l$. This graph clearly shows that the error in $\eta$, at a fixed
$l$, appears to vary linearly with the error in $R$. Since we have previously
established that the error in $R$ varies as ${\cal O}(L^2)$ we can infer that
the global discretization error in $\eta$ appears to be ${\cal O}(L^2)$.

It should be noted that the metrics generated for various values of $R$ are not
isometric to each other. To see this note that \eqnrfr{ZeroThreeR} is symmetric
under the transformation $R\rightarrow k^2R,\>\eta\rightarrow k\eta$ for any
$k$. Thus we can always scale $R$ to $R=2$. However the resulting re-scaling of
$\eta$ is, through the boundary condition \eqnrfr{BC}, equivalent to changing
the ADM mass. This proves the assertion. The upshot of this that there is no
single method to compare two metrics with differing values of $R$. The method
described above is just one of many possible ways to compare our approximate
metric against the exact metric. If one is not careful it is possible to
generate an apparently acceptable definition of the error which displays, for
fixed $R$, an error that diverges as $l\rightarrow\infty$. As an example,
suppose we wrote our exact metric in the form
$$
d{\tilde s}^2 = \rho^4(r)\left( dr^2 + r^2\, d\Omega^2\right)
$$
where $\rho(r) = 1+(m/2r)$. To compare this metric with our approximate metric
\eqnrfr{ApproxG} we could generate a transformation between the $r$ and $l$
coordinates by integrating, in parallel with the main equation 
\eqnrfr{ZeroThreeR},
$$
{dr\over dl} = {r\over \eta}
\eqno\eqndef{RandL}
$$
starting from $r=m/2$ at $l=0$. The error could then be defined as
$$
e(l,R) = -1 + {\eta(l) \over r(l) \rho(r(l))^2}
\eqno\eqndef{SecondError}
$$
The results are displayed in Figure (\figrfr{TwoDPlotC}) and clearly display the
stated divergence. However, at a fixed value of $l$ we find, see Figure
(\figrfr{TwoDPlotD}), that the error vanishes as $R\rightarrow2$, ie. at each
physical point the approximate metric converges to the exact metric.

Note that in this example all that we have changed is the way in which the two
metrics are compared. Why did the first method work so much better than the
second? The answer lies in the choice of the transformation \eqnrfr{RandL}. This
contains the function $\eta$ which is only an approximation to the exact
$\tilde\eta$. The result is that $r$ and $l$ are not properly aligned. What
this means is the following. To each metric we can compute the proper distance
from the throat to any point in question. For $ds$ the distance to the point
with coordinate $l$ is just $l$. For the exact metric $d\tilde s$ the distance
is $\tilde l(r) = \int_{m/2}^r\>\rho^2(u)\>du$ for the point with coordinate $r$.
The above transformation \eqnrfr{RandL} does not produce
$l = {\tilde l}(r)$. In fact, when both $l$ and $\tilde l$ are large,
$$
{d\eta\over dl} \sim \sqrt{R\over2}\ ,\hskip 2cm
{d\tilde\eta\over d\tilde l} \sim 1
$$
which when combined with the above \eqnrfr{RandL} and
$dr/d{\tilde l} = r/{\tilde\eta}$ leads to
$$
\tilde l \sim l^{\sqrt{2/R}}
$$
Thus $\tilde l - l$ diverges (for $R\neq2$) as $l\rightarrow\infty$. This is the
source of the apparent divergence in the metrics (for fixed $R$). In contrast,
the original definition of the error \eqnrfr{FirstError} does not suffer from
this problem because it compares the metric coefficients $\eta$ and
$\tilde\eta$ at the same proper distance from their respective throats.

An honest assessment of the above example is that even though it has given
the correct answers it must be viewed as a very benign test of the smooth
lattice method. The sole contribution of the smooth lattice was to aid in the
computation of the scalar 2-curvature, which was already known to be $2$. One
could probably concoct any number of schemes which spit out the magic number
$2$. A far better test would be to use the smooth lattice approach to fully
discretise the 3-metric. This brings us to our second example.

\beginsection{Fully discretised 3-metric}
Our aim in this example is to subject the smooth lattice approach to a much
more stringent test than that used in the previous example. We will base our
test on a fully discretised three dimensional lattice. Our plan of attack is as
follows. First, we will choose the structure of our lattice. Coordinate and
gauge conditions that reflect the desired spherical symmetries will then be
imposed after which the smooth lattice equations will be written out in
full. Finally we will argue that the Bianchi identities will need to be imposed
so as to produce a set of equations from which the correct metric can be
obtained.

Consider now the construction of a lattice for a spherically symmetric space.
An extreme example would be to choose a lattice in which the vertices have been
randomly scattered throughout the space. This is not only impractical but it
also fails to take advantage of the obvious symmetries of the space. We will
instead choose a lattice (see Figures
(\figrfr{ThreeLatticeA},\figrfr{ThreeLatticeB})) built from a single tube
stretching from the throat out to the distant flat regions of the space. The
tube has a rectangular cross-section and it is subdivided into a sequence of
cube like cells. Each cell stretches from one 2-sphere to the next and each
successive pair of cells defines the region of each RNC. The four edges of the
tube, not just the radial edges of the cells, will be required to be global
radial geodesics. It is important to note that the rectangular tiles such as
that defined by the vertices 1,2,3 and 4, do not lie in the 2-spheres (except
at the throat). The edges such as $(1,2)$ are geodesic segments of the full
3-metric and thus cannot also be geodesic segments of the 2-spheres.

Though we have only constructed one tube we will assume, on the basis of
spherical symmetry, that it captures all of the important geometric information
of the 3-metric. Thus there is no need to replicate this tube through out the
space.

Let us now turn to the issue of coordinate and gauge conditions. This
will entail arranging our RNC in each pair of cells, imposing restrictions on
the curvature components and imposing the gauge condition that the radial edges
of the tube are global geodesics.

We are free to choose an orthonormal RNC frame at the origin.
Thus $g_{\mu\nu} = {\rm diag}(1,1,1)$ at the origin. The Riemann normal
coordinates for the vertices in each pair of cells were chosen as per Table
\tblrfr{CoordsRect}. This choice can be achieved as follows. First align the
$z$-axis with the radial geodesic running up the centre of the tube. The origin
can then be slid up and down this geodesic to set the $z$ coordinates of
vertices 1,2,3 and 4 to zero. Finally, by a suitable rotation about the z-axis,
the remaining pattern amongst the $x$ and $y$ coordinates can be achieved.

\bgroup

\def\A#1{\hbox to 8mm{$\displaystyle\hfil#1$}}
\def\B#1{\hbox to 5mm{$\displaystyle\hfil#1$}}
\def\C#1{\hbox to 6mm{$\displaystyle\hfil#1$}}
\def\D#1{\hbox to 4mm{$\displaystyle\hfil#1$}}
\def\E#1{\hbox to 8mm{$\displaystyle\hfil#1$}}
\def\F#1{\hbox to 5mm{$\displaystyle\hfil#1$}}
%
% two types of horizontal lines
%
\def\maindivider{\leaders\hrule height 1.5pt depth 0pt\hfil}
\def\medmdivider{\leaders\hrule height 1.0pt depth 0pt\hfil}

\def\Stroke{\vrule width 1.0pt}
\def\MyStroke{\omit\Stroke}
%
% a bit of vertical space
%

%
% table = \hbox { left hand edge, \vbox, right hand edge }
% the \vbox contains the body of the table
%
% \tabskip is the inter-column hglue. setting it in one column
% will only have effect in subsequent columns
%
% \offinterlineskip is used in conjunction with the \vrule entry
% in \halign to force each line in the table to be of a specific
% height and depth
%
% NOTE : At least one line in the \halign must have an entry for ALL
%        of the fields declared in the template (including the final
%        vrule entry, I do this by ending some lines with &\cr not \cr).
% NOTE : The integer parameter in the multispan equals the number of
%        columns in the table plus two.
%
{\offinterlineskip
\vskip 18pt plus 5pt minus 5pt
\centerline{\hfil\vrule width 0pt%
\vtop{\tabskip 0.0cm\halign{%
\vrule height 16pt depth 7pt width 0pt#\tabskip=0.0cm&%
\hskip0.5cm\hfil{#}\hfil&#\vrule width 0pt&%
\hfil{#}\hfil\hskip0.5cm&#\vrule width 0pt&%
\hskip0.5cm\hfil{#}\hfil&#\vrule width 0pt&%
\hfil{#}\hfil\hskip0.5cm&#\vrule width 0pt&%
\hskip0.5cm\hfil{#}\hfil&#\vrule width 0pt&%
\hfil{#}\hfil\hskip0.5cm&%
\vrule height 16pt depth 7pt width 0pt#\tabskip=0.0cm\cr
%--- end of template
\multispan{13}\maindivider\cr
&\multispan{11}
\hfil \bf Table \tbldef{CoordsRect}. Coordinates for the vertices in
Figure \figrfr{ThreeLatticeB}\hfil&\cr
\multispan{13}\maindivider\cr
&\bf Vertex\ &&$(x^\mu)$&\MyStroke%
&\bf Vertex\ &&$(x^\mu)$&\MyStroke%
&\bf Vertex\ &&$(x^\mu)$&\cr
\multispan{13}\medmdivider\cr
&$1^\m$&&$(\A{a^\m},\A{a^\m},\B{b^\m})$&\MyStroke%
&$1$&&$(\C{a},\C{a},\D{0})$&\MyStroke%
&$1^\p$&&$(\E{a^\p},\E{a^\p},\F{b^\p})$&\cr
&$2^\m$&&$(\A{a^\m},\A{-a^\m},\B{b^\m})$&\MyStroke%
&$2$&&$(\C{a},\C{-a},\D{0})$&\MyStroke%
&$2^\p$&&$(\E{a^\p},\E{-a^\p},\F{b^\p})$&\cr
&$3^\m$&&$(\A{-a^\m},\A{-a^\m},\B{b^\m})$&\MyStroke%
&$3$&&$(\C{-a},\C{-a},\D{0})$&\MyStroke%
&$3^\p$&&$(\E{-a^\p},\E{-a^\p},\F{b^\p})$&\cr
&$4^\m$&&$(\A{-a^\m},\A{a^\m},\B{b^\m})$&\MyStroke%
&$4$&&$(\C{-a},\C{a},\D{0})$&\MyStroke%
&$4^\p$&&$(\E{-a^\p},\E{a^\p},\F{b^\p})$&\cr
\multispan{13}\maindivider\cr}}\vrule width 0pt\hfil}}

\egroup

\vskip\parskip

The spherical symmetry of the space must impose some restrictions on the
curvature components in the Riemann normal coordinates. Recall that the general
form of a RNC metric is
$$
g_{\mu\nu}(x) = g_{\mu\nu} - {1\over3}R_{\mu\alpha\nu\beta}\> x^\alpha x^\beta
              + \Oeps(3)
$$
The only parameters which we can play with are the $R_{\mu\alpha\nu\beta}$.
Thus to respect the required symmetries we must restrict these parameters
accordingly. To do this we revert for the moment to a generic spherically
symmetric metric
$$
ds^2 = dr^2 + \eta(r)^2\left(d\theta^2 + \sin^2\theta\, d\phi^2\right)
$$
where $\eta(r)$ is any smooth function and $r$ is the radial proper distance
measured from the throat (this was written as $l$ in the previous sections). In
these coordinates there are only two non-trivial frame components,
\setbox0=\vbox{\openup10pt%
\halign{$\displaystyle\hfil#\hfil$\cr
R_{\hat r\hat\theta\hat r\hat\theta} = R_{\hat r\hat\phi\hat r\hat\phi}
= -{1\over\eta} {d^2\eta\over dr^2}\cr
R_{\hat\theta\hat\phi\hat\theta\hat\phi} = 
{1\over\eta^2}\left( 1 - \left({d\eta\over dr}\right)^2 \right)\cr
}}
$$
\vcenter{\box0}
\eqno\eqndef{FrameRiem}
$$
where, for example, $R_{\hat\theta\hat\phi\hat\theta\hat\phi} 
= R_{\mu\nu\alpha\beta}\>
e{}^\mu_\theta e{}^\nu_\phi e{}^\alpha_\theta e{}^\beta_\phi$.
Thus we expect only two non-trivial curvature components in the RNC frame. If
we align the $r$ axis with the $z$ axis then it is easy to
see that the corresponding non-trivial RNC components must be $R_{xyxy}$ and
$R_{xzxz} = R_{yzyz}$. Thus, the RNC metric can be reduced to just
$$
ds^2 = dx^2 + dy^2 + dz^2
-{1\over3} R_x (xdy - ydx)^2
-{1\over3} R_z (xdz - zdx)^2
-{1\over3} R_z (ydz - zdy)^2
$$
where $R_x = R_{xyxy}$ and $R_z = R_{xzxz} = R_{yzyz}$. Incidently, for any pair
of vectors $u^\mu$ and $v^\mu$ we have
$$
R_{\mu\alpha\nu\beta}\> u^\mu u^\nu v^\alpha v^\beta
 = R_x (v^x u^y - v^y u^x)^2
 + R_z (v^x u^z - v^z u^x)^2
 + R_z (v^y u^z - v^z u^y)^2
$$
Using this metric and the above choice of coordinates we obtain, from the smooth
lattice equations,
$$\eqalignno{%
0&= 3L^2 - 12 a^2 + 4 R_x a^4
&\eqndef{EqtnA}\cr
0&= 3(L^\m)^2 - 12 (a^\m)^2 + 4 R_x(a^\m)^4 + 4 R_z (a^\m b^\m)^2
&\eqndef{EqtnB}\cr
0&= 3(L^\p)^2 - 12 (a^\p)^2 + 4 R_x(a^\p)^4 + 4 R_z (a^\p b^\p)^2
&\eqndef{EqtnC}\cr
0&= 3(d^\m)^2 - 6(a-a^\m)^2 - 3(b^\m)^2 + 2 R_z (ab^\m)^2
&\eqndef{EqtnD}\cr
0&= 3(d^\p)^2 - 6(a-a^\p)^2 - 3(b^\p)^2 + 2 R_z (ab^\p)^2
&\eqndef{EqtnE}\cr
0&= L^2\cos\beta + 4 R_x a^4
&\eqndef{EqtnF}\cr
3Ld^\m\cos\alpha^\m &= 6a(a-a^\m) - 4 R_x a^3(a-a^\m) - 2 R_z (ab^\m)^2
\cr
3Ld^\p\cos\alpha^\p &= 6a(a-a^\p) - 4 R_x a^3(a-a^\p) - 2 R_z (ab^\p)^2
\cr}
$$
We have previously stated that the radial edges of our tube must be global
geodesics. Thus there can be no kink in the incoming and outgoing edges
at vertices 1,2,3 and 4. This implies that
$$\alpha^\p + \alpha^\m = \pi$$
which using the above leads to
$$\eqalign{%
0 =&\phantom{+}\>
     d^\p\left( 6a(a-a^\m) - 4 R_x a^3(a-a^\m) - 2 R_z (ab^\m)^2\right)\cr
   &\phantom{+}\llap{+}\>
     d^\m\left( 6a(a-a^\p) - 4 R_x a^3(a-a^\p) - 2 R_z (ab^\p)^2\right)\cr}
\eqno\eqndef{EqtnK}
$$
There are seven quantities that we must compute $a,a^\p,a^\m,b^\p,b^\m,R_x$ and
$R_z$. We thus need seven equations which we take to be equations
(\eqnEqtnA--\eqnEqtnF) and (\eqnEqtnK). Though
this may seem like a well defined system with seven equations for seven unknowns
there is a serious problem. Suppose we were to freely choose all of the
$d^\p,d^\m,L^\p,L,L^\m$ and $\cos\beta$. We could then solve the lattice
equations and thus obtain the curvatures in each RNC frame along the tube. Since
this is a well defined set of equations we see that this is equivalent to having
an arbitrarily specified set of curvatures along the tube. However, in the
continuum metric the two curvatures are derived from one function and therefore
they cannot be freely specified. Indeed from \eqnrfr{FrameRiem} we see that
$$
0 = {d\over dr}\left(\eta^2 R_{\hat x}\right)
  - \left({d\eta^2\over dr}\right) R_{\hat z}
\eqno\eqndef{FirstBianchi}
$$
where $R_{\hat x} = R_{\hat\theta\hat\phi\hat\theta\hat\phi}$ and
$R_{\hat z} = R_{\hat r \hat\theta\hat r \hat\theta}$. This is nothing other
than the standard Bianchi identity for the continuum metric. We can expect that
a similar but discrete version of this equation must also exist in our smooth
lattice model.

\beginsubsection{Bianchi Identities}
It is well known \Ref(6) that a necessary and sufficient condition for the
existence of a metric $g_{\mu\nu}(x)$ from which a given curvature tensor
$R_{\mu\nu\alpha\beta}$ can be derived is just the Bianchi identities
$$0 = R_{\mu\nu\alpha\beta;\rho}
     +R_{\mu\nu\beta\rho;\alpha}
     +R_{\mu\nu\rho\alpha;\beta}
$$
In each of our RNC frames this can be written as
$$0 = R_{\mu\nu\alpha\beta,\rho}
     +R_{\mu\nu\beta\rho,\alpha}
     +R_{\mu\nu\rho\alpha,\beta}
$$
since $\Gamma^\mu_{\alpha\beta}(x) = \Oeps(2)$ throughout each RNC (in the
conformal metric, see the Appendix). Unfortunately we cannot make much use of
this equation as it stands. The problem is that the inversion of the smooth
lattice equations returns the curvatures but not their derivatives. One might
object by arguing that for the RNC metric we can evaluate the curvature and its
derivatives to any order. This may be so but such calculations cannot be
expected to be accurate estimates of those quantities for the smooth metric
(that the lattice is interpolating). The same situation arises in the much
simpler case of a piecewise quadratic interpolation of a function of one
variable. The third derivatives of each interpolant is zero yet it is most
unlikely that the function has a zero third derivative at each point. The normal
practice is to use the second derivative of the interpolant as an estimate at
just one point, usually at the central point. Estimates of higher derivatives
can then be obtained by interpolation of this derived data.

This same philosophy seems appropriate for our lattice. The estimates for the
curvatures are valid at just one point, the origin of the RNC. Derivatives of
the curvatures should be estimated by differentiation of a local interpolation
of the point estimates of the curvatures. There is one slight subtlety here --
the curvatures are defined in different RNC frames. This will require some
coordinate transformations to bring the local curvatures into one common frame
before the interpolation and differentiation can be performed. This is actually
not very difficult. Suppose for example that the origins of two neighbouring
frames are $O$ and $O'$. Suppose that the associated metric and curvatures are
$g_{\mu\nu},R_{\mu\nu\alpha\beta}$ and $g_{\mu'\nu'},R_{\mu'\nu'\alpha'\beta'}$
respectively. We seek the representation of $R_{\mu'\nu'\alpha'\beta'}$ in the
RNC at $O$. To do this start with the set of coordinate basis vectors at $O'$
and parallel transport them to $O$. They will be related to the basis vectors at
$O$ by some transformation matrix $\Lambda^{\mu'}{}_\nu$. Then the transformed
value of the curvature will be $R_{\mu'\nu'\alpha'\beta'}
\Lambda^{\mu'}{}_\mu\Lambda^{\nu'}{}_\nu
\Lambda^{\alpha'}{}_\alpha\Lambda^{\beta'}{}_\beta$.
For this calculation we can, to leading order in $\eps$, evaluate
$\Lambda^{\mu'}{}_\nu$ solely from the flat space parts of the metrics
$g_{\mu\nu}$ and $g_{\mu'\nu'}$.

Though the procedure just described may be suitable for a generic lattice there
is a simpler approach for our symmetric lattice. The idea is to use the
equivalent integral representation of the Bianchi identities. In three dimensions
this happens to be
$$\eqalign{%
0 &= \int_M\>
 \left( R_{\mu\nu\alpha\beta,\rho}
       +R_{\mu\nu\beta\rho,\alpha}
       +R_{\mu\nu\rho\alpha,\beta} \right) \> dx^\alpha dx^\beta dx^\rho\cr
  &= \int_{\partial M}\>
      R_{\mu\nu\alpha\beta} \> dx^\alpha dx^\beta\cr}
$$
where $M$ is an arbitrary three dimensional region with boundary $\partial M$.
To leading order in
$\eps$ the limits of integration can be set as if $M$ was everywhere flat.
This later form is much easy to apply since it does not require estimates of the
derivatives of the curvatures. For our lattice we will choose $M$ to be the
pair of cells that define the typical RNC frame. Though this consists of two
cells and thus ten boundary faces we can treat them, to leading order in
$\eps$, as just one larger cell (ie. a linear cell, with six faces, that expands
from $L^\m$ to $L^\p$ over a length $d^\m+d^\p$). The above equation can then be
written as
$$\eqalign{%
0 =&\phantom{+}\>
    \int_{S^\m_x} \> R_{\mu\nu\alpha\beta} \> dx^\alpha dx^\beta
   +\int_{S^\p_x} \> R_{\mu\nu\alpha\beta} \> dx^\alpha dx^\beta\cr
   &\phantom{+}\llap{+}\>
    \int_{S^\m_y} \> R_{\mu\nu\alpha\beta} \> dx^\alpha dx^\beta
   +\int_{S^\p_y} \> R_{\mu\nu\alpha\beta} \> dx^\alpha dx^\beta\cr
   &\phantom{+}\llap{+}\>
    \int_{S^\m_z} \> R_{\mu\nu\alpha\beta} \> dx^\alpha dx^\beta
   +\int_{S^\p_z} \> R_{\mu\nu\alpha\beta} \> dx^\alpha dx^\beta\cr}
$$
where $S^\m_x,S^\p_x$ are the faces of the cells on the $+x$ and $-x$
axes respectively (with similar definitions for the remaining four faces). We
will now compute the curvatures on each of these faces in this RNC frame. This
information can be obtained by an interpolation and coordinate transformation of
the curvatures from neighbouring cells.

At this point, it is appropriate to represent all of the curvature tensors in
bivector form. Let  $U_{\mu\nu},V_{\mu\nu}$ and $W_{\mu\nu}$ be an orthonormal
set of normalised bivectors ($1=U_{\mu\nu}U^{\mu\nu}$, $0=U_{\mu\nu}V^{\mu\nu}$
etc.) associated with the $xy,xz$ and $yz$ planes respectively. Then the
curvature tensor at $O$ is
$$
R_{\mu\nu\alpha\beta} =  R_x U_{\mu\nu}U_{\alpha\beta}
                       + R_z V_{\mu\nu}V_{\alpha\beta}
                       + R_z W_{\mu\nu}W_{\alpha\beta}
$$
while at $O^\p_z$ we would have, in the coordinates for $O^\p_z$,
$$
R^\p_{\mu\nu\alpha\beta} =  R^\p_x U^\p_{\mu\nu}U^\p_{\alpha\beta}
                          + R^\p_z V^\p_{\mu\nu}V^\p_{\alpha\beta}
                          + R^\p_z W^\p_{\mu\nu}W^\p_{\alpha\beta}
$$
However, it is clear that, to first order in $\eps$, the transformation matrix
$\Lambda^{\mu'}{}_\nu$ for these two frames is just the identity matrix. Thus 
the linear interpolant for the curvature, along the $z$ axis between $O$ and
$O^\p_z$ is
$$\eqalign{%
R_{\mu\nu\alpha\beta}(d) =&
    \left({d\over d^\p}\right)R^\p_{\mu\nu\alpha\beta}
  + \left(1-{d\over d^\p}\right)R_{\mu\nu\alpha\beta}\cr
=&\phantom{+}\>
    \left( \left({d\over d^\p}  \right)R^\p_x 
         + \left(1-{d\over d^\p}\right)R_x \right) U_{\mu\nu}U_{\alpha\beta}\cr
 &\phantom{+}\llap{+}\>
    \left( \left({d\over d^\p}  \right)R^\p_z 
         + \left(1-{d\over d^\p}\right)R_z \right) V_{\mu\nu}V_{\alpha\beta}\cr
 &\phantom{+}\llap{+}\>
    \left( \left({d\over d^\p}  \right)R^\p_z 
         + \left(1-{d\over d^\p}\right)R_z \right) W_{\mu\nu}W_{\alpha\beta}\cr}
$$
where $d$ is the proper distance measured along the $z$ axis from $z=0$. For
the face $S^\p_z$ we have $d=d^\p$.

For the interpolation between $O$ and $O^\p_x$, the procedure is much
the same with the exception that the transformation matrix
$\Lambda^{\mu'}{}_\nu$ is now a rotation. Indeed it is easy to see that the
bivectors in the two frames are related by
$$\eqalign{%
U^x_{\mu\nu} &= U_{\mu\nu}\cos\Delta\rho + W_{\mu\nu}\sin\Delta\rho\cr
V^x_{\mu\nu} &= V_{\mu\nu}\cr
W^x_{\mu\nu} &= W_{\mu\nu}\cos\Delta\rho - U_{\mu\nu}\sin\Delta\rho\cr}
$$
where $2\sin(\Delta\rho/2) = (L^\p-L^\m)/(d^\p+d^\m)$. The interpolation can then
be obtained by generalising these equations to be active transformations of the
original bivectors $U,V$ and $W$ into a moving set of bivectors, namely
$$
\eqalign{%
U^x_{\mu\nu}(\rho) &= U_{\mu\nu}\cos\rho + W_{\mu\nu}\sin\rho\cr
V^x_{\mu\nu}(\rho) &= V_{\mu\nu}\cr
W^x_{\mu\nu}(\rho) &= W_{\mu\nu}\cos\rho - U_{\mu\nu}\sin\rho\cr}
$$
leading to
$$
R_{\mu\nu\alpha\beta}(\rho) = R_x U^x_{\mu\nu}(\rho)U^x_{\alpha\beta}(\rho)
                            + R_z V^x_{\mu\nu}(\rho)V^x_{\alpha\beta}(\rho)
                            + R_z W^x_{\mu\nu}(\rho)W^x_{\alpha\beta}(\rho)
$$
The interpolation to the face $S^\p_x$ can now be obtained by setting
$\rho=\Delta\rho/2$. Note that we can use the approximation
$\Delta\rho/2 \approx \sin(\Delta\rho/2)$ since the four radial geodesic edges of the tube
must be close in order for the smooth lattice to be a good approximation to the
smooth continuum metric.

Similar interpolations can be constructed for the remaining faces. These can
then be substituted into the Bianchi identity with the result, assuming each
$R_{\mu\nu\alpha\beta}$ is constant on each face,
$$\eqalign{%
0 =&\phantom{+}\>
    \left( R^\p_x A^\p_z - R^\m_x A^\m_z \right) U_{\mu\nu}\cr
   &\phantom{+}\llap{+}\>
       R_z A^\p_x 
       \Big(  V^y_{\mu\nu}(+\Delta\rho/2) 
            + W^x_{\mu\nu}(+\Delta\rho/2)  \Big)
     - R_z A^\m_x 
       \Big(  V^y_{\mu\nu}(-\Delta\rho/2)
            + W^x_{\mu\nu}(-\Delta\rho/2) \Big)\cr}
$$
where $A^\p_z,A^\m_z$ and $A^\p_x,A^\m_x$ are the areas of $S^\p_z,S^\m_z$ and
$S^\p_x$ respectively. The areas are easily seen to be
$$\eqalign{%
A^\p_z &=\left( L^\p \right)^2\cr
A^\m_z &=\left( L^\m \right)^2\cr
A^\p_x = A^\m_x &={1\over2}
                   \left( L^\p + L^\m \right)\left( d^\p + d^\m\right)\cr}
$$
Contracting the above equation with $U_{\mu\nu}$ and using
$\Delta\rho = (L^\p-L^\m)/(d^\p+d^\m)$ leads directly to
$$
0 = \left(L^2 R_x\right)^\p - \left(L^2 R_x\right)^\m
  - \left(L^\p + L^\m\right)\left(L^\p-L^\m\right) R_z
$$
This is our discrete version of the Bianchi identity. It clearly resembles the
previous Bianchi identity \eqnrfr{FirstBianchi} when expressed in finite
difference form. Note that contractions with the other bivectors, $V$ and $W$,
leads only to trivial equations.

\beginsubsection{The smooth lattice equations}
Its been a rather long road but we can now write down all of the smooth lattice
equations, namely,
$$\eqalignno{%
0 = & 12 a^2 - 4 R_x a^4 - 3L^2
&\eqndef{BigA}\cr
0 = & 12 (a^\m)^2 - 4 R_x(a^\m)^4 - 4 R_z (a^\m b^\m)^2 - 3(L^\m)^2
&\eqndef{BigB}\cr
0 = & 12 (a^\p)^2 - 4 R_x(a^\p)^4 - 4 R_z (a^\p b^\p)^2 - 3(L^\p)^2
&\eqndef{BigC}\cr
0 = & 6(a-a^\m)^2 + 3(b^\m)^2 - 2 R_z (ab^\m)^2 - 3(d^\m)^2
&\eqndef{BigD}\cr
0 = & 6(a-a^\p)^2 + 3(b^\p)^2 - 2 R_z (ab^\p)^2 - 3(d^\p)^2
&\eqndef{BigE}\cr
0 = & 4 R_x a^4 + L^2\cos\beta
&\eqndef{BigF}\cr
0 = &\phantom{+}\>
     d^\p\left( 6a(a-a^\m) - 4 R_x a^3(a-a^\m) - 2 R_z (ab^\m)^2\right)\cr
    &\phantom{+}\llap{+}\>
     d^\m\left( 6a(a-a^\p) - 4 R_x a^3(a-a^\p) - 2 R_z (ab^\p)^2\right)
&\eqndef{BigG}\cr
0 = & R_x + 2R_z
&\eqndef{BigH}\cr
0 = & \left(L^2 R_x\right)^\p - \left(L^2 R_x\right)^\m
  -   \left(L^\p + L^\m\right)\left(L^\p-L^\m\right) R_z
&\eqndef{BigI}\cr
}
$$
The second last equation is just the Hamiltonian constraint ${}^{(3)}R=0$.

\beginsubsection{Solution strategy}
The smooth lattice was constructed by starting from the throat and successively
adding cells to the end of the tube. There are two parts to this scheme, first
the choice of initial data at the throat and second, the repeated solution of
the smooth lattice equations.

Consider first the generic step in which one extra cell is added to the tube.
Suppose we are given values for $R^\m_x$ and $R_x$ (from previous
calculations). Then we can solve the nine equations (\eqnBigA--\eqnBigI) for the
nine quantities $a,a^\p,a^\m,b^\p,b^\m,R_z,R^\p_x,(L^\p)^2$ and $\cos\beta$.
Then the $R^\m_x$ and $R_x$ of the next RNC frame can be taken to be $R_x$ and
$R^\p_x$ of the current RNC frame (the transformation matrix between the
pair of frames is just the identity matrix, to first order in $\eps$). This
process can then be repeated as many times as might be required.

All of the equations were solved using a Newton-Raphson method with initial
guesses chosen to be flat space. That is, given $d^\m,d^\p,L^\m$ and $L$, all of
the parameters were estimated using $R_x=R_z=0$. This is very easy to do,
leading to
$$\openup10pt\displaylines{%
\cos\beta = 0\cr
L^\p = L + \left({d^\p\over d^\m}\right)\left(L - L^\m\right)\cr
a ={1\over2}L\>,\hskip 1cm
a^\m = {1\over2}L^\m\>,\hskip 1cm
a^\p = {1\over2}L^\p\cr
b^\m = -\left( \left(d^\m\right)^2 - 
                 {1\over2}\left( L^\m - L \right)^2\right)^{(1/2)}\cr
b^\p = +\left( \left(d^\p\right)^2 - 
                 {1\over2}\left( L^\p - L \right)^2 \right)^{(1/2)}\cr
}
$$
At the throat we need to make a few minor alterations to the above. Our
boundary condition is that the throat must be a minimal surface. This constraint
can be imposed via a standard reflection symmetric condition. That is, for the
RNC centred on a point on the throat we demand that cells on either side of the
throat be mirror images of each other. This leads to
$$\eqalignno{%
0 &= a^\p - a^\m
&(\eqnBigB')\cr
0 &= b^\p + b^\m
&(\eqnBigD')\cr
\noalign{and}
L^\p &= L^\m\cr
d^\p &= d^\m}
$$
This however leads immediately to a problem with the Bianchi identity at the
throat in that it cannot be solved for a unique value for $R^\p_x$. There are
two solutions to this problem. First, one can set
$$
0 = R^\p_x - R_x
\eqno\eqndef{AltBianchi}
$$
This follows from the observation that in the continuum limit, as
$d^\p\rightarrow0$ and $d^\m\rightarrow0$, the Bianchi identity can be reduced to
$$
0 = {d\over dl}\left(L^2 R_x\right) - \left({dL^2\over dl}\right) R_z
$$
which, with $0=dL/dl$ at the throat, leads to $0=dR_x/dl$ and hence the estimate
$R^\p_x=R_x$.

The second solution is to combine the above the differential equation with the
constraint $0=R_x+2R_z$. This leads directly to
$$
0 = {d\over dl}\left(L^3 R_x\right)
$$
and thus
$$
0 = \left(L^3 R_x\right)^\p - \left(L^3 R_x\right) 
\eqno\eqndef{SecondBianchi}
$$
Though this is valid along the whole tube we only apply it at the throat so as
not to compromise the independence of the smooth lattice equations (we want the
method to stand on its own and not to be supported or dragged along by results
from the continuum).

At the throat we replaced the equations (\eqnBigB,\eqnBigD) and (\eqnBigI) with
the equations $(\eqnBigB'$,\penalty-250$\eqnBigD')$ and (\eqnSecondBianchi). The
nine equations were then solved for the same nine quantities using as initial
guesses
$$\openup10pt\displaylines{%
\cos\beta = 0\cr
L^\m = L^\p = L\cr
a^\m = a = a^\p = {1\over2}L\cr
-b^\m = b^\p = d^\p\cr
R_x = R_z = 0\cr
}
$$
Using these initial guesses, the Newton-Raphson method for both the
generic and initial problems converged in less then five iterations.

Now let us return to the first point raised at the beginning of this section,
namely, the choice of initial data $L,L^\p$ and $R_x$ on the throat.
Collectively this corresponds to setting the ADM mass and the fraction of a
2-sphere covered by each tile (such as $(1,2,3,4)$ in Figure
(\figrfr{ThreeLatticeB})). These could be chosen randomly but instead we chose
to compute them from the known 3-metric so that we could explore the convergence
properties of the smooth lattice solutions. Indeed the easiest way to compare
the smooth lattice solutions against the exact 3-metric is to mimic the
construction of the lattice using the exact 3-metric. Thus we construct a series
of concentric 2-spheres and four radial geodesics. The vertices of the lattice
arise as the intersections of the radial geodesics with the 2-spheres. All of
the leg lengths, angles and curvatures can then be computed and compared with
those from the smooth lattice.

The exact metric in isotropic coordinates is
$$
ds^2 = \rho(r)^4
       \left(dr^2 + r^2(d\theta^2 + \sin^2\theta\, d\phi^2)\right)
\eqno\eqndef{ExactMetric}
$$
where $\rho(r) = 1+m/(2r)$. For all of our calculations we chose
$m=1$.

We can chose the coordinates of the vertices according to the following table.

\bgroup

\def\A#1{#1}
\def\B#1{#1}
\def\C#1{\hbox to 9mm{$\displaystyle\hfil#1$}}
%
% two types of horizontal lines
%
\def\maindivider{\leaders\hrule height 1.5pt depth 0pt\hfil}
\def\medmdivider{\leaders\hrule height 1.0pt depth 0pt\hfil}

\def\Stroke{\vrule width 1.0pt}
\def\MyStroke{\omit\Stroke}
%
% a bit of vertical space
%

%
% table = \hbox { left hand edge, \vbox, right hand edge }
% the \vbox contains the body of the table
%
% \tabskip is the inter-column hglue. setting it in one column
% will only have effect in subsequent columns
%
% \offinterlineskip is used in conjunction with the \vrule entry
% in \halign to force each line in the table to be of a specific
% height and depth
%
% NOTE : At least one line in the \halign must have an entry for ALL
%        of the fields declared in the template (including the final
%        vrule entry, I do this by ending some lines with &\cr not \cr).
% NOTE : The integer parameter in the multispan equals the number of
%        columns in the table plus two.
%
{\offinterlineskip
\vskip 18pt plus 5pt minus 5pt
\centerline{\hfil\vrule width 0pt%
\vtop{\tabskip 0.0cm\halign{%
\vrule height 16pt depth 7pt width 0pt#\tabskip=0.0cm&%
\hskip0.5cm\hfil{#}\hfil&#\vrule width 0pt&%
\hfil{#}\hfil\hskip0.5cm&#\vrule width 0pt&%
\hskip0.5cm\hfil{#}\hfil&#\vrule width 0pt&%
\hfil{#}\hfil\hskip0.5cm&#\vrule width 0pt&%
\hskip0.5cm\hfil{#}\hfil&#\vrule width 0pt&%
\hfil{#}\hfil\hskip0.5cm&%
\vrule height 16pt depth 7pt width 0pt#\tabskip=0.0cm\cr
%--- end of template
\multispan{13}\maindivider\cr
&\multispan{11}
\hfil \bf Table \tbldef{CoordsPolar}. Coordinates for the vertices of the exact
lattice.\hfil&\cr
\multispan{13}\maindivider\cr
&\bf Vertex\ &&$(r,\theta,\phi)$&\MyStroke%
&\bf Vertex\ &&$(r,\theta,\phi)$&\MyStroke%
&\bf Vertex\ &&$(r,\theta,\phi)$&\cr
\multispan{13}\medmdivider\cr
&$1^\m$&&$(\A{r^\m},\B{\Delta\theta},\C{0})$&\MyStroke%
&$1$&&$(\A{r},\B{\Delta\theta},\C{0})$&\MyStroke%
&$1^\p$&&$(\A{r^\p},\B{\Delta\theta},\C{0})$&\cr
&$2^\m$&&$(\A{r^\m},\B{\Delta\theta},\C{\pi/2})$&\MyStroke%
&$2$&&$(\A{r},\B{\Delta\theta},\C{\pi/2})$&\MyStroke%
&$2^\p$&&$(\A{r^\p},\B{\Delta\theta},\C{\pi/2})$&\cr
&$3^\m$&&$(\A{r^\m},\B{\Delta\theta},\C{\pi})$&\MyStroke%
&$3$&&$(\A{r},\B{\Delta\theta},\C{\pi})$&\MyStroke%
&$3^\p$&&$(\A{r^\p},\B{\Delta\theta},\C{\pi})$&\cr
&$4^\m$&&$(\A{r^\m},\B{\Delta\theta},\C{3\pi/2})$&\MyStroke%
&$4$&&$(\A{r},\B{\Delta\theta},\C{3\pi/2})$&\MyStroke%
&$4^\p$&&$(\A{r^\p},\B{\Delta\theta},\C{3\pi/2})$&\cr
\multispan{13}\maindivider\cr}}\vrule width 0pt\hfil}}

\egroup

\vskip\parskip

The size of the patch on the initial 2-sphere is determined by the value of
$\Delta\theta$. We chose $\Delta\theta = \eps(\pi/40)$ where $\eps$ is a scale
parameter in the range $1/256\leq\eps\leq2$. The effective number of tiles on
each 2-sphere can be estimated as $N=4\pi/(\Delta\theta)^2$ which for the chosen
range, corresponds to $509<N<134\times10^6$ tiles.

The values of $r^\p$ are obtained by solving, via a
Newton-Raphson method, the equation
$$
\eqalignno{%
d^\p &= \int_r^{r^\p}\>\left(1+{m\over 2u}\right)^2\>du\cr
     &= F(r^\p) - F(r)
&\eqndef{RadiusEqtn}\cr}
$$
where $F(u) = u + m\log(2u/m) - m^2/(4u)$. On the throat we chose $r=m/2$ and
$r^\m$ so that $d^\p = d^\m = F(r) - F(r^\m)$.

The exact leg lengths, which we will write as $L^\m_\star,L_\star$ and
$L^\p_\star$, require a little bit of work to compute. They are the lengths of
the geodesic segments joining the various vertices. This entails the solution
of a two point boundary value problem
\setbox0=\vbox{\openup10pt%
\halign{$\displaystyle\hfil#\hfil$\cr
0 = {d^2x^\mu\over d\lambda^2} + \Gamma^\mu_{\alpha\beta}(x)
                                 {dx^\alpha\over d\lambda}
                                 {dx^\beta\over d\lambda}\cr
x^\mu(0) = x^\mu_a\>,\hskip1.5cm x^\mu(1) = x^\mu_b\cr}}
$$
\vcenter{\box0}
\eqno\eqndef{TwoPointBVP}
$$
for the geodesics $x^\mu(\lambda)$ for which we used a simple
shooting method built on a 4-th order Runge-Kutta routine. The lengths were then
computed by evaluating $\int ds$ along the geodesic. This was done by
adding one extra differential equation to the Runge-Kutta routine, namely,
$${dL\over d\lambda} = \left( g_{\mu\nu}(x) 
                       {dx^\alpha\over d\lambda}
                       {dx^\beta\over d\lambda}\right)^{(1/2)}
\eqno\eqndef{GeodesicLen}
$$
In solving the above two point boundary value problem we used isotropic $x,y,z$
coordinates so as to avoid the coordinate singularity at $\theta=0$. Though we
did not need to integrate through this singularity its presence degrades the
quality of our estimates for the geodesics and the leg lengths.

The frame components of the curvature are easily computed from the above metric
\eqnrfr{ExactMetric} with the result that
$$\eqalignno{%
(R_x)_\star &\equiv  R_{\hat\alpha\hat\beta\hat\alpha\hat\beta}
                 = {2m\over r^3\rho(r)^6}
&\eqndef{ExactRx}\cr
(R_z)_\star &\equiv  R_{\hat r\hat\alpha\hat r\hat\alpha}
                  =  R_{\hat r\hat\beta\hat r\hat\beta}
                  =  {-m\over r^3\rho(r)^6}
&\eqndef{ExactRz}\cr}
$$
where $\rho(r) = 1+m/(2r)$.

The above equations (\eqnRadiusEqtn--\eqnExactRx) were used to compute initial
values for $L_\star,L^\p_\star$ and $(R_x)_\star$. These were then assigned to
$L,L^\p$ and $R_x$. The radial lengths leg lengths $d^\p$ were chosen to be
$d^\p = \gamma L$ where $\gamma$ is a fixed constant. This choice ensures that
the cells do not become too fat nor too elongated as successive cells are added
to the tube. This scheme provides for an adjustable step length, short steps
near the throat where the curvatures are large but longer steps away from the
throat where the curvatures are weak. In all of our calculations we set
$\gamma=0.1$ which we arrived at by direct experimentation (this point will be
discussed further in the following section).

\beginsubsection{Results}
The smooth lattice equations (\eqnBigA--\eqnBigI) were
repeatedly solved to generate data out to $l=50$. As each new cell of the
lattice was constructed, the equation \eqnrfr{RadiusEqtn} was solved so that
the smooth lattice quantities $L,R_x$ and $R_z$ could be compared with the exact
values $L_\star,(R_x)_\star$ and $(R_z)_\star$.

The errors between the exact and smooth lattices depend upon the distance along
the edge of the tube measured from the throat $l=\sum d^\p$ and upon the choice
of $\Delta\theta$. The fractional errors were computed as
$$\eqalign{%
EL(l,\eps) & = -1 + {L\over L_\star}\cr
ER_x(l,\eps) & = -1 + {R_x\over (R_x)_\star}\cr
ER_z(l,\eps) & = -1 + {R_z\over (R_z)_\star}\cr}
$$
where $\eps = (40/\pi)\Delta\theta$ is the scale parameter.
These errors are displayed in Figure (\figrfr{ThreeDPlotA})
and clearly show that the errors vanish,
at fixed $l$, as $\Delta\theta\rightarrow0$. However, for a fixed $\Delta\theta$,
the errors appear to grow without bound for increasing values of $l$. This is
much like the behaviour we encountered in section (\secrfr{Two}) (though much
less pronounced). We can infer
from that section that the divergence in the errors in this example may also be due
to an ambiguity in correlating the radial positions between the numerical and
exact metrics. For our smooth lattice there are at least two ways to measure the
radial proper distance. One can measure it along the outer edges of the tube, as
we have done, or along the central geodesic (ie. along the $z$ axis). It is an
arbitrary choice as to which of these is to be used to define a radial position
in the exact metric. Both choices agree at the throat but progressively diverge
as we move away from the throat (because the tube is always increasing in cross
section).

We have also estimated the rate of convergence by plotting the errors
as function of $\eps$ for a series of fixed values of $l$. These are
displayed in Figure (\figrfr{ThreeDPlotC}) and clearly show
that the convergence of each of the quantities, $L,R_x,R_z$ to their continuum
values is quadratic in the lattice spacing. The wiggles in the lower left corner
are probably due to round-off errors becoming significant.

\beginsubsection{Variations}
Three variants of the above smooth lattice method were tried. First we repeated
the above calculations but this time using equation \eqnrfr{AltBianchi} instead
of \eqnrfr{SecondBianchi} as our boundary condition on the throat. The results
are plotted in Figures (\figrfr{ThreeDPlotE},\figrfr{ThreeDPlotG}) and
the only significant difference is the appearance of an
instability in the errors of the curvatures. The amplitude and wavelength of
the oscillations in the curvatures were both found to decrease with decreasing
$\gamma$. For $\gamma\approx2$ the errors peaked at about $40$ (very large
indeed!) with only about two cycles occurring. By trial and error we judged
$\gamma=0.1$ to give the most pleasing results (to the eye). Note also that 
in the regions where the {\sl relative errors} $ER_x$ are large and oscillatory,
the {\sl absolute errors} $R_x - (R_x)_\star$ are of order $10^{-5}$. Thus this
instability is not a cause for great concern. Indeed it appears to have little
effect on the errors in the leg lengths (compare Figures \figrfr{ThreeDPlotA} and
\figrfr{ThreeDPlotE}).

For our second variation we took advantage of the integrated form of the Bianchi
identity \eqnrfr{SecondBianchi} to replace the smooth lattice equation
\eqnrfr{BigI} with the equation
$$
k = L^3 R_x
\eqno(\eqnBigI')
$$
with $k$ a constant computed on the throat from the exact metric. The nine
equations $(\eqnBigA-\eqnBigH)$ and $(\eqnBigI')$ were solved for the nine
quantities $a,a^\p,a^\m,b^\p,b^\m,R_z,R_x,(L^\p)^2$ and $\cos\beta$. The results
for this variation were much the same as for the original method. However, this
variation showed no signs of instability in the errors in the curvatures.

Our third and final variation was designed so as to remove the need for the
Bianchi identities. Recall that the Bianchi identity was required to constrain
the two degrees of freedom in the curvatures $R_x$ and $R_z$ so that they were
derivable from one metric function. We know that in the exact metric each
tile such as $(1,2,3,4)$, {\it when viewed as a tile on the 2-sphere}, is a
scaled image of the tile on the throat. It is not unreasonable to expect that
when the tiles are built with respect to the full 3-metric a similar though
approximate scaling could be used and that the associated errors would be
negligible for small tiles. To investigate this opinion we altered the structure
of the lattice to include a diagonal brace joining vertices 1 to 3. Similar
braces were also included for the pairs of vertices $1^\m,3^\m$ and $1^\p,3^\p$.
The lengths of these diagonals will be denoted by $m,m^\m$ and $m^\p$
respectively. The $\cos\beta$'s are discarded in favour of these diagonals. It is
a straightforward calculation, from the main smooth lattice equations
\eqnrfr{LegsEqtn}, to show that
$$
0 = 8a^2 - m^2
\eqno\eqndef{DiagEqtnA}
$$
The condition that each tile is a scaled version of the tile on the throat can be
written as
$$
0 = L^2 - (k' m)^2
\eqno\eqndef{DiagEqtnB}
$$
where $k'$ is constant along the tube. There are two apparently reasonable
choices for the constant $k'$. It could be calculated either from the exact
3-metric on the throat or from the distant regions, far from the throat, where
the metric is flat. In fact both choices are unacceptable.
The former choice, setting $k'$ on the throat, cannot be used because it does
not allow the lattice (with our assumed coordinates, see Table
(\tblrfr{CoordsRect})) to become asymptoticly flat. For smaller and smaller sizes
of the initial tile on the throat we can expect that the asymptotic curvature
might also get smaller and smaller yet it seems unlikely that this could save the
day. In the second choice, which requires $(k')^2=1/2$, we find by direct
inspection of the smooth lattice equations that $0=R_x=R_z$ everywhere (which
when coupled with the boundary condition on the throat then forces $L^\m=L=L^\p$
along the tube). Thus this second choice certainly cannot be used. We thus
opted for the first choice (but with few expectations).

In a typical RNC we computed, for this variation,
$a,a^\p,a^\m,b^\p,b^\m,R_z,R_x,(L^\p)^2$ and $m^2$. This required nine equations
which we took to be the smooth lattice equations (\eqnBigA--\eqnBigI) with
\eqnrfr{BigF} (which defines $\cos\beta$) and \eqnrfr{BigI} (the Bianchi
identity) replaced by \eqnrfr{DiagEqtnA} and \eqnrfr{DiagEqtnB}. The results are
displayed in Figure (\figrfr{ThreeDPlotM})
and are as expected -- the method does not converge
in any sense. It appears that forcing the tiles of the smooth lattice to observe
the same symmetries as the tiles of the 2-spheres is incorrect. We conclude that
the Bianchi identities cannot be ignored -- they are essential in providing the
correct constraint on the $L$'s and $\cos\beta$'s (or the $L$'s and $m$'s).

\beginsubsection{Are the Bianchi identities essential?}
Much has been made of the Bianchi identities and the role that they play in the
smooth lattice equations. Yet does it not seem odd that in the smooth lattice
approach we find ourselves computing one of the curvatures by integrating a
differential equation yet all standard formulas give the curvatures as
explicit functions? As another explanation as to why this is so consider the
frame components of the curvature of the exact metric. These are
$$\openup10pt\displaylines{%
(R_z)_\star
= -{1\over\beta} {d^2\eta\over dr^2}\cr
(R_x)_\star
= {1\over\eta^2}\left( 1 - \left({d\eta\over dr}\right)^2 \right)\cr }
$$
If we now choose the leg lengths according to $L(r)=\eta(r)\Delta\theta$
then we get
$$\openup10pt\displaylines{%
(R_z)_\star
= -{1\over L} {d^2L\over dr^2}\cr
(R_x)_\star
= {1\over L^2}\left( \left(\Delta\theta\right)^2 
                    -\left({dL\over dr}\right)^2 \right)\cr }
$$
Now suppose that all we are given are the discrete leg-lengths and that we wish
to extract the curvatures from the discrete lattice. We can expect that in some
continuum limit we would recover the above formulae. But where in this process
would the number $\Delta\theta$ come into the game? How is it extracted from
the discrete lattice (prior to the continuum limit) and how do we know that it
is a constant along the tube? The answer must lie partly in the constraint
that the radial edges be global geodesics and partly in the use of the Bianchi
identities. In fact it is easy to show (see section (\secrfr{ContinuumLimit})
below) that, in the continuum limit, equations (\eqnBigA--\eqnBigG), can be
reduced to just
$$
R_z = -{1\over L} {d^2L\over dr^2}
$$
while from the remaining equations (\eqnBigH--\eqnBigI) we can obtain
$$
R_x = {1\over L^2}\left(  k'' - \left({dL\over dr}\right)^2 \right)
$$
where $k''$ is a constant along the tube. That this constant equals
$(\Delta\theta)^2$ can be seen by evaluating this expression for large $r$
where the curvatures vanish. Thus using a metric such as \eqnrfr{ExactMetric}
introduces global information that is simply not accessible from the purely
local calculations used in one RNC of the smooth lattice.

This again shows the essential nature of the Bianchi identities in producing the
correct estimates for the curvatures.

\beginsection{The Continuum Limit}\secdef{ContinuumLimit}
Though the numerical results presented in the previous sections provide strong
evidence that the smooth lattice equations are a convergent approximation to
the correct differential equations it would be nice to establish this by direct
analytical methods.

The most direct route to the continuum is to introduce a scaling for
all the leg lengths such as $L\rightarrow\eps L$, $d^\p\rightarrow\eps d^\p$
etc., and to then express the solutions for $a,a^\p,\cdots$ as power series in
$\eps$. The terms in the power series can be obtained by substitution into the
smooth lattice equations. From equations (\eqnBigA--\eqnBigE) we obtain,
assuming $R_x$ and $R_z$ are known,
$$
\eqalign{%
a&={L\over2}\left(1+{1\over24} R_x L^2 + \Oeps(4)\right)\cr
a^\p&={L^\p\over2}\left(1+{1\over24}R_x(L^\p)^2 
  + {1\over6}R_z\left( (d^\p)^2 -{1\over2}(L-L^\p)^2 \right) + \Oeps(4)\right)\cr
a^\m&={L^\m\over2}\left(1+{1\over24}R_x(L^\m)^2 
  + {1\over6}R_z\left( (d^\m)^2 -{1\over2}(L-L^\m)^2 \right) + \Oeps(4)\right)\cr
(b^\p)^2 &= (d^\p)^2 - {1\over2} (L-L^\p)^2 + \Oeps(4)\cr
(b^\m)^2 &= (d^\m)^2 - {1\over2} (L-L^\m)^2 + \Oeps(4)\cr
}
$$
The expansions have been carried as far as is necessary to obtain the continuum
limit. These can be substituted into equation \eqnrfr{BigG} which can then
be reduced to a differential equation by employing the usual
Taylor series expansions
$$
\eqalign{%
L^\p = L + \left({dL\over dl}\right) d^\p 
         + \left({d^2 L\over dl^2}\right) {(d^\p)^2\over2} + \cdots\cr
L^\m = L - \left({dL\over dl}\right) d^\m 
         + \left({d^2 L\over dl^2}\right) {(d^\m)^2\over2} + \cdots\cr
}
$$
where $l$ is the proper distance measured from the throat.
The result, after cancelling a common factor of $-3(d^\p + d^\m)L/2$, is
$$
0 = \left({d^2L\over dl^2} + R_z L\right) 
                           + \Oeps(3)
                           + \One(d^\p - d^\m)
$$
The first error term $\Oeps(3)$ arises from the curvature terms in the
smooth lattice equations while the second error term
$\One(d^\p - d^\m)$ arises from the Taylor series expansion in $L$.
Clearly in the limit when $\eps\rightarrow0$ we recover the
correct differential equation. We can also see that when the successive $d^\p$
are chosen to be a smooth function of $L$ then $d^\p-d^\m = \Oeps(2)$ and thus
{\sl both} error terms are $\Oeps(3)$.

Equations (\eqnBigH--\eqnBigI) can be treated in a similar fashion leading to
$$
0 = {d\over ds} \left(L^2 R_x\right) + \left({d L^2\over ds}\right) R_z
+\Oeps(2)(d^\p - d^\m)
$$

In both of the above equations the errors terms are $\Oeps(2)$ {\sl relative}
to the main terms. Thus we can infer that the relative errors $EL(l,\eps)$ and
$ER_x(l,\eps)$ should both be $\Oeps(2)$. This is exactly what we observed in
our numerical studies.

We have not used equation \eqnrfr{BigF} since it can be viewed as a definition
of $\cos\beta$.

\beginsection{The Regge calculus}
The Regge calculus \Ref(1) is another lattice theory of General relativity.
Though it employs a lattice it differs significantly from our method in the way
in which the field equations are computed from the lattice variables. In the
Regge calculus the metric interior to each pair of cells is deemed, a-priori, to
be flat. This precludes the use of standard methods for computing the curvatures
(such as pointwise differentiation of a metric). Instead one is forced to look
at integral quantities, in particular the action integral
$$I = \int\>R\sqrt{-g}\>d^4x$$
The curvature for the piecewise flat metric can be interpreted in the sense of
distributions as a series of Dirac delta functions on the 2-dimensional
sub-spaces of the lattice (usually the triangles of a simplicial lattice).
The action integral can then be evaluated as
$$
I = 2\sum_i\>\theta_i\,A_i
$$
where the sum includes all of the triangles and $A_i$ and $\theta_i$ are the
areas and defects on these triangles. The Regge field equations are then
obtained by extremising the action with respect to the leg lengths, leading to
$$
0 = \sum_i\>\theta_i\,{\partial A_i\over \partial L_j}
$$
There is one such equation for each leg $L_j$.

Though this derivation mimics that used in Einstein's theory one cannot state,
by this fact alone, that the Regge calculus is a consistent discretisation of
Einstein's equations. By consistent we mean that as the Regge lattice is
successively refined the solutions of the Regge equations converge to solutions
of Einstein's equations. Proving this is extremely difficult primarily because
the metric is not differentiable as a point function. Another complication is
that any metric can be approximated by many different sequences of lattices yet
one would want to develop theorems that account for this vast array of lattices. 
One firm result has however been proven, \Ref (7), that the Regge and
Hilbert actions converge when evaluated over a fixed region of the limit
smooth space. The proof is very detailed and far from trivial. The state of
play is (i) that any smooth metric can be approximated, to any accuracy, by a
simplicial metric (such metrics need not be solutions of their equations),
(ii) the Regge and Hilbert actions converge when evaluated on a fixed region of
the limit smooth space (again, for any smooth metric) but (iii) their remains an
open question as to whether or not the limit metric of a sequence of solutions
of Regge's equations is also a solution of Einstein's equations.

Of course the later question can be tackled via direct numerical experiments.
In all cases reported to date it appears that the Regge solutions do appear to
converge to solutions of Einstein's equations. However Brewin \Ref(8) has
argued that this may be because the experiments have been conducted on highly
symmetric spacetimes with the Regge lattices constrained by those symmetries. It
could be that the symmetries are so strong as to bring about a concordance
between the two theories. Brewin argued that in less symmetric spacetimes the
distinction between the two theories may become apparent. He went further to
show that even for the symmetric spacetimes, such as Schwarzschild, a lattice
that was built purely from 4-simplices and the known exact metric, did not appear
to have the appropriate truncation errors when evaluated on the Regge
equations. This and other examples were used to argue that the Regge equations
are fundamentally not a consistent discretisation of Einstein's equations. This
is not a view shared by the rest of the Regge calculus community. 

In contrast there can be little argument regarding the correct field equations
for the smooth lattice approach. They are just the usual Einstein equations.

Of particular note for the current discussion are the calculations of Wong
\Ref(9). He used the Regge calculus to solve the time symmetric initial value
equations for a Schwarzschild spacetime. He chose two methods, one in which the
2-spheres were triangulated by icosahedra and a second in which he used
infinitesimal tiles to generate a rectangular tube similar to our second main
example. In this second method Wong proceeded to the limit of vanishingly
small cross section in the tube which he called the continuum method.
However, in both cases Wong chose to retain a fully discrete radial
sub-division. The results for both methods were very good with the typical
errors in the range of 1-10\% for the icosahedral method and 0.001-1\% for the
continuum method (for quantities such as volumes and distances from the throat).

It is tempting to compare these results with our own. In the icosahedral method
there are 20 triangles per 2-sphere. This lies between the second and third
sub-divisions of our first example and to a choice of $\eps\approx10$
in our second example. In both cases we find the error in the leg lengths to be
approximately 20\% while the errors in the curvatures are approximately 50\%
for the first example and 100\% (yikes!) for the second example.
These are larger than Wong's results which is slightly surprising
since fitting a smooth metric to a lattice could be expected to be more
accurate than fitting a piecewise flat metric to the same lattice.

Wong's continuum can only be compared with our tube method. We can see from
Figure \figrfr{ThreeDPlotC} that our errors are around $10^{-5}$\%
whereas Wong's best results are around $10^{-3}$\%.
This comparison is a bit unfair on both methods. In all of our calculations
the edges of our cells were scaled via the parameter $\eps$ whereas Wong chose
to retain a discrete radial sub-division. This gives an advantage to our method.
However where Wong obtains an advantage over our method is in his extensive
use of the target continuum metric. Indeed he writes the metric in the
form
$$
ds^2 = dr^2 + \eta(r)^2\left(d\theta^2 + \sin^2\theta\,d\phi^2\right)
$$
(with a slight change in replacing his $\rho(r)$ with our $\eta(r)$) and
estimates all of the leg lengths in the Regge lattice in terms of the unknown
function $\eta(r)$ and freely chosen quantities such as $\Delta\theta$ and
$\Delta\phi$. In regulating the choice of leg lengths via this metric he has
avoided the problem of having too many degrees of freedom in the lattice. We had
the same problem but we resolved it by way of the Bianchi identities.

It is unlikely that such a metric ansatz will be available in less symmetric
spaces forcing one to look at fully discrete ways in which to impose the
desired symmetries. Yet an overly enthusiastic imposition of the symmetries on
the space of possible lattices can restrict that space to a set of measure
zero. For example, suppose we set out to use the Regge calculus with tubes much
like that used in our second example. We could take four such tubes and join
them together along one common radial edge. In this model we would not need the
$\cos\beta$'s. All of the defects can then be easily calculated. In fact we find
that they are {\bf all} non-positive. The only solution of Regge's equations is
therefore flat space. This can even be seen without doing any calculations --
try fitting the four tubes together in the distant flat regions where the
defects should all be close to zero while $dL/dr$ is constant but non-zero. Thus
some of the symmetries amongst the legs must be relaxed. But this then introduces
extra degrees of freedom. How are they to be constrained? This is where Wong's
ansatz saves the day. It produces sufficient variation in the leg lengths of the
four tubes so as to obtain meaningful solutions of the Regge equations while
retaining the desired spherical symmetries.

It is hard to compare Wong's continuum method and our second example on a level
playing field as Wong's method obtains significant advantage in using a method
not generally available in a fully discrete implementation of the Regge calculus.

The smooth lattice method could have been applied directly to Wong's
icosahedral method. We chose not do so because it seemed unlikely to add
anything new to our investigations.

It is important to note that the smooth lattice method uses the same
lattice structures as used in the Regge calculus. Thus it inherits all of the
attractive features of the Regge calculus, in particular the natural division
between the topological and metric information and that the metric is recorded
in pure scalar form (eg. the leg lengths). But more importantly, the smooth
lattice method has the additional advantage that it extracts a smooth metric
from the lattice and thus that the Einstein equations may be applied directly to
the lattice. There is also a practical advantage of the smooth lattice method
-- the equations are much easier to construct and evaluate than those of the
Regge calculus (witness some of the unwieldy calculations of defect angles
with their attendant inverse trigonometric and hyperbolic functions, its a
mess!). The smooth lattice equations can reasonably be expected to be much
quicker to evaluate and to solve on a computer than the Regge equations.

\beginsection{Discussion}
On the limited evidence provided here it is hard to state with any confidence
whether or not the smooth lattice method will be of any use in numerical
relativity. However the results are very encouraging. Furthermore the strong
theoretical basis for the method, its central use of the equivalence principle
and its computational
simplicity gives us strong reasons to believe that the method will prove to be
useful. Of course much more work needs to be done.

Currently we are adapting these ideas to a 3+1 formulation in which the smooth
lattice method is applied to the 3-geometry while retaining a continuous time
coordinate. We will employ a dynamical 3-dimensional lattice whose evolution
will be controlled by the standard ADM 3+1 equations (which will be applied
directly to the 3-lattice). The smooth lattice equations will be used in each
time evolution step to aid in the calculation of the source terms. For example,
at each time step we would be given a discrete 3-metric on a 3-lattice from
which we could estimate the 3-curvatures by way of the smooth lattice equations.
There is no role for the Bianchi identities in this phase of the integration, we
simply have to live with whatever estimates we get for the curvatures. However
we can expect the Bianchi identities to play an important role in the solution of
the initial value problem where the initial 3-metric is being constructed. We
can speculate that the constraints associated with the Bianchi identities on the
3-metric will be preserved by the 3+1 evolution equations. This is pure
speculation at the moment and may in fact prove to be wrong. We will be testing
this 3+1 method on the evolution of the Schwarzschild initial data presented
here and we hope to report on this work in the near future.

Another interesting question is how much mismatch in the metric do we incur (or
how much can we allow) between pairs of adjacent RNC frames? Recall that each RNC
frame is constructed by choosing its metric such that it agrees with a finite
set of geodesic leg lengths. Consider now two adjacent frames that share a
common cell. For a typical curve in that cell this pair of frames will not yield
the same estimates for the length of that curve (except on the geodesic edges of
that cell). By how much do these estimates differ and is there an optimal choice
of curvatures that minimises the difference? If such a minimising relation exists
it must constrain the choice of curvatures in each frame. However, we already
know that the curvatures must be constrained by the Bianchi identities. How then
are these constraints related? This too is currently being investigated.

In both of our examples we chose to solve the Hamiltonian constraint by a
radial integration. However the constraints are known to be elliptic equations
and so should be properly solved as a boundary value problem. This leads us to
an interesting problem. How can asymptotic conditions such as {\sl the
conformal factor behaves like $1+k/r$ for large $r$} be imposed using local
Riemann normal coordinates? The difficulty is that this asymptotic condition is
a statement about global properties of the space whereas the Riemann normal
coordinates are purely local. Another way of looking at this is to ask how can
two observers, Joe and Mary, each in their own RNC frame with Mary centred on
the throat and Joe in the distant flat regions, determine which observer is in
the asymptoticly flat part of the space. Recall that each observer is in a
locally freely falling frame and the only geometric quantities that they can
measure are the curvature components. Thus the statement that observer Joe is
in the asymptoticly flat region must say something about the curvatures
measured by Joe relative to those measured by Mary. It would seem that the
asymptotic boundary conditions would need to be formulated in terms of the
curvatures. Yet such a conclusion seems at odds with the standard theory of
elliptic pde's. We seem to be imposing boundary conditions on the {\sl second}
derivatives of a function which is a solution of a {\sl second} order elliptic
equation. This is indeed odd and needs to be resolved if this method is to be
applied to general initial value problems.

\beginsection{Appendix}
This appendix, which is part of a larger report \Ref(5) on Riemann normal
coordinates, is included here to provide a self-contained derivation of the two
smooth lattice equations \eqnrfr{LegsEqtn} and \eqnrfr{AnglesEqtn}. This will
require some preliminary discussion on Riemann normal coordinates. In deriving
our equations we will make repeated use of series expansions in the curvature
(and its derivatives) around flat space. We have already seen this in action
when we wrote the metric as
$$
g_{\mu\nu}(x) = g_{\mu\nu} - {1\over3}R_{\mu\alpha\nu\beta}\> x^\alpha x^\beta
                + \Oeps(3)
\eqno\eqnrfr{RNCmetric}
$$
We shall prove this in section (\secrfr{TheMetric}) below.

This expansion can be interpreted as a flat space part plus a (presumably) small
quadratic perturbation. Naturally one might have reservations about the
convergence of such an expansion. We shall first establish that for sufficiently
small regions in which this RNC is defined the truncation errors are negligible.
We will do this by introducing a conformal transformation of the original
metric. This will prove to be extremely useful. All of the results in this
appendix will subsequently be quoted in terms of these conformal coordinates
whereas in the body of this paper our results are expressed in the physical
coordinates.

For the remainder of this appendix we will be concerned with one RNC frame. We
will also refer to the region over which the RNC is defined as the patch (which
in the body of the paper consisted of two consecutive cells of the tube or the
set of triangles attached to a vertex).

\beginsubsection{Conformal coordinates}
Let the typical length scale of the patch containing $O$ be $\eps$. Let
the coordinates of the patch be $x^\mu$ and let the coordinates of $O$ be
$x^\mu_{\star}$. Now define a new set of coordinates $y^\mu$ by
$$x^\mu = x^\mu_{\star} + \eps y^\mu$$
Then
$$\eqalign{%
  ds^2 &= g_{\mu\nu}(x)dx^\mu dx^\nu\cr
       &= \eps^2 g_{\mu\nu}(x_{\star} + \eps y) dy^\mu dy^\nu\cr}$$
Define the conformal metric $d\tilde s$ by
$$\eqalign{%
d{\tilde s}^2 &= g_{\mu\nu}(x_{\star} + \eps y) dy^\mu dy^\nu\cr
              &= {\tilde g}_{\mu\nu}(y,\eps) dy^\mu dy^\nu\cr}$$

From the above it is easy to see that, at $O$,
$$
\left.\eqalign{%
{\tilde g}_{\mu\nu} & = g_{\mu\nu}\cr
{\tilde g}_{\mu\nu,\alpha} &= \eps g_{\mu\nu,\alpha}\cr
{\tilde g}_{\mu\nu,\alpha\beta} &= \eps^2 g_{\mu\nu,\alpha\beta}\cr}
\quad\quad\right\rbrace
\eqno\eqndef{EpsMetric}
$$
where the partial derivatives on the left are with respect to $y$ and those on
the right are with respect to $x$. For each higher derivative an extra power of
$\eps$ will appear.

From the above we immediately obtain
$${\tilde R}_{\mu\nu\alpha\beta} = \eps^2 R_{\mu\nu\alpha\beta}$$
and as $R_{\mu\nu\alpha\beta}$ is independent of $\eps$,
we see that
$${\tilde R}_{\mu\nu\alpha\beta} = \Oeps(2)$$
for $\eps<<1$.

Clearly, then, as $\eps\rightarrow0$ the conformal metric is flat.

There are now two ways to look at the patch. We can view it (in the original
coordinates $x^\mu$) as a patch of length scale $\eps$ with a curvature
independent of $\eps$. Or we can view it (in the conformal coordinates $y^\mu$)
as a patch of fixed size but with a curvature that varies as $\eps^2$. This
later view is useful since in using it we can be sure that the series expansions
around flat space are convergent (for a sufficiently small $\eps$).

We will use these conformal coordinates for the remainder of this Appendix.
We will also drop the tilde and revert to $x^\mu$ as the generic coordinates
(even while working in the conformal frame.)

\beginsubsection{Riemann normal coordinates}
In Riemann normal coordinates the geodesics through $O$ must all be of the form
$$x^\mu(s) = a^\mu_1 s$$
for some set of numbers $a^\mu_1$.
By direct substitution into the geodesic equation
$$ 
0 = {d^2 x^\mu\over ds^2}
  +  \Gamma^\mu_{\alpha\beta}(x) {d x^\alpha\over ds}
                                 {d x^\beta\over ds}
\eqno\eqndef{GeodesicEqtn}
$$
and its derivatives, one obtains, at the origin $O$,
$$
\eqalignno{%
0 &= \Gamma^{\mu}_{\alpha\beta}
&\eqndef{ZeroGamma}\cr
0 &= \Gamma^{\mu}_{\alpha\beta,\nu} +
     \Gamma^{\mu}_{\beta\nu,\alpha} +
     \Gamma^{\mu}_{\nu\alpha,\beta}
&\eqndef{ZeroGammaDeriv}\cr}$$
It is easy to see, by continuing in this way, that all symmetric derivatives
of the connection vanish at the origin in Riemann normal coordinates.

\beginsubsection{Metric}\secdef{TheMetric}
Consider a Taylor series expansion of the metric around the origin $O$, namely,
$$
g_{\mu\nu}(x) =  g_{\mu\nu}
               + g_{\mu\nu,\alpha\beta} {x^\alpha x^\beta\over 2}
               + \Oeps(3)
$$
There is no linear term because $g_{\mu\nu,\alpha} =0$ at the origin. It is
a simple algebraic exercise to show, given
\eqnrfr{ZeroGamma} and \eqnrfr{ZeroGammaDeriv}, that
$$
\Gamma^\mu_{\alpha\beta,\nu} = 
                     - {1\over3}\left( R^\mu{}_{\alpha\beta\nu}
                                      +R^\mu{}_{\beta\alpha\nu}\right)
\eqno\eqndef{GammaDerivRiem}
$$
from which it follows that
$$
g_{\mu\nu,\alpha\beta} = - {1\over3}\left( R_{\mu\alpha\nu\beta}
                                          +R_{\mu\beta\nu\alpha}\right)
\eqno\eqndef{MetricDerivRiem}
$$
and finally
$$R_{\mu\nu\alpha\beta} =  g_{\alpha\nu,\mu\beta}
                         - g_{\alpha\mu,\nu\beta}
$$
Substituting these into the above we obtain
$$
g_{\mu\nu}(x) = g_{\mu\nu} - {1\over3} R_{\mu\alpha\nu\beta}\>x^\alpha x^\beta
                            + \Oeps(3)
\eqno\eqnrfr{RNCmetric}
$$

\beginsubsection{Connection}
We can also propose a Taylor series expansion for the connection about
the origin $O$, namely,
$$
\Gamma^\mu_{\alpha\beta}(x) = 
      \Gamma^\mu_{\alpha\beta}
    + \Gamma^\mu_{\alpha\beta,\rho} x^\rho
    + \Gamma^\mu_{\alpha\beta,\rho\tau} {x^\rho x^\tau\over2}
    + \cdots
$$
Our first observations are that
$$\eqalign{%
\Gamma^\mu_{\alpha\beta} &= 0\cr
\Gamma^\mu_{\alpha\beta,\rho} &= \Oeps(2)\cr
\Gamma^\mu_{\alpha\beta,\rho\tau} &= \Oeps(3)\cr}
$$
and in general the $n$-th derivative of $\Gamma$ will be $\Oeps(n+1)$ at $O$.
This follows by simple inspection of the standard formula for computing the
metric connection and the previously stated asymptotic behaviour of the conformal
metric \eqnrfr{EpsMetric}.

We are only interested in the leading term in the above expansion, and so after
using \eqnrfr{GammaDerivRiem} we obtain
$$
\Gamma^\mu_{\alpha\beta}(x) = 
              - {1\over3}\left( R^\mu{}_{\alpha\beta\nu}
                               +R^\mu{}_{\beta\alpha\nu}\right)x^\nu
              + \Oeps(3)
\eqno\eqndef{ExpGamma}
$$

\beginsubsection{Geodesics}\secdef{Geodesics}
For the geodesics, we employ a series expansion in $s$, the distance measured
along the geodesic,
$$
x^\mu(s) = a^\mu_0
          + a^\mu_1  s
          + a^\mu_2 {s^2\over 2}
          + a^\mu_3 {s^3\over 6}
          + \cdots
\eqno\eqndef{ExpXmu}
$$
We will defer for the moment stating the nature of the truncation error.
Our primary aim here is to determine as many of the $a^\mu_i$ as we can in
terms of just $g_{\mu\nu}$ and $R_{\mu\nu\alpha\beta}$. This can be done by
demanding that the above expansion for $x^\mu(s)$ is a solution of the
geodesic equation \eqnrfr{GeodesicEqtn}.

The basic steps are to substitute \eqnrfr{ExpXmu} into \eqnrfr{ExpGamma} and to
then substitute
all of these quantities into the geodesic equation \eqnrfr{GeodesicEqtn}.
The result is a
polynomial in $s$ and as this must be identically zero for all $s$,
we equate the separate
coefficients of powers of $s$ to zero. For the first two terms
$s^0$ and $s^1$ we obtain, respectively,
$$\eqalign{%
0 &= a^\mu_2 + \Gamma^\mu_{\alpha\beta,\rho}
                 a^\rho_0 a^\alpha_1 a^\beta_1
             + \Gamma^\mu_{\alpha\beta,\rho\tau}
                 a^\rho_0 a^\tau_0 a^\alpha_1 a^\beta_1\cr
0 &= a^\mu_3 + \Gamma^\mu_{\alpha\beta,\rho}
               \left(  a^\rho_1 a^\alpha_1 a^\beta_1
                     +2 a^\rho_0 a^\alpha_1 a^\beta_2 \right)
             + \Gamma^\mu_{\alpha\beta,\rho\tau}
               \left( 2 a^\rho_1 a^\tau_0 a^\alpha_1 a^\beta_1
                     +2 a^\rho_0 a^\tau_0 a^\alpha_2 a^\beta_1 \right)}
$$
The term $\Gamma^\mu_{\alpha\beta,\rho} a^\rho_1 a^\alpha_1 a^\beta_1$
is zero in view of \eqnrfr{ZeroGammaDeriv}.
Thus, to order $\eps^3$, we obtain
$$\eqalign{%
a^\mu_2 & = - \Gamma^\mu_{\alpha\beta,\rho}
                a^\rho_0 a^\alpha_1 a^\beta_1 + \Oeps(3)\cr
a^\mu_3 & = \Oeps(3)\cr}
$$
Clearly this process can be developed in full to obtain recurrence
relations amongst all of the remaining $a^\mu_i$. We will not need these but what
we do require is their behaviour in $\eps$. It is not hard to see that in the
generic equation for $a^\mu_n$ the leading terms will be
$$
0 = a^\mu_n + \Gamma^\mu_{\alpha\beta,\rho}
                a^\rho_{n-2} a^\alpha_1 a^\beta_1 + \cdots
$$
Since we have already established that $a^\mu_2$ is $\Oeps(2)$ it follows that
$a^\mu_n$ will be $\Oeps(n)$ for $n\ge2$.

Assembling the above results leads finally to
$$
x^\mu(s) = a^\mu_0 + a^\mu_1 s
                         + {1\over3} R^\mu{}_{\alpha\beta\rho}\>
                           a^\rho_0 a^\alpha_1 a^\beta_1 s^2
                         + \Oeps(3)
\eqno\eqndef{ExpXFinal}
$$
Notice that $a^\mu_0$ and $a^\mu_1$ remain undetermined -- they can only be
computed from appropriate boundary or initial conditions.

\beginsubsection{Geodesic boundary value problem}\secdef{GeodesicBVP}
In this case we are looking for the geodesic which passes through two given
points. Let the coordinates of initial point be $x^\mu_i$ and those for the
final point be $x^\mu_j$. Suppose the geodesic distance between the two points
is $L_{ij}$. The $L_{ij}$ cannot be freely specified as they must be derivable
from the metric and the coordinates. A equation for $L_{ij}$  will be given in a
later section (\secrfr{GeodesicLenSq}).

Our aim is to solve for $a^\mu_0$ and $a^\mu_1$ such that
$$\eqalign{%
x^\mu(s=0) &= x^\mu_i = a^\mu_0 + \Oeps(3)\cr
x^\mu(s=L_{ij}) &= x^\mu_j = a^\mu_0 + a^\mu_1 L_{01}
                                 + {1\over3} R^\mu{}_{\alpha\beta\rho}\>
                                    a^\rho_0 a^\alpha_1 a^\beta_1 L^2_{01}
                                 + \Oeps(3)\cr}
$$
The first equation is easy to solve, namely, $a^\mu_0 = x^\mu_i + \Oeps(3)$.
However, the second equation does appear to pose a bit of a problem -- it looks
like a nasty quadratic equation for each of the $a^\mu_1$.
Fortunately this equation can be solved by an iterative
method to within $\Oeps(3)$. The starting point is to first substitute for
$a^\mu_0$ to obtain
$$
x^\mu_j = x^\mu_i + a^\mu_1 L_{01}
             + {1\over3} R^\mu{}_{\alpha\beta\rho}\>
                x^\rho_i a^\alpha_1 a^\beta_1 L^2_{01}
             + \Oeps(3)
$$
We can then re-write this as
$$
a^\mu_1 = {1\over L_{01}}\left( x^\mu_j - x^\mu_i
             - {1\over3} R^\mu{}_{\alpha\beta\rho}\>
                x^\rho_i a^\alpha_1 a^\beta_1 L^2_{01} \right)
           + \Oeps(3)
$$
which we will evaluate as a fixed point iteration scheme.

Since $R^\mu{}_{\alpha\beta\rho} = \Oeps(2)$ we obtain the first approximation
$$a^\mu_1 =  {1\over L_{01}}\left(x^\mu_j - x^\mu_i\right) + \Oeps(2)$$
This can now be substituted back into the previous equation leading to the
second approximation
$$
a^\mu_1 = {1\over L_{ij}}\left(\Delta x^\mu_{ij}
        - {1\over3} R^\mu{}_{\alpha\beta\rho}\>
          x^\rho_i \Delta x^\alpha_{ij} \Delta x^\beta_{ij}\right)
        + \Oeps(3)
\eqno\eqndef{TermA}
$$
where $\Delta x^\mu_{ij} = x^\mu_j - x^\mu_i$.
This is as far as we can proceed with this iteration scheme because the errors
in this approximation for $a^\mu_1$ and those in the equation we are iterating
on are both $\Oeps(3)$. Combining
these results for $a^\mu_0$ and $a^\mu_1$ and substituting into
\eqnrfr{ExpXFinal} leads to the following equation for the geodesic passing
through the two points $x^\mu_i$ and $x^\mu_j$
$$
x^\mu(s) = x^\mu_i + \lambda \Delta x^\mu_{ij}
           - {\lambda(1-\lambda)\over3}R^\mu{}_{\alpha\beta\rho}\>
              x^\rho_i \Delta x^\alpha_{ij} \Delta x^\beta_{ij}
           +\Oeps(3)
\eqno\eqndef{PathXa}
$$
where $\lambda = s/L_{ij}$.

\beginsubsection{Geodesic distance}\secdef{GeodesicLenSq}
Consider two points with coordinates $x^\mu_i$ and $x^\mu_j$. Since there
exists, by assumption, a unique geodesic joining this pair of points, the
distance between them should also be uniquely defined in terms of their
coordinates and the metric.

Our aim is to evaluate, along the geodesic,
$$L_{ij} = \int^1_0\>\left( g_{\mu\nu}(x) {dx^\mu\over d\lambda}
                                      {dx^\nu\over d\lambda} \right)^{1/2}\>
           d\lambda
$$
The equation for $x^\mu(\lambda)$ is simply \eqnrfr{PathXa} for $0<\lambda<1$.
This can be substituted into the expansion \eqnrfr{RNCmetric} for $g_{\mu\nu}(x)$ with
the result
$$
g_{\mu\nu}(x(\lambda)) = g_{\mu\nu}
   - {1\over3} R_{\mu\alpha\nu\beta}\>
     \left( x^\alpha_i + \lambda \Delta x^\alpha_{ij} \right)
     \left( x^\beta_i + \lambda \Delta x^\beta_{ij} \right)
   + \Oeps(3)
$$
It is a simple matter to substitute these into the integrand, leading to
$$
\left( dL\over d\lambda \right)^2 =
      g_{\mu\nu} \Delta x^\mu_{ij}\Delta x^\nu_{ij}
   - {1\over3} R_{\mu\alpha\nu\beta}\>
     x^\alpha_i x^\beta_i \Delta x^\mu_{ij} \Delta x^\nu_{ij}
   + \Oeps(3)
$$
The important point to note is that this result does not depend on $\lambda$.
Thus the integrand is constant and so the integration is trivial. The result
follows immediately,
$$
L^2_{ij} = g_{\mu\nu} \Delta x^\mu_{ij}\Delta x^\nu_{ij}
          - {1\over3} R_{\mu\alpha\nu\beta}\>
            x^\alpha_i x^\beta_i \Delta x^\mu_{ij} \Delta x^\nu_{ij}
          + \Oeps(3)
\eqno\eqndef{LenSqFinal}
$$
From this result it is easy to establish, using the symmetries of
$R_{\mu\nu\alpha\beta}$, the following equivalent equations for $L^2_{ij}$
$$\eqalignno{%
L^2_{ij} &= (g_{\mu\nu} - {1\over3} R_{\mu\alpha\nu\beta}\>
                                  {{\Bar x}^\alpha_{ij}}\>
                                  {{\Bar x}^\alpha_{ij}})\Delta x^\mu_{ij}
                                                         \Delta x^\nu_{ij}
          + \Oeps(3)\cr
         &= g_{\mu\nu} \Delta x^\mu_{ij}\Delta x^\nu_{ij}
          - {1\over3} R_{\mu\alpha\nu\beta}\>
            x^\alpha_i x^\beta_i x^\mu_j x^\nu_j
          + \Oeps(3)\cr
         &= g_{\mu\nu} \Delta x^\mu_{ij}\Delta x^\nu_{ij}
          - {1\over3} R_{\mu\alpha\nu\beta}\>
            \Delta x^\alpha_{0i} \Delta x^\beta_{0i}
            \Delta x^\mu_{0j} \Delta x^\nu_{0j}
          + \Oeps(3)
&\eqndef{LenSqAlt}\cr}
$$
where ${\Bar x}^\mu_{ij} = (x^\mu_i+x^\mu_j)/2$ and where $x^\mu_0$ are the
coordinates of the origin (which in our case are zero).

Note that when these equations are re-expressed in terms of the physical metric
all that changes is that the error term $\Oeps(3)$ is replaced with $\Oeps(5)$.

\beginsubsection{Generalised Cosine law}
Consider a geodesic triangle with vertices $i,j$ and $k$. We would like to be
able to compute the angles subtended at each vertex in terms of the usual
quantities, the metric, the coordinates etc. We will develop the appropriate
equations in two stages. First, we will consider the simple case of computing
the angle at a vertex coincident with the origin. We shall then generalise this
result to the case were all three vertices are distinct from the origin.

To start the ball rolling consider a geodesic triangle with vertices $i,j$ and
$O$, the origin. We seek an equation for the angle between the geodesic
segments joining $O$ to $i$ and $O$ to $j$. The unit tangent vectors to these
geodesic segments at $O$ are, from equation \eqnrfr{PathXa},
$$\eqalign{%
v^\mu_{oi} & = \Delta x^\mu_{oi}/L_{oi}\cr
v^\mu_{oj} & = \Delta x^\mu_{oj}/L_{oj}\cr}
$$
Now let $\theta_{ij}$ be the angle subtended at $O$. Then
$$
\cos\theta_{ij} = g_{\mu\nu} v^\mu_{oi} v^\nu_{oj}
                = g_{\mu\nu} \Delta x^\mu_{oi} \Delta x^\nu_{oj}
                        /(L_{oi}L_{oj})
$$
We can obtain two useful variants of this equation by writing, first,
$\Delta x^\mu_{oi} = \Delta x^\mu_{oj} - \Delta x^\mu_{ij}$ and second,
$\Delta x^\mu_{oj} = \Delta x^\mu_{oi} + \Delta x^\mu_{ij}$. This gives
$$\eqalign{%
L_{oi} L_{oj} \cos\theta_{ij}
  &= g_{\mu\nu}\left(\Delta x^\mu_{oj} - \Delta x^\mu_{ij}\right)
               \Delta x^\nu_{oj}\cr
  &= g_{\mu\nu}\Delta x^\mu_{oi}
               \left(\Delta x^\nu_{oi} + \Delta x^\nu_{ij}\right)\cr}
$$
Adding these two equations leads to
$$
2L_{oi} L_{oj} \cos\theta_{ij} = L^2_{oi} + L^2_{oj}
                               - g_{\mu\nu} \Delta x^\mu_{ij} \Delta x^\nu_{ij}
\eqno\eqndef{Cosa}
$$
However, from equation \eqnrfr{LenSqAlt} we see that
$$
g_{\mu\nu} \Delta x^\mu_{ij} \Delta x^\nu_{ij} = L^2_{ij}
   + {1\over3} R_{\mu\alpha\nu\beta}\>
     \Delta x^\mu_{oi} \Delta x^\nu_{oi}
     \Delta x^\alpha_{oj} \Delta x^\beta_{oj} +\Oeps(3)
$$
Thus we have
$$
2L_{oi} L_{oj} \cos\theta_{ij} = L^2_{oi} + L^2_{oj} - L^2_{ij}
   - {1\over3} R_{\mu\alpha\nu\beta}\>
     \Delta x^\mu_{oi} \Delta x^\nu_{oi}
     \Delta x^\alpha_{oj} \Delta x^\beta_{oj} +\Oeps(3)
\eqno\eqndef{Cosb}
$$
With this equation we have achieved our first aim : to obtain an equation
when the vertex resides at the origin. To obtain an equation applicable
to the general case, where the vertex is not at the origin, we can imagine
transforming to a second set of Riemann normal coordinates with an origin at
some other point, say $O'$. We can do this simply by shifting the coordinates,
eg. $x^\mu \rightarrow x^\mu + c^\mu$. The coordinates, metric and Riemann
components at the respective origins will therefore be related by
$$\openup10pt\displaylines{%
x'^\mu = x^\mu + c^\mu\cr
g'_{\mu\nu}(O') = g_{\mu\nu}(O) - {1\over3} R_{\mu\alpha\nu\beta}(O)\>
                                   c^\alpha c^\beta + \Oeps(3)\cr
R'_{\mu\nu\alpha\beta}(O') = R_{\mu\alpha\nu\beta}(O) + \Oeps(1)\cr}
$$
Now the important observation is that the above equation \eqnrfr{Cosb} is
covariant with respect to this transformation (whereas \eqnrfr{Cosa} is not).
That is, it applies to any three vertices of a geodesic triangle. Let us now
relabel the vertices as $i,j$ and $k$. Then the angle subtended at vertex $k$
can be computed from
$$
2L_{ik} L_{jk} \cos\theta_{ij} = L^2_{ik} + L^2_{jk} - L^2_{ij}
   - {1\over3} R_{\mu\alpha\nu\beta}\>
     \Delta x^\mu_{ik} \Delta x^\nu_{ik}
     \Delta x^\alpha_{jk} \Delta x^\beta_{jk} +\Oeps(3)
\eqno\eqndef{CosFinal}
$$
As with the geodesic length equation \eqnrfr{LenSqFinal}, one need only replace
the $\Oeps(3)$ error term with $\Oeps(5)$ to obtain the correct equation in the
original physical coordinates.

\beginsection{Acknowledgements}
I am pleased to acknowledge the many illuminating discussions I have had with
Joe Monaghan and Tony Lun on this and many other topics in numerical relativity.
I would also like to thank Adrian Gentle and Tony Lun for their careful reading
of this paper.

{
\beginsection{References}
\parskip=5pt plus 0pt minus 4pt
\R 1!Regge, Tullio. (1961)!
     General Relativity without Coordinates.!
     Il Nuovo Cimento. Vol.XIX,No.3(1961) pp.558-571.\par

\R 2!Petrov, A.Z. (1969)!
     Einstein Spaces. (Chp.1 sec.7)!
     Pergamon Press, Oxford. 1969\par

\R 3!Eisenhart, L.P. (1926)!
     Riemannian geometry.(Sec.18 and App.3)!
     Princeton University Press, Princeton 1926\par

\R 4!Misner, C.W, Thorne, K. and Wheeler, J.A. (1972)!
     Gravitation (pp.285-287)!
     W.H. Freeman. San Francisco. 1973.\par

\R 5!Brewin, L.C. (1997)!
     Riemann Normal Coordinates.!
     Department of Mathematics Preprint,
     Monash University, Clayton, Vic. 3168. Also available from\hfil\break
    {\tt http://newton.maths.monash.edu.au:8000/papers/rnc-notes.ps.gz}\par

\R 6!Penrose, R. and Rindler (1984)!
     Spinors and Spacetime. Vol.1(p.364)!
     Cambridge University Press. 1984.\par

\R 7!Cheeger, J, M\"uller, W and Schrader, R. (1984)!
     On the Curvature of Piecewise Flat Spaces.!
     Comm. Math. Phys. Vol.92(1984) pp.405-454.\par

\R 8!Brewin, L.C. (1995)!
     The Regge Calculus is not an approximation to
     General Relativity.!
     Department of Mathematics Preprint,
     Monash University, Clayton, Vic. 3168. Also available from
     {\tt gr-qc/9502043}\ and\hfil\break
     {\tt http://newton.maths.monash.edu.au:8000/papers/truncation.ps.gz}\par

\R 9!Cheuk-Yin Wong (1971)!
     Application of Regge Calculus to the Schwarzschild and Reisner-Nordstr{\o}m
     Geometries at the Moment of Time Symmetry.!
     J.Math.Phys. Vol.12(1971) pp.70-78.\par}

\eject
% =============================================================================
\input gr-qc-figs

\HideDisplacementBoxes
\SetRokickiEPSFSpecial
\twelvepointsgl
\nopagenumbers
\overfullrule=0pt
\def\Bar{\overline}

\def\MyBig{\seventeenpointsgl}

\def\eps{\epsilon}
\def\Oeps(#1){{\cal O}(\eps^{#1})}
\def\m{{\hbox{\tt-}}}
\def\p{{\hbox{\tt+}}}
\advance\baselineskip 2pt
\advance\vsize 1.0cm\relax
\advance\voffset+1.0cm\relax
%
%-----------------------------------------------------------------------------
\def\stomp#1{\setbox0=\hbox{#1}\dp0=0pt\ht0=0pt\box0}
%
% This version of the \at routine ensures that its boxe's
% have no depth, width or height.
%
\def\at(#1,#2)#3{\vbox to 0pt{\kern#2cm%
                 \hbox to 0pt{\kern#1cm\stomp{#3}\hss}\vss}\nointerlineskip}
%-----------------------------------------------------------------------------
%
\parindent=0pt
\newbox\boxa
\newbox\image
\newbox\xlabel
\newbox\ylabel
\newbox\zlabel
\newbox\caption
\newbox\imagea
\newbox\imageb
\newbox\xlabela
\newbox\ylabela
\newbox\zlabela
\newbox\xlabelb
\newbox\ylabelb
\newbox\zlabelb
\newbox\capta
\newbox\captb
%
%-----------------------------------------------------------------------------
%
\setbox\capta=\hbox to 8cm{\hsize=8cm\vtop{%\raggedright%
{\bf Figure \figdef{FigTetA}}\ A simple approximation to a 2-sphere. Though the
legs of the tetrahedron appear straight they are taken to be geodesic segments
of the 2-sphere.}}
\setbox\captb=\hbox to 8cm{\hsize=8cm\vtop{%\raggedright%
{\bf Figure \figdef{FigTetB}}\ Better approximations for the 2-sphere are
obtained by sub-dividing each triangle according to this pattern. Newly created
vertices are placed at the centre of old legs and are later displaced radially
so that they, the new vertices, touch the 2-sphere.}}
\setbox\boxa=%
\hbox to 18.0cm{% Specifies the width of the box
\vtop to 18.5cm{% Specifies the depth of the box, the box has zero height.
\at(  1,10.5){\bBoxedEPSF{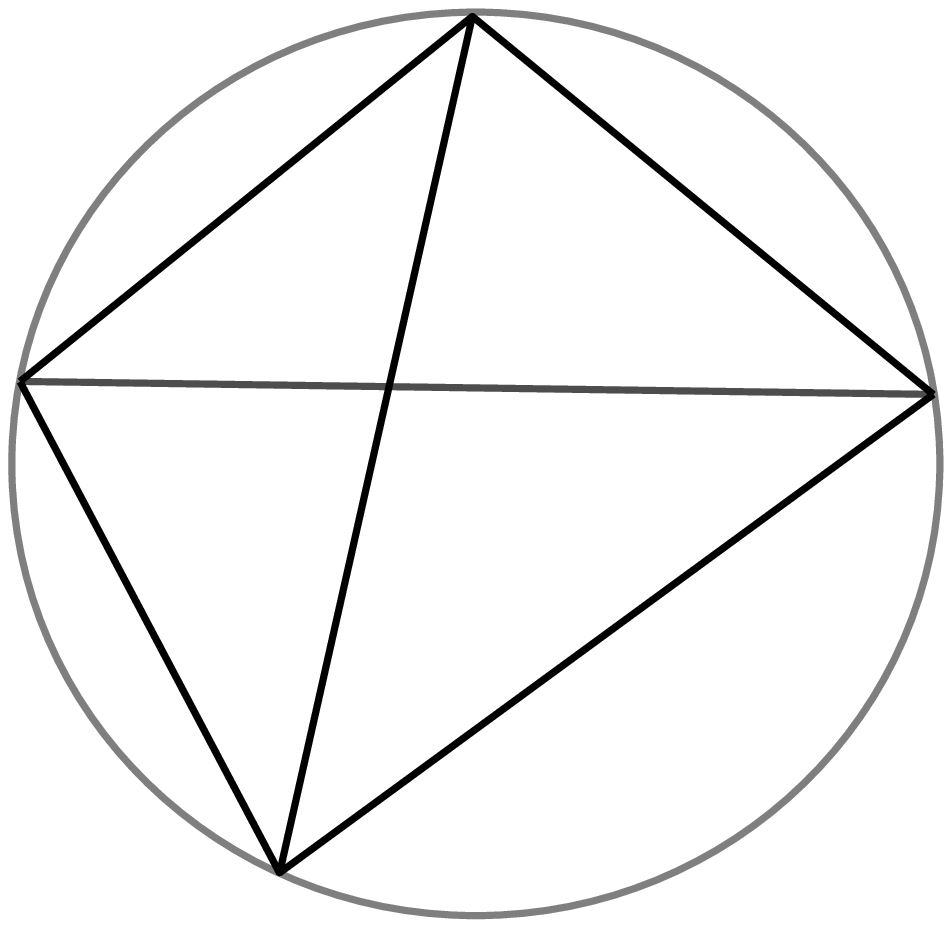 scaled 700}}
\at(  1,19.0){\bBoxedEPSF{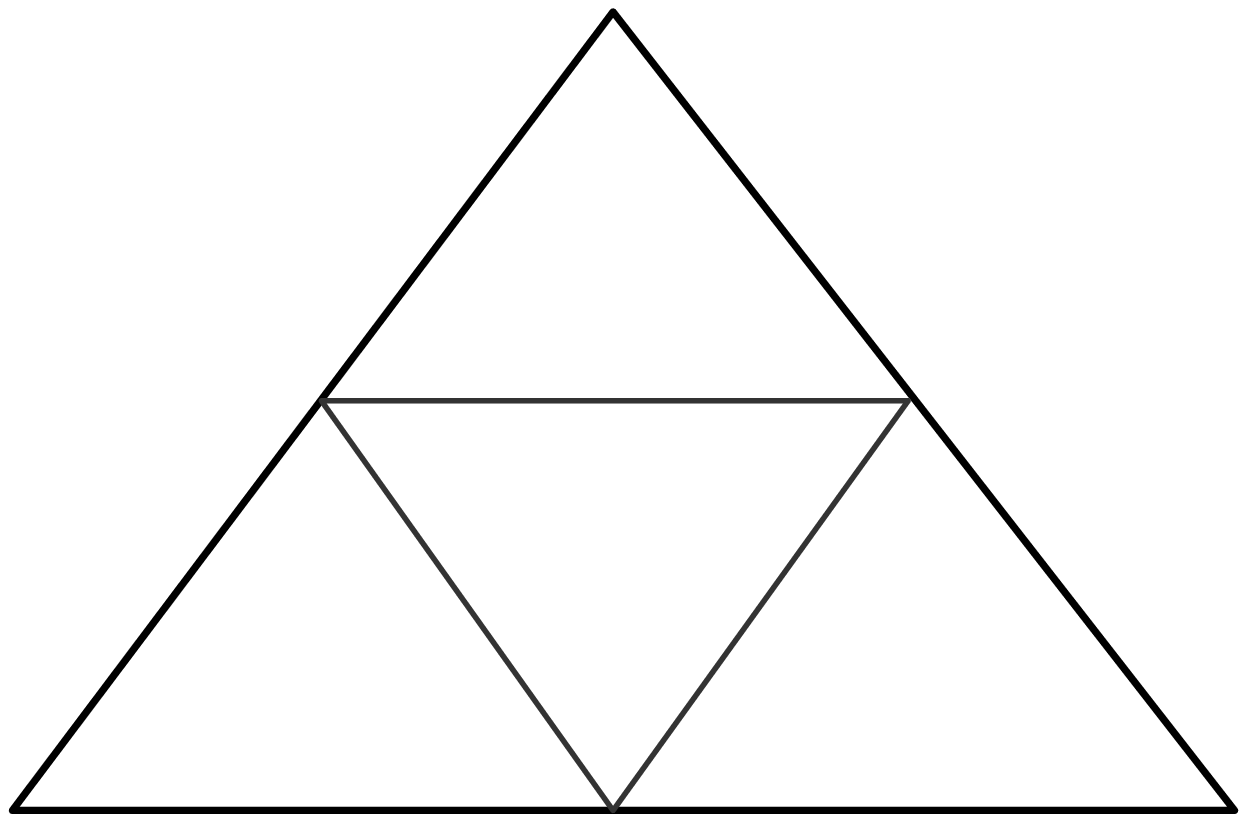 scaled 700}}
\at(  9, 7){\box\capta}
\at(  9, 13){\box\captb}
\vfill}\hfill}
%
% Now place the boxes on the page.
%
\centerline{\box\boxa}\vfill\eject
%-----------------------------------------------------------------------------
\setbox\caption=\hbox to 15cm{\hsize=15cm\vtop{%\raggedright%
{\bf Figure \figdef{TwoDPlotA}}\ The relative error in $\eta$ as a function
of the distance $l$ from the throat and of the curvature $R$. Note that the
errors vanish as $R\rightarrow2$ while they remain bounded for increasing $l$.}}
\setbox\image=\hbox{\bBoxedEPSF{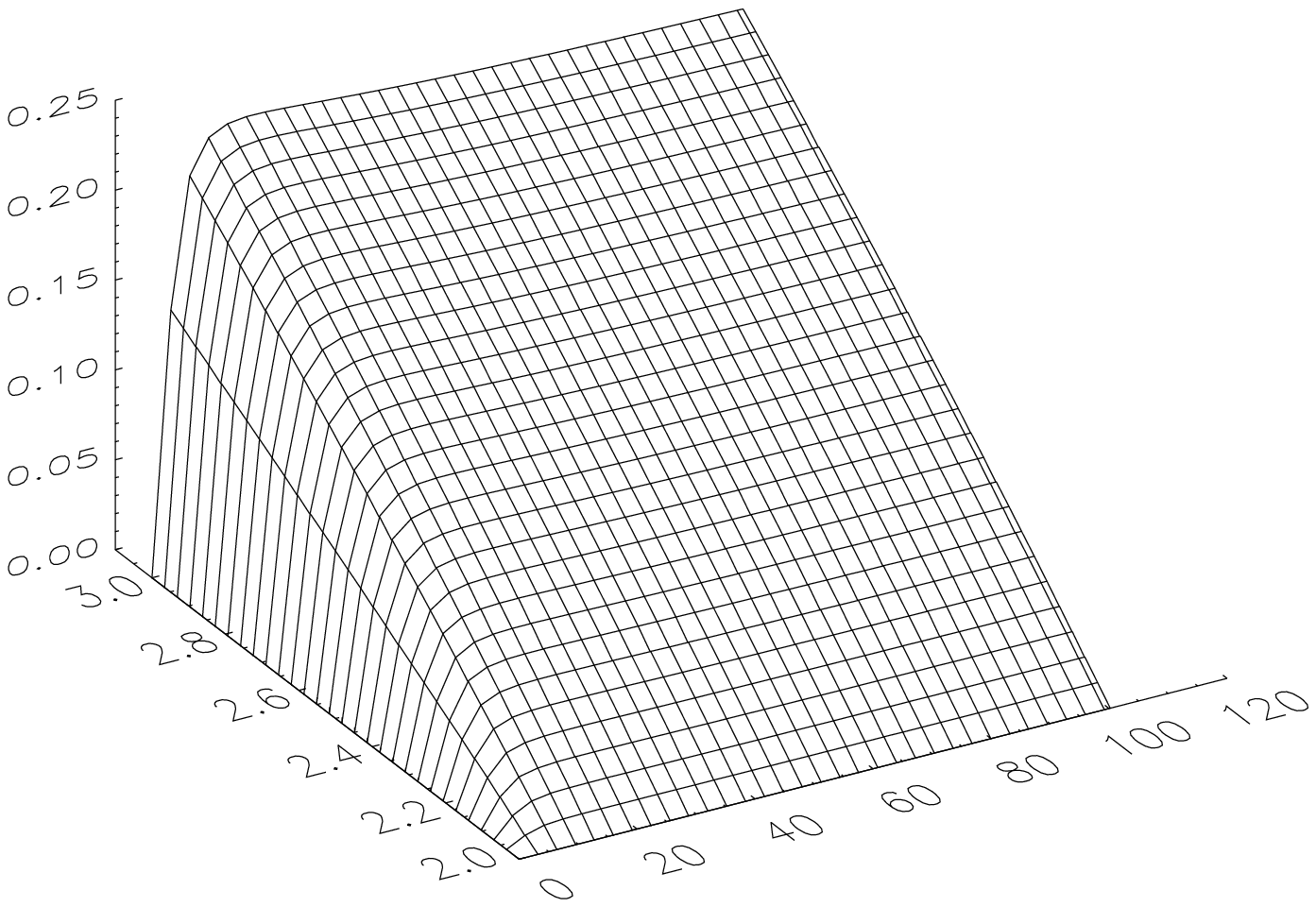 scaled 1000}}
\setbox\xlabel=\hbox{\MyBig $l$}
\setbox\ylabel=\hbox{\MyBig $R$}
\setbox\zlabel=\hbox{\MyBig $-1+\eta/{\tilde\eta}$}
\setbox\zlabel=\hbox{\rotl\zlabel}
\setbox\boxa=%
\hbox to 18.0cm{% Specifies the width of the box
\vtop to 18.5cm{% Specifies the depth of the box, the box has zero height.
\at(  0.0,20.0){\box\image}
\at(  1.0,22.0){\box\caption}
\at(  4.0,18.0){\box\ylabel}
\at( 13.0,19.0){\box\xlabel}
\at(  0.5,14.0){\box\zlabel}
\vfill}\hfill}
%
% Now place the boxes on the page.
%
\centerline{\box\boxa}\vfill\eject
%-----------------------------------------------------------------------------
\setbox\caption=\hbox to 15cm{\hsize=15cm\vtop{%\raggedright%
{\bf Figure \figdef{TwoDPlotB}}\ A plot of the relative error
in $\eta$ for various fixed values of $l$. This plot shows that the error
in the metric varies linearly with the error in the curvature.}}
\setbox\image=\hbox{\bBoxedEPSF{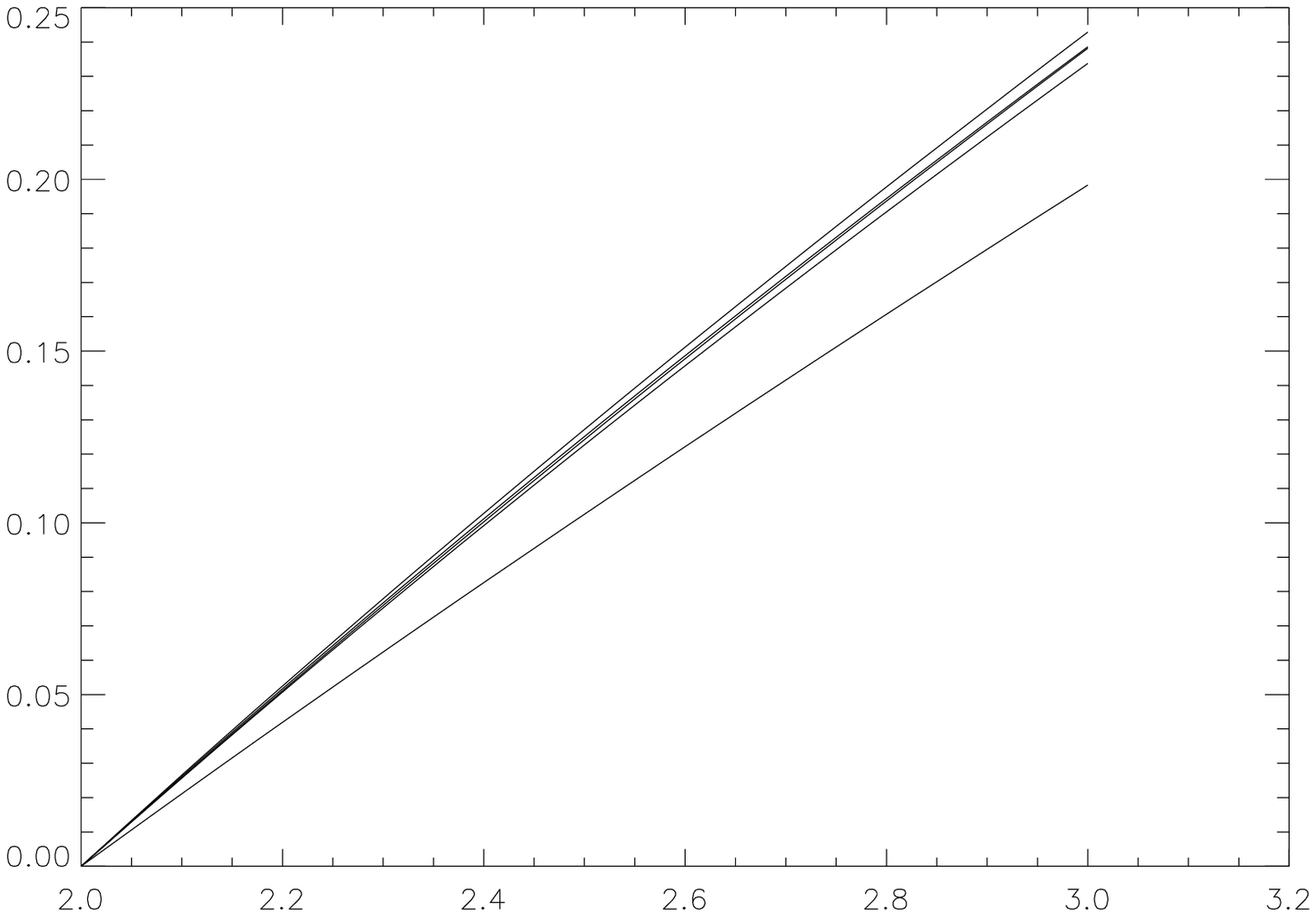 scaled 900}}
\setbox\xlabel=\hbox{\MyBig$R$}
\setbox\ylabel=\hbox{\MyBig$-1+\eta/{\tilde\eta}$}
\setbox\ylabel=\hbox{\rotl\ylabel}
\setbox\boxa=%
\hbox to 18.0cm{% Specifies the width of the box
\vtop to 18.5cm{% Specifies the depth of the box, the box has zero height.
\at( 0.0,20.0){\box\image}
\at( 1.0,22.0){\box\caption}
\at( 8.5,20.5){\box\xlabel}
\at( 0.0,15.0){\box\ylabel}
\at( 13.5,11.5){$l=5$}
\at(  7.5,11.5){$l=10,20,40,80$}
\vfill}\hfill}
%
% Now place the boxes on the page.
%
\centerline{\box\boxa}\vfill\eject
%-----------------------------------------------------------------------------
\setbox\caption=\hbox to 15cm{\hsize=15cm\vtop{%\raggedright%
{\bf Figure \figdef{TwoDPlotC}}\ This is similar to Figure \figrfr{TwoDPlotA}
but using the alternative definition of the error, equation \eqnrfr{SecondError}.
Notice that the errors are not bounded for increasing $l$.}}
\setbox\image=\hbox{\bBoxedEPSF{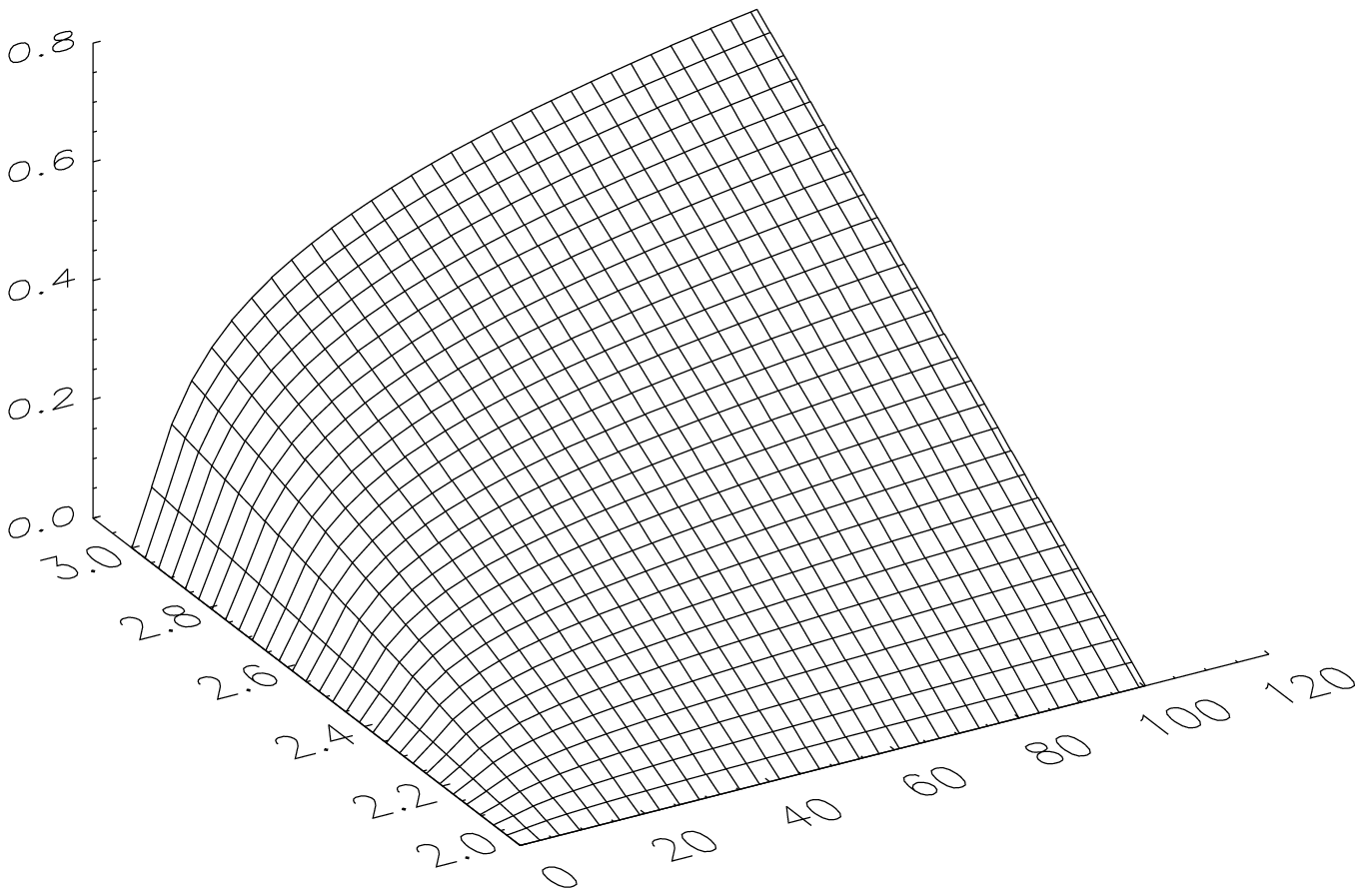 scaled 1000}}
\setbox\xlabel=\hbox{\MyBig$l$}
\setbox\ylabel=\hbox{\MyBig$R$}
\setbox\zlabel=\hbox{\MyBig $-1+\eta/(r\rho^2)$}
\setbox\zlabel=\hbox{\rotl\zlabel}
\setbox\boxa=%
\hbox to 18.0cm{% Specifies the width of the box
\vtop to 18.5cm{% Specifies the depth of the box, the box has zero height.
\at(  0.0,20.0){\box\image}
\at(  1.0,22.0){\box\caption}
\at(  4.0,18.0){\box\ylabel}
\at( 13.0,19.0){\box\xlabel}
\at(  0.5,14.0){\box\zlabel}
\vfill}\hfill}
%
% Now place the boxes on the page.
%
\centerline{\box\boxa}\vfill\eject
%-----------------------------------------------------------------------------
\setbox\caption=\hbox to 15cm{\hsize=15cm\vtop{%\raggedright%
{\bf Figure \figdef{TwoDPlotD}}\ This demonstrates that even though the relative
errors in Figure \figrfr{TwoDPlotC} are unbounded for large $l$, they do
vanish as $R\rightarrow2$ at fixed $l$.}}
\setbox\image=\hbox{\bBoxedEPSF{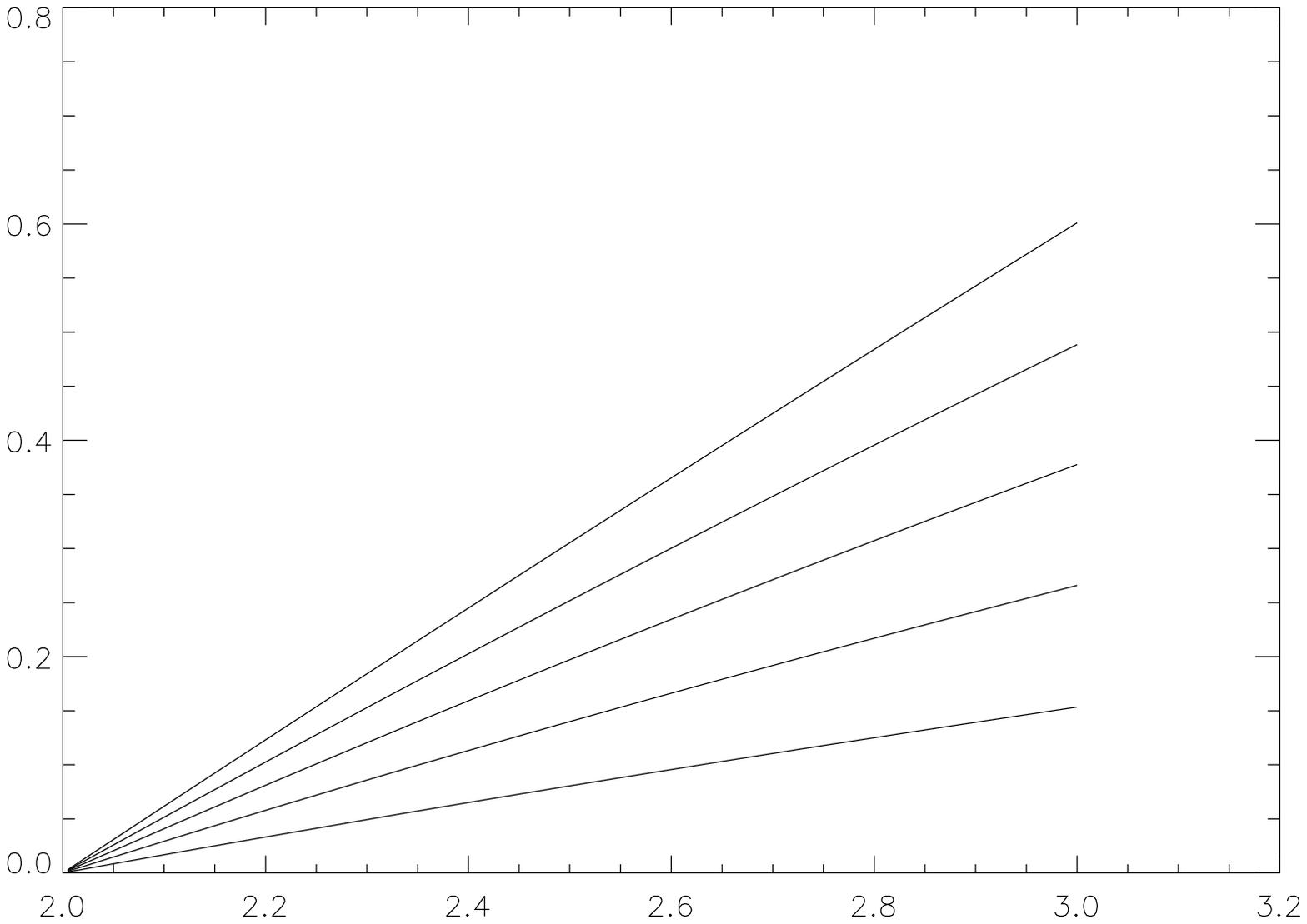 scaled 900}}
\setbox\xlabel=\hbox{\MyBig$R$}
\setbox\ylabel=\hbox{\MyBig$-1+\eta/(r\rho^2)$}
\setbox\ylabel=\hbox{\rotl\ylabel}
\setbox\boxa=%
\hbox to 18.0cm{% Specifies the width of the box
\vtop to 18.5cm{% Specifies the depth of the box, the box has zero height.
\at( 0.0,20.0){\box\image}
\at( 1.0,22.0){\box\caption}
\at( 8.5,20.5){\box\xlabel}
\at( 0.0,15.0){\box\ylabel}
\at( 13.5,17.4){$l=5$}
\at( 13.5,16.0){$l=10$}
\at( 13.5,14.7){$l=20$}
\at( 13.5,13.4){$l=40$}
\at( 13.5,12.1){$l=80$}
\vfill}\hfill}
%
% Now place the boxes on the page.
%
\centerline{\box\boxa}\vfill\eject
%-----------------------------------------------------------------------------
\setbox\caption=\hbox to 15cm{\hsize=15cm\vtop{%\raggedright%
{\bf Figure \figdef{ThreeLatticeA}}\ A tube of cells that extends from the throat
to the far flat regions of the space. Each pair of cells will be covered by a
Riemann normal coordinate frame. The edges are geodesic segments of the local
smooth lattice metric. The four radial edges are global geodesics.}}
\setbox\image=\hbox{\bBoxedEPSF{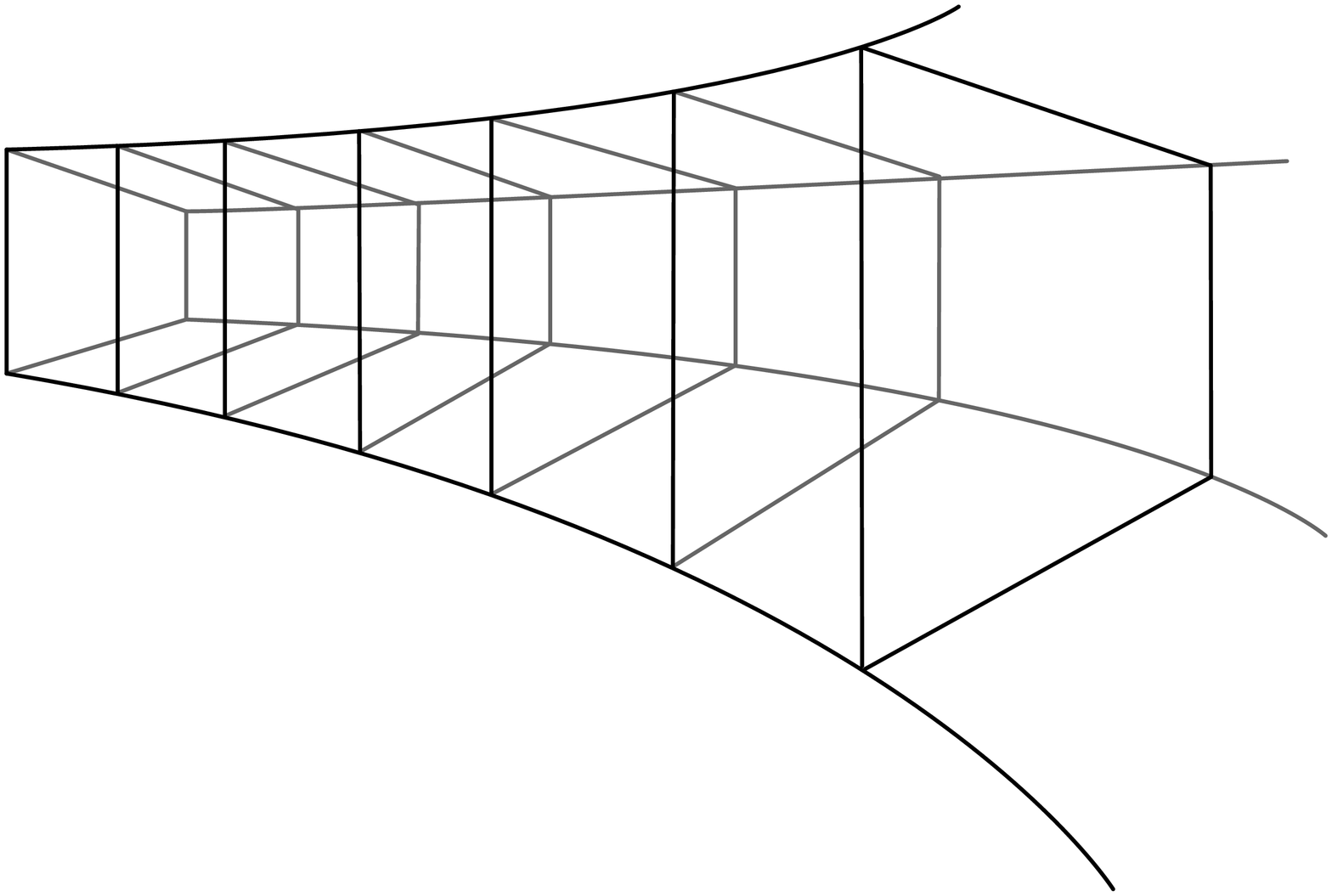 scaled 1000}}
%\setbox\image=\hbox{\rotl\image}
%
\setbox\boxa=%
\hbox to 24.0cm{% Specifies the width of the box
\vtop to 15.0cm{% Specifies the depth of the box, the box has zero height.
\at( 0,15.0){\box\image}
\at( 1,14.0){\box\caption}
\vfill}\hfill}
%
% Now place the boxes on the page.
%
\setbox\boxa=\hbox{\rotl\boxa}
\centerline{\box\boxa}\vfill\eject
%-----------------------------------------------------------------------------
\setbox\caption=\hbox to 17cm{\hsize=17cm\vtop{%\raggedright%
{\bf Figure \figdef{ThreeLatticeB}}\ The generic pair of cells for one Riemann
normal coordinate frame. The $z$-axis runs up the middle of the tube.}}
\setbox\image=\hbox{\bBoxedEPSF{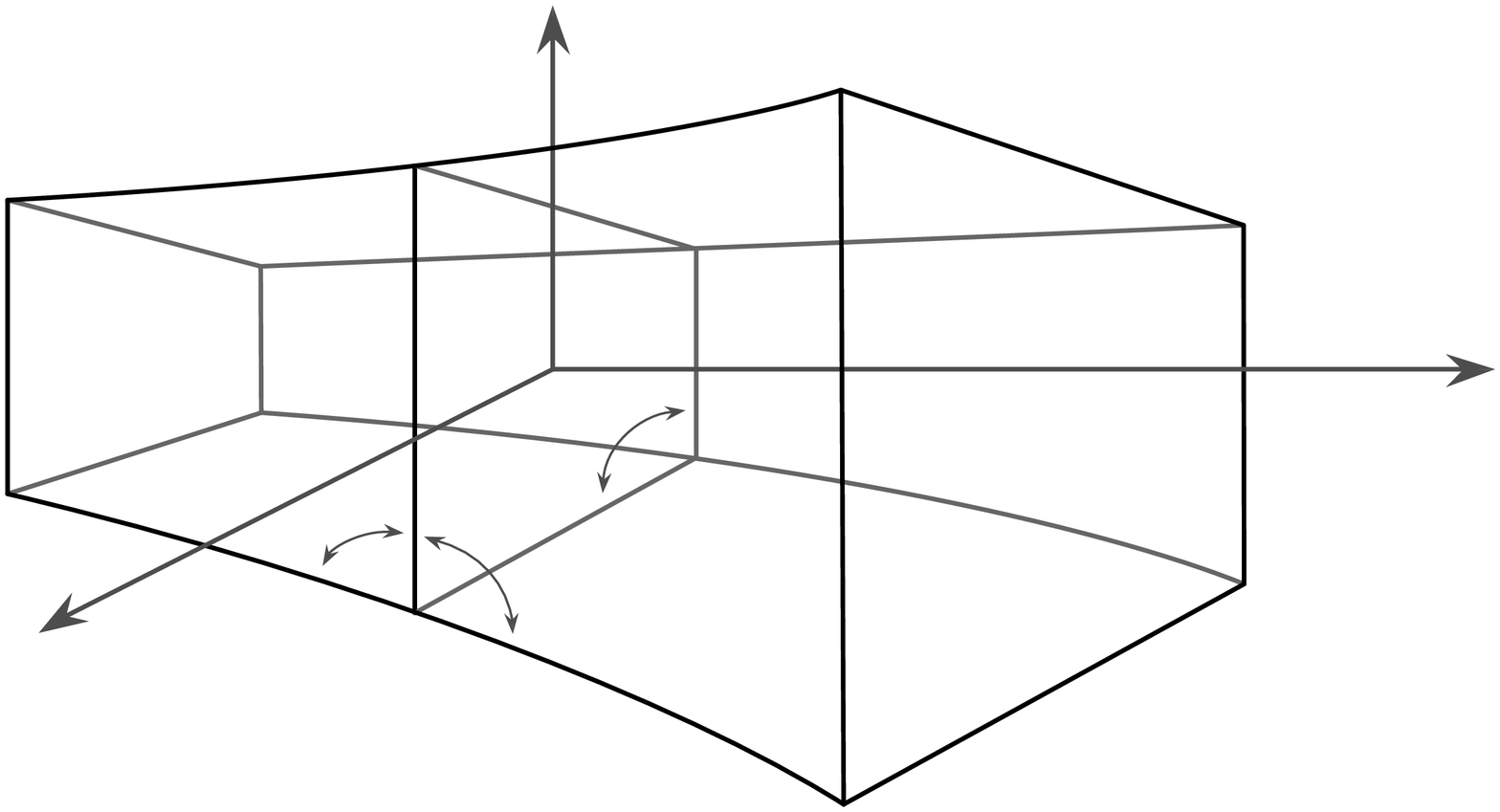 scaled 1000}}
\setbox\boxa=%
\hbox to 24.0cm{% Specifies the width of the box
\vtop to 15.0cm{% Specifies the depth of the box, the box has zero height.
\at( 0.0,12.0){\box\image}
\at( 1.0,14.0){\box\caption}
%\at( 2.5, 5.0){\MyBig$\beta^\m$}
\at( 9.5, 5.9){\MyBig$\beta$}
%\at(18.3, 7.5){\MyBig$\beta^\p$}
\at( 5.5, 7.5){\MyBig$\alpha^\m$}
\at( 7.5, 7.7){\MyBig$\alpha^\p$}
\at( 2.9, 6.9){\MyBig$L^\m$}
\at( 9.6, 8.3){\MyBig$L$}
\at(17.2,11.2){\MyBig$L^\p$}
\at( 3.3, 1.6){\MyBig$d^\m$}
\at(10.7, 0.7){\MyBig$d^\p$}
\at( 9.5,-0.5){\MyBig$X$}
\at( 1.0,10.5){\MyBig$Y$}
\at(23.0, 4.2){\MyBig$Z$}
\at( 6.5,10.2){\MyBig$1$}
\at( 6.5, 1.3){\MyBig$2$}
\at(11.4, 2.7){\MyBig$3$}
\at(11.4, 7.3){\MyBig$4$}
\at(13.2,13.0){\MyBig$1^\p$}
\at(13.2, 0.3){\MyBig$2^\p$}
\at(20.0, 2.5){\MyBig$3^\p$}
\at(20.0, 9.0){\MyBig$4^\p$}
\at( 0.0, 8.0){\MyBig$1^\m$}
\at( 0.0, 1.8){\MyBig$2^\m$}
\at( 4.2, 3.0){\MyBig$3^\m$}
\at( 4.2, 6.5){\MyBig$4^\m$}
\vfill}\hfill}
%
% Now place the boxes on the page.
%
\setbox\boxa=\hbox{\rotl\boxa}
\centerline{\box\boxa}\vfill\eject
%-----------------------------------------------------------------------------
\setbox\caption=\hbox to 15cm{\hsize=15cm\vtop{%\raggedright%
{\bf Figure \figdef{ThreeDPlotA}}\ The relative errors $EL(l,\eps)$ and
$ER_x(l,\eps)$. Note that the errors die quickly as $\eps\rightarrow0$ yet
they are unbounded for large $l$. The errors $ER_z$ are identical to $ER_x$.}}
\setbox\imagea=\hbox{\bBoxedEPSF{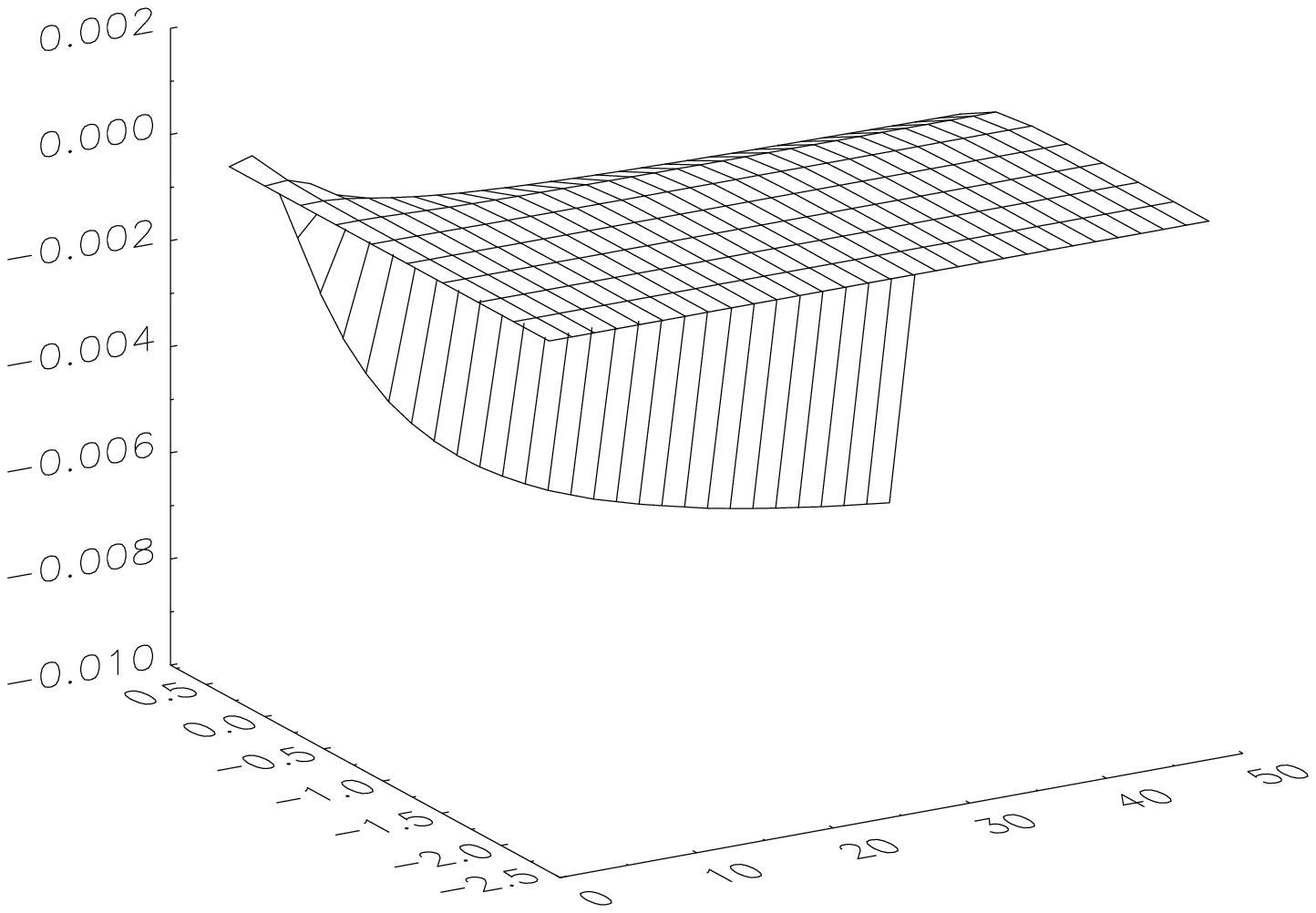 scaled 1000}}
\setbox\imageb=\hbox{\bBoxedEPSF{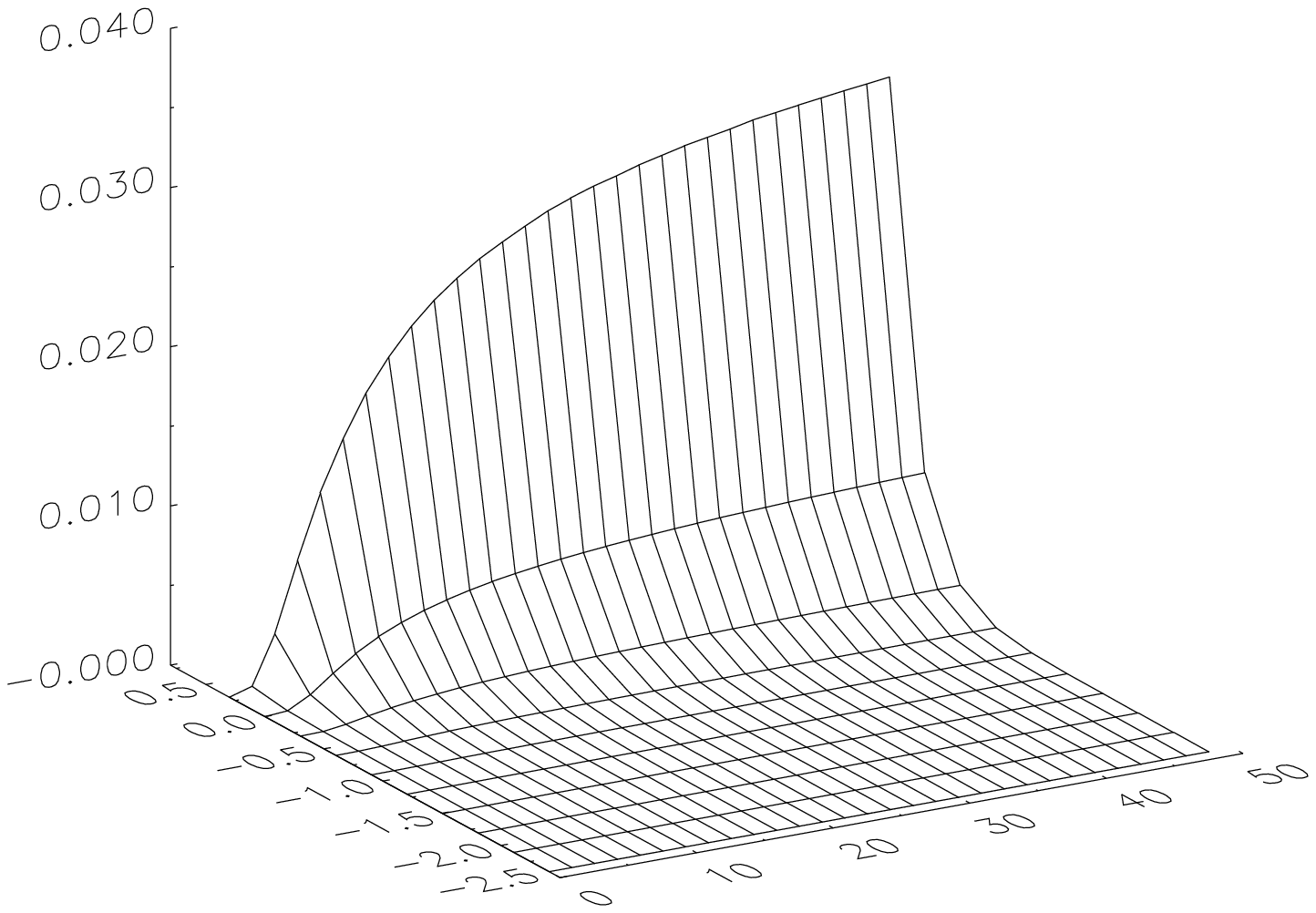 scaled 1000}}
\setbox\xlabela=\hbox{\MyBig $l$}
\setbox\ylabela=\hbox{\MyBig $\log \eps$}
\setbox\zlabela=\hbox{\MyBig $EL(l,\eps)$}
\setbox\zlabela=\hbox{\rotl\zlabela}
\setbox\xlabelb=\hbox{\MyBig $l$}
\setbox\ylabelb=\hbox{\MyBig $\log \eps$}
\setbox\zlabelb=\hbox{\MyBig $ER_x(l,\eps)$}
\setbox\zlabelb=\hbox{\rotl\zlabelb}
\setbox\boxa=%
\hbox to 18.0cm{% Specifies the width of the box
\vtop to 22.0cm{% Specifies the depth of the box, the box has zero height.
\at( 0.0,10.0){\box\imagea}
\at(13.5, 9.2){\box\xlabela}
\at( 3.0, 8.5){\box\ylabela}
\at( 0.5, 4.0){\box\zlabela}
\at( 0.0,21.5){\box\imageb}
\at(13.5,20.7){\box\xlabelb}
\at( 3.0,20.0){\box\ylabelb}
\at( 0.5,15.5){\box\zlabelb}
\at( 1.0,22.0){\box\caption}
\vfill}\hfill}
%
% Now place the boxes on the page.
%
\centerline{\box\boxa}\vfill\eject
%-----------------------------------------------------------------------------
\setbox\caption=\hbox to 15cm{\hsize=15cm\vtop{%\raggedright%
{\bf Figure \figdef{ThreeDPlotC}}\ Cross sections of Figure
\figrfr{ThreeDPlotA} for various values of $l$. In each
case the error converges quadraticly with respect to $\eps$. The irregular
behaviour in $ER_x$ is probably due to round-off error.
The curves correspond to $l=5,10,20,40$ and $80$ with the error generally
increasing with $l$. This applies also to Figure \figrfr{ThreeDPlotG}.}}
\setbox\imagea=\hbox{\bBoxedEPSF{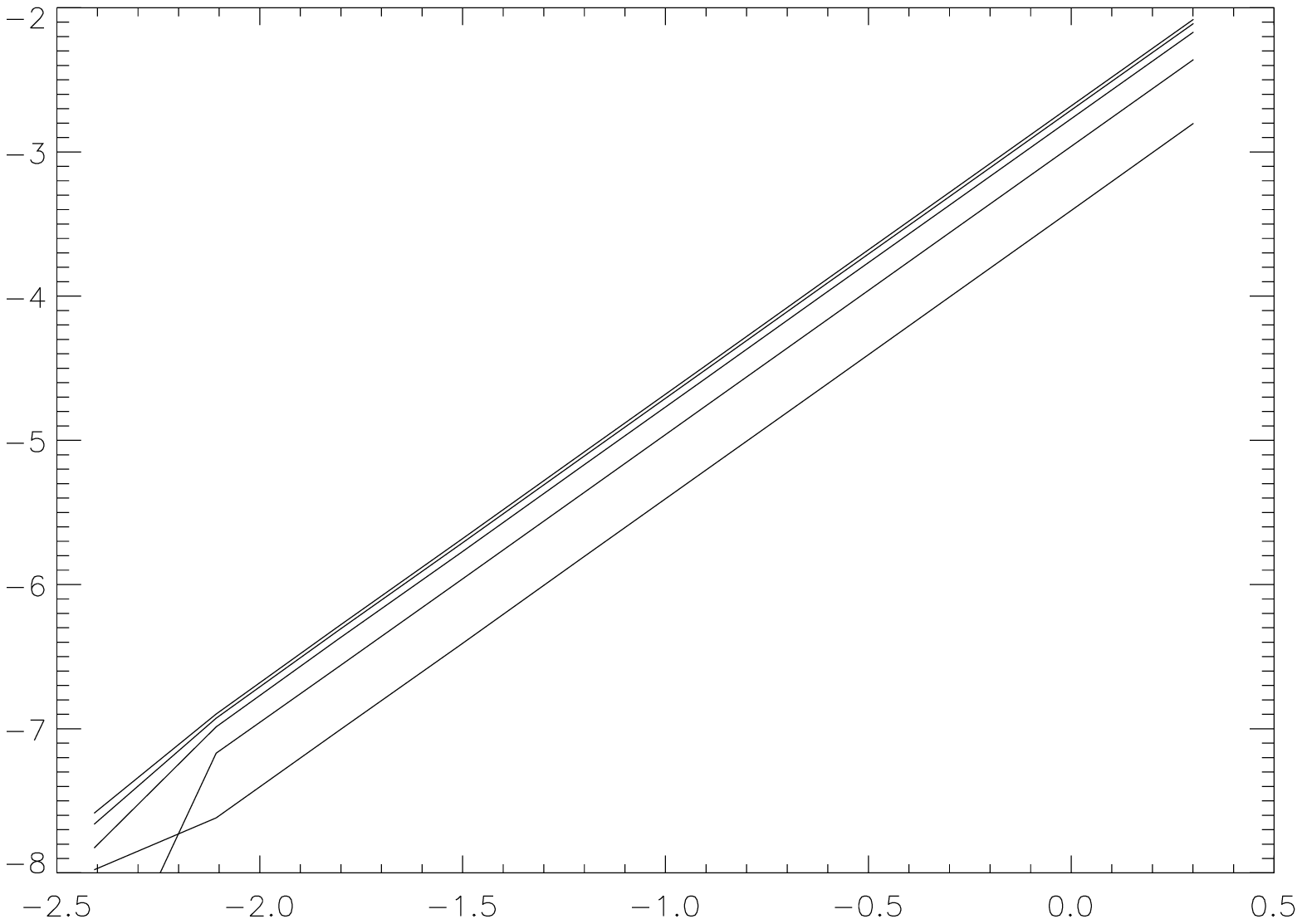 scaled 900}}
\setbox\imageb=\hbox{\bBoxedEPSF{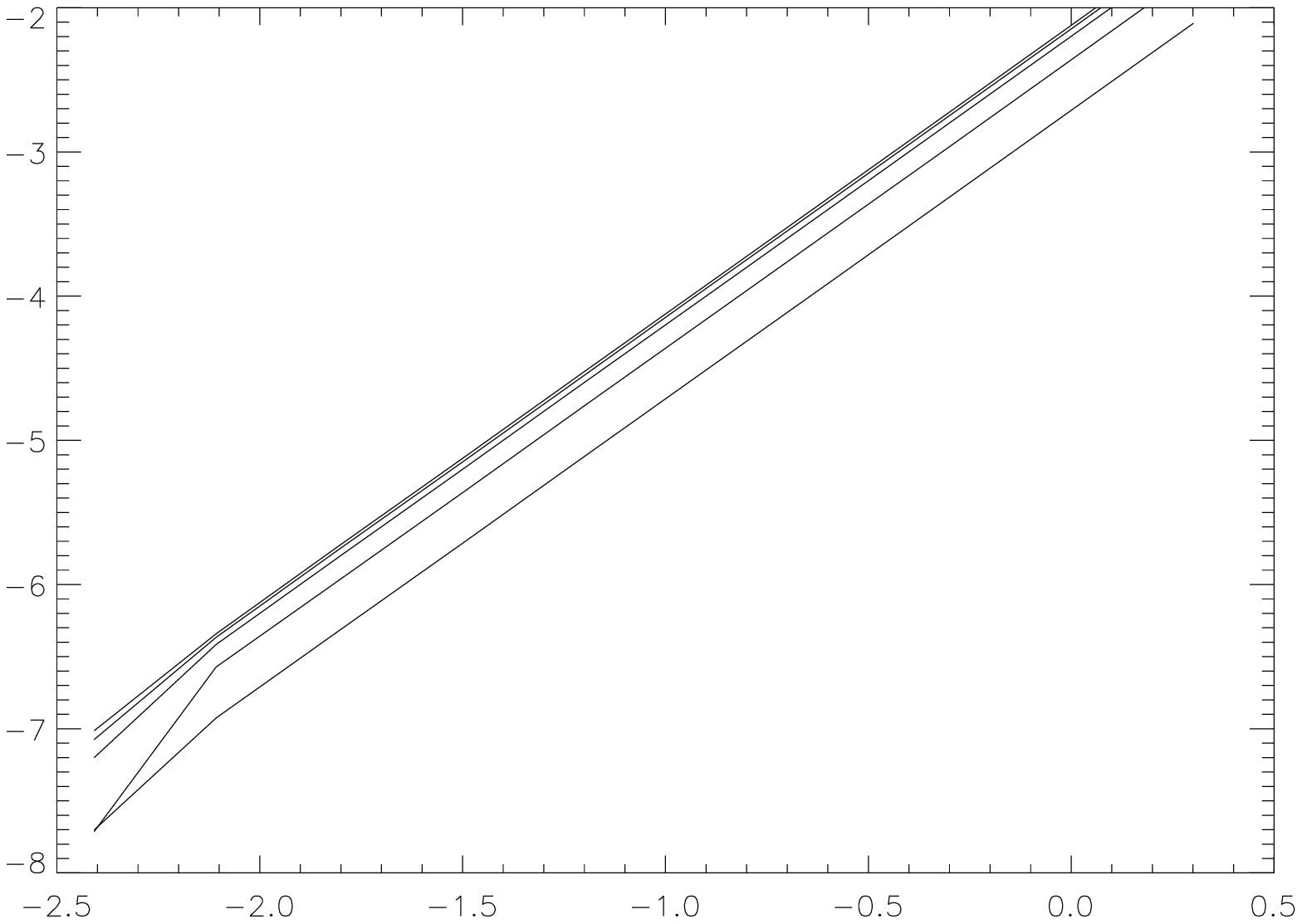 scaled 900}}
\setbox\xlabel=\hbox{\MyBig$\eps$}
\setbox\ylabela=\hbox{\MyBig$EL(l,\eps)$}
\setbox\ylabelb=\hbox{\MyBig$ER_x(l,\eps)$}
\setbox\ylabela=\hbox{\rotl\ylabela}
\setbox\ylabelb=\hbox{\rotl\ylabelb}
\setbox\boxa=%
\hbox to 18.0cm{% Specifies the width of the box
\vtop to 22.0cm{% Specifies the depth of the box, the box has zero height.
\at( 0, 9.0){\box\imagea}
\at( 0,20.0){\box\imageb}
\at( 8.5,20.5){\box\xlabel}
\at( 0, 4.0){\box\ylabela}
\at( 0,15.0){\box\ylabelb}
\at( 1,22.0){\box\caption}
\vfill}\hfill}
%
% Now place the boxes on the page.
%
\centerline{\box\boxa}\vfill\eject
%-----------------------------------------------------------------------------
\setbox\caption=\hbox to 15cm{\hsize=15cm\vtop{%\raggedright%
{\bf Figure \figdef{ThreeDPlotE}}\ Errors $EL(l,\eps)$ and $ER_x(l,\eps)$ for
the first variation in which the alternative inner boundary condition
\eqnrfr{AltBianchi} was used. The oscillations in $ER_x$ are not significant
since $R_x\approx10^{-5}$ at $l=50$ and the absolute errors are
very small.}}
\setbox\imagea=\hbox{\bBoxedEPSF{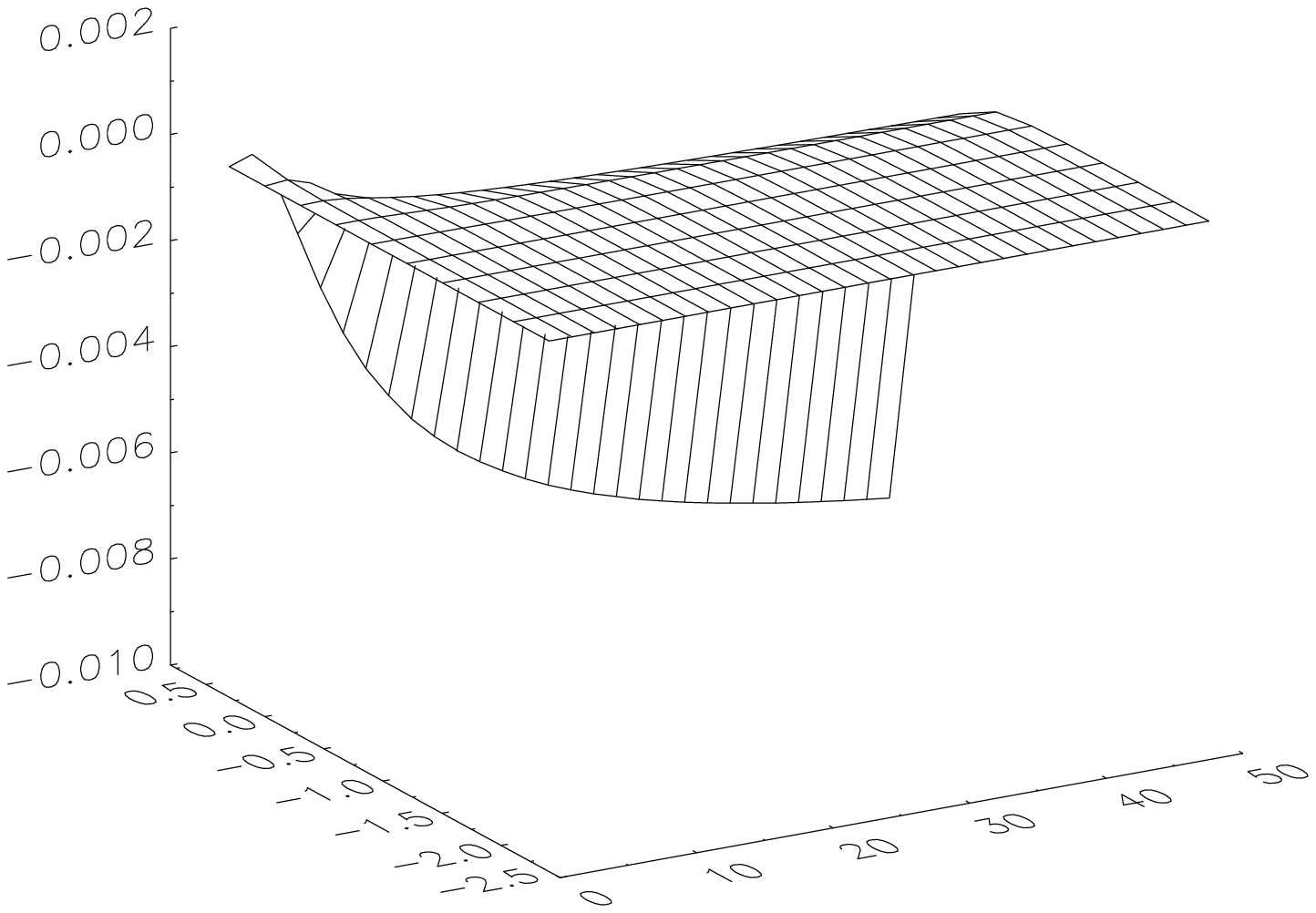 scaled 1000}}
\setbox\imageb=\hbox{\bBoxedEPSF{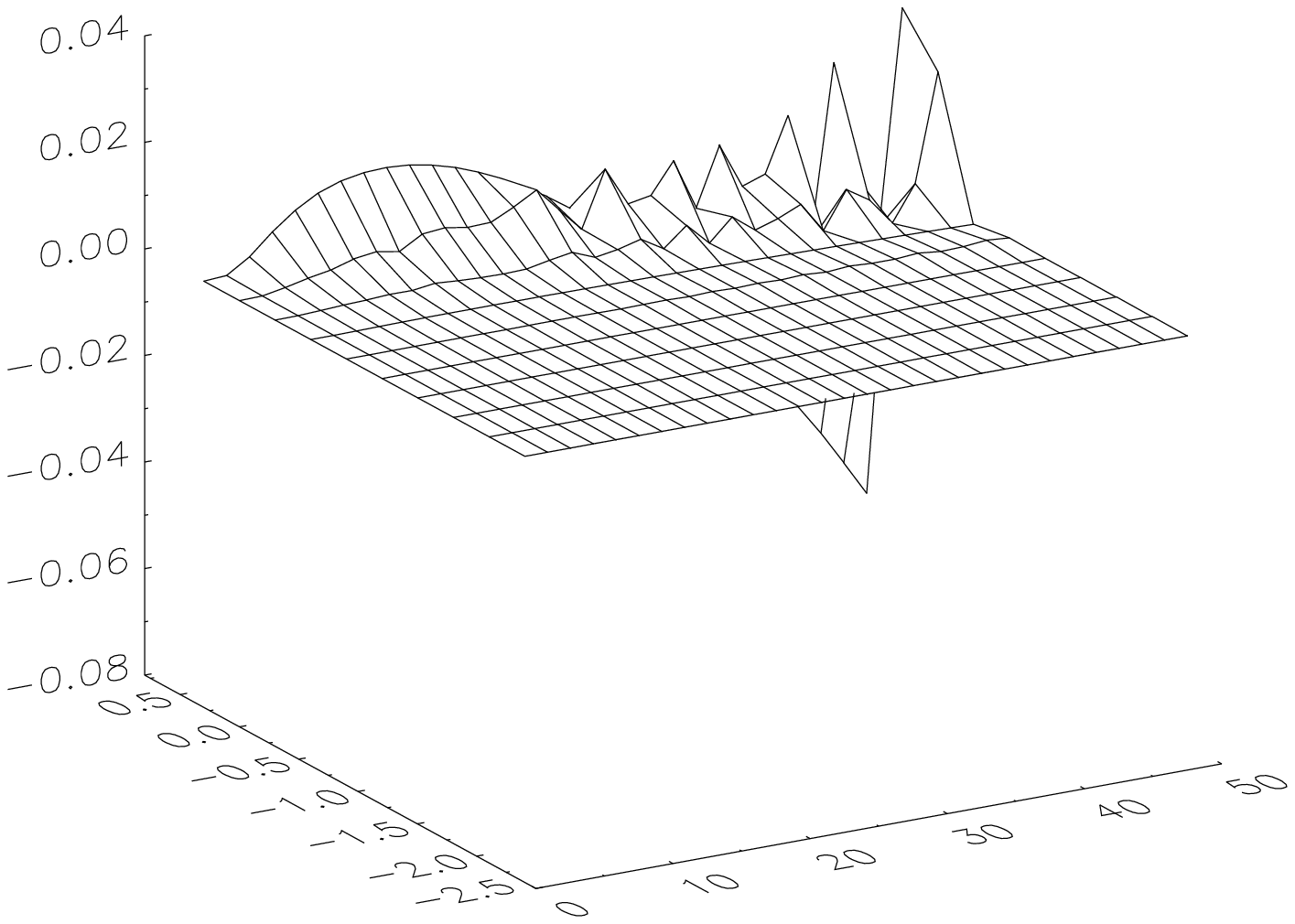 scaled 1000}}
\setbox\xlabela=\hbox{\MyBig $l$}
\setbox\ylabela=\hbox{\MyBig $\log \eps$}
\setbox\zlabela=\hbox{\MyBig $EL(l,\eps)$}
\setbox\zlabela=\hbox{\rotl\zlabela}
\setbox\xlabelb=\hbox{\MyBig $l$}
\setbox\ylabelb=\hbox{\MyBig $\log \eps$}
\setbox\zlabelb=\hbox{\MyBig $ER_x(l,\eps)$}
\setbox\zlabelb=\hbox{\rotl\zlabelb}
\setbox\boxa=%
\hbox to 18.0cm{% Specifies the width of the box
\vtop to 22.0cm{% Specifies the depth of the box, the box has zero height.
\at( 0.0,10.0){\box\imagea}
\at(13.5, 9.2){\box\xlabela}
\at( 3.0, 8.5){\box\ylabela}
\at( 0.5, 4.0){\box\zlabela}
\at( 0.0,21.5){\box\imageb}
\at(13.5,20.7){\box\xlabelb}
\at( 3.0,20.0){\box\ylabelb}
\at( 0.5,15.5){\box\zlabelb}
\at( 1.0,22.0){\box\caption}
\vfill}\hfill}
%
% Now place the boxes on the page.
%
\centerline{\box\boxa}\vfill\eject
%-----------------------------------------------------------------------------
\setbox\caption=\hbox to 15cm{\hsize=15cm\vtop{%\raggedright%
{\bf Figure \figdef{ThreeDPlotG}}\ Cross sections of Figure
\figrfr{ThreeDPlotE}. In each case the error converges quadraticly with
respect to $\eps$. The irregular behaviour in $ER_x$ is due to the
oscillations in the curvatures.}}
\setbox\imagea=\hbox{\bBoxedEPSF{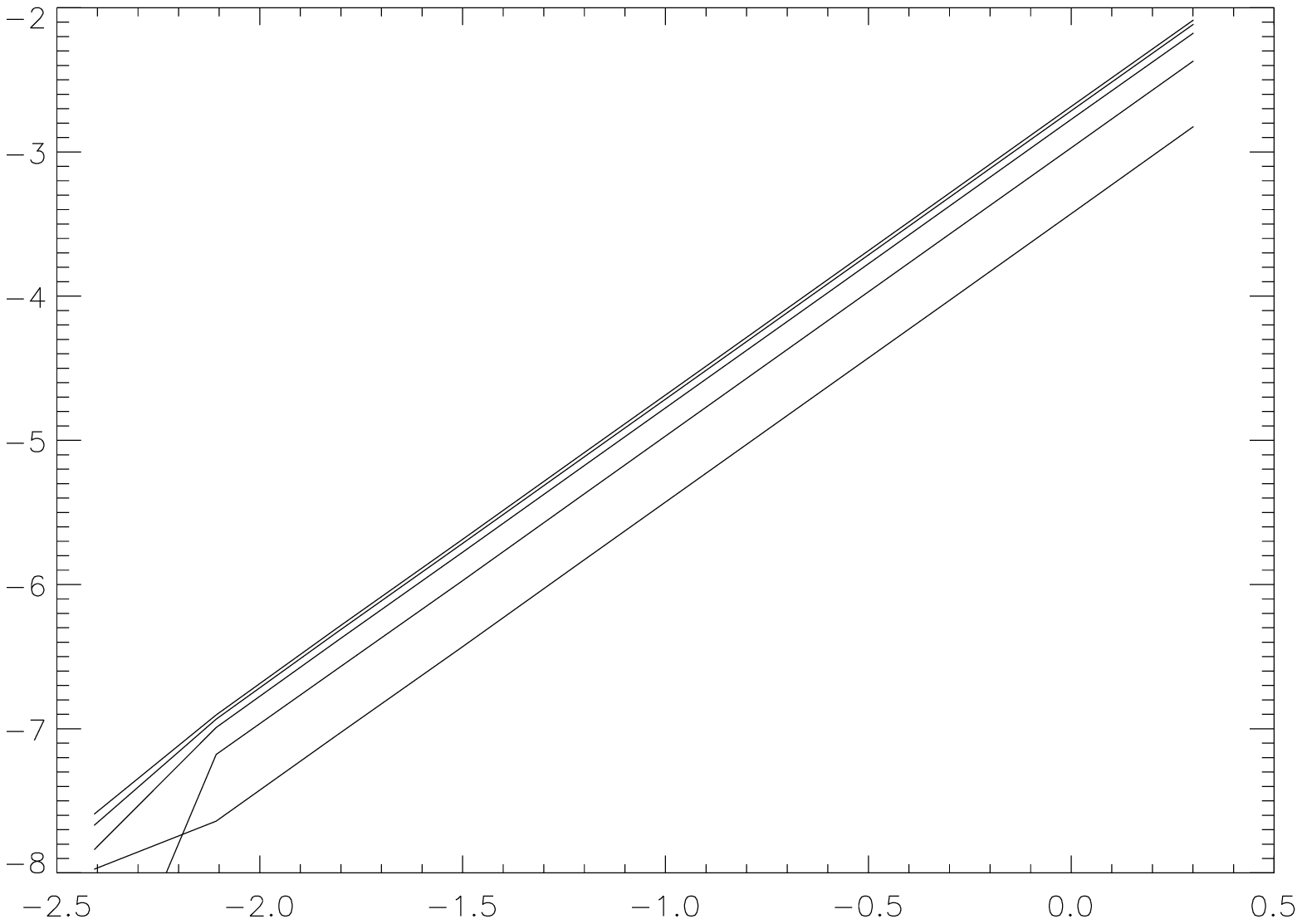 scaled 900}}
\setbox\imageb=\hbox{\bBoxedEPSF{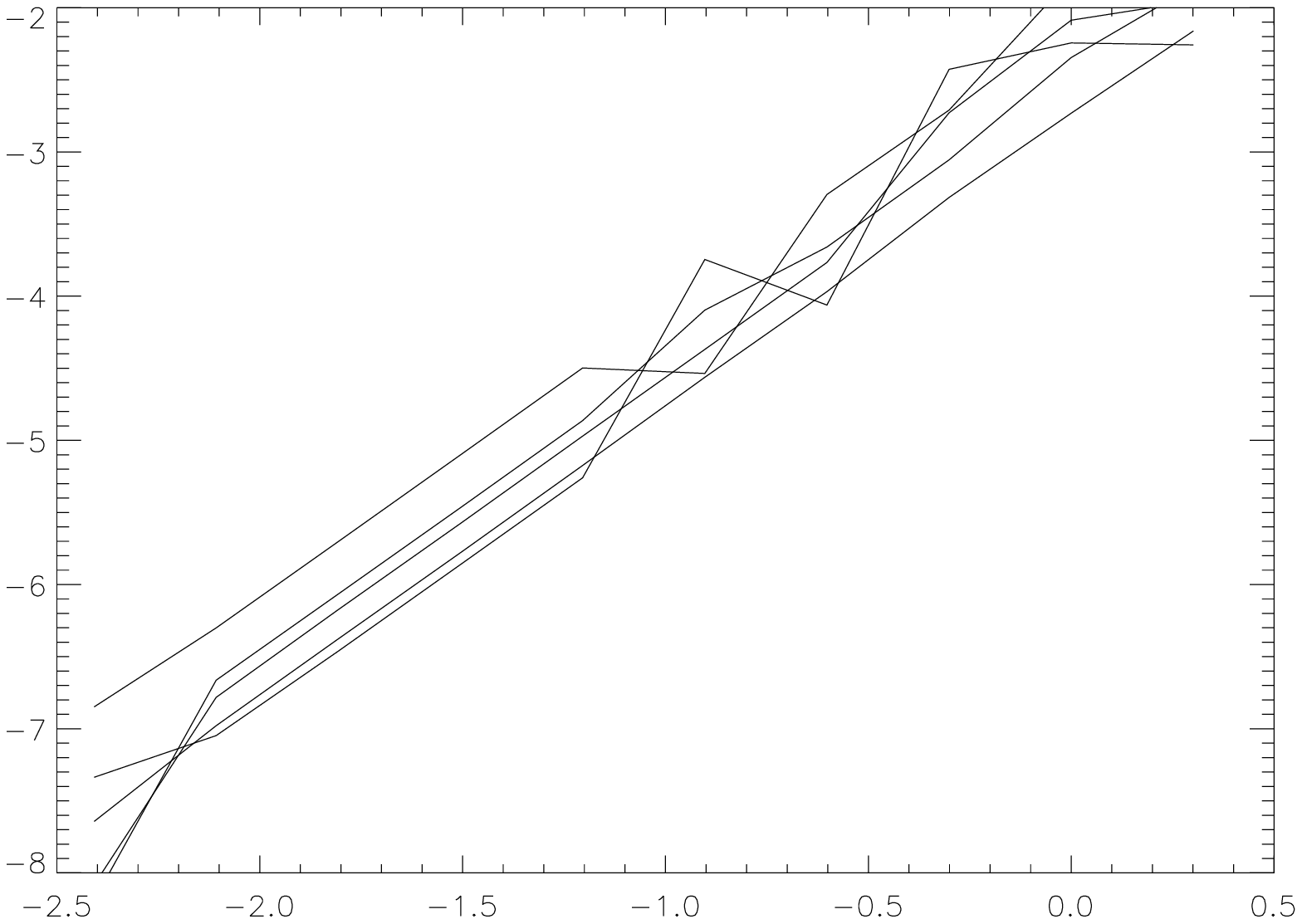 scaled 900}}
\setbox\xlabel=\hbox{\MyBig$\eps$}
\setbox\ylabela=\hbox{\MyBig$EL(l,\eps)$}
\setbox\ylabelb=\hbox{\MyBig$ER_x(l,\eps)$}
\setbox\ylabela=\hbox{\rotl\ylabela}
\setbox\ylabelb=\hbox{\rotl\ylabelb}
\setbox\boxa=%
\hbox to 18.0cm{% Specifies the width of the box
\vtop to 22.0cm{% Specifies the depth of the box, the box has zero height.
\at( 0, 9.0){\box\imagea}
\at( 0,20.0){\box\imageb}
\at( 8.5,20.5){\box\xlabel}
\at( 0, 4.0){\box\ylabela}
\at( 0,15.0){\box\ylabelb}
\at( 1,22.0){\box\caption}
\vfill}\hfill}
%
% Now place the boxes on the page.
%
\centerline{\box\boxa}\vfill\eject
%-----------------------------------------------------------------------------
\setbox\caption=\hbox to 15cm{\hsize=15cm\vtop{%\raggedright%
{\bf Figure \figdef{ThreeDPlotM}}\ Errors $EL(l,\eps)$ and $ER_x(l,\eps)$ for
the third variation in which
successive tiles were scaled versions of previous tiles. There is no convergence
in either $\eps$ or $l$.}}
\setbox\imagea=\hbox{\bBoxedEPSF{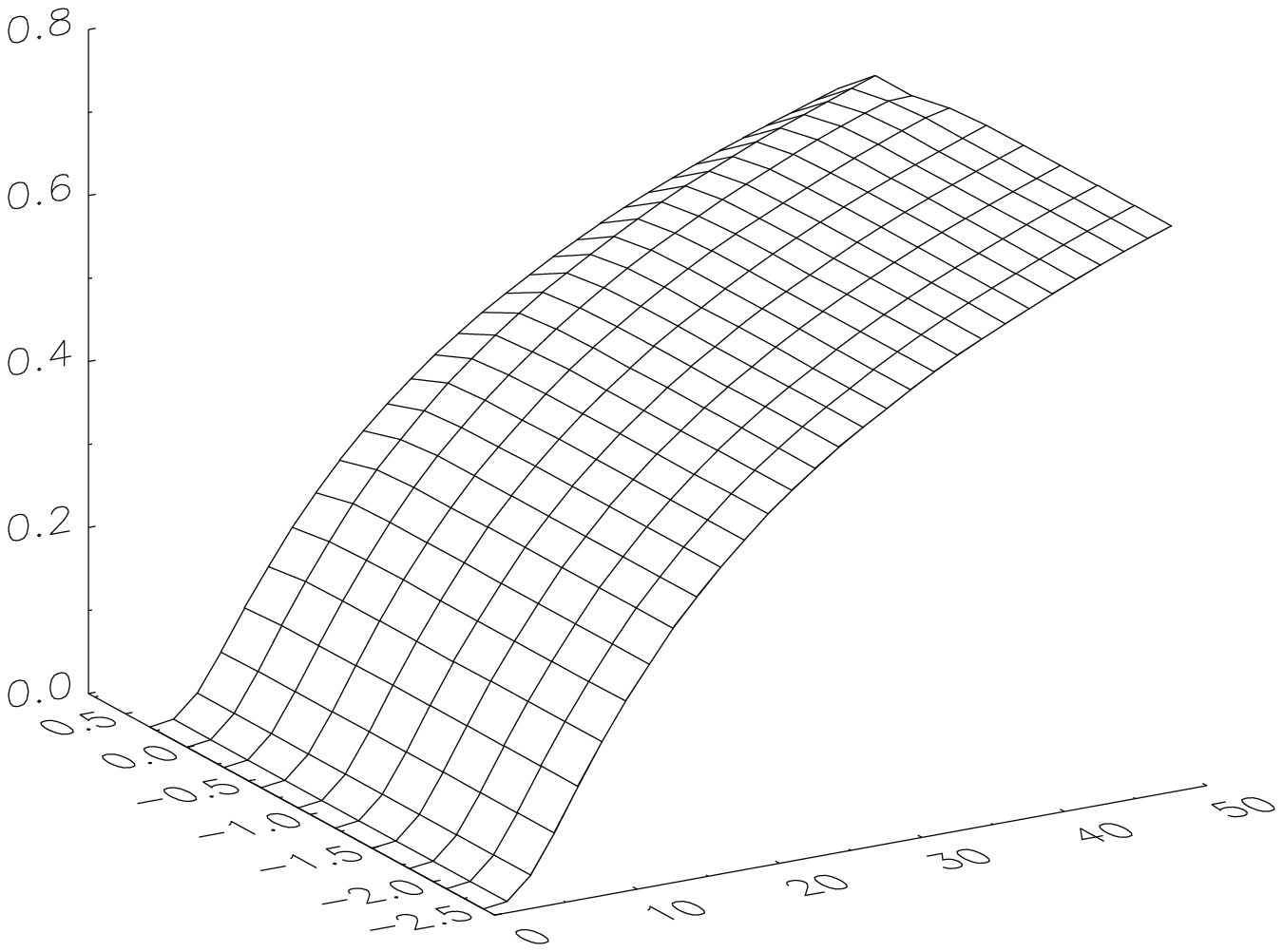 scaled 1000}}
\setbox\imageb=\hbox{\bBoxedEPSF{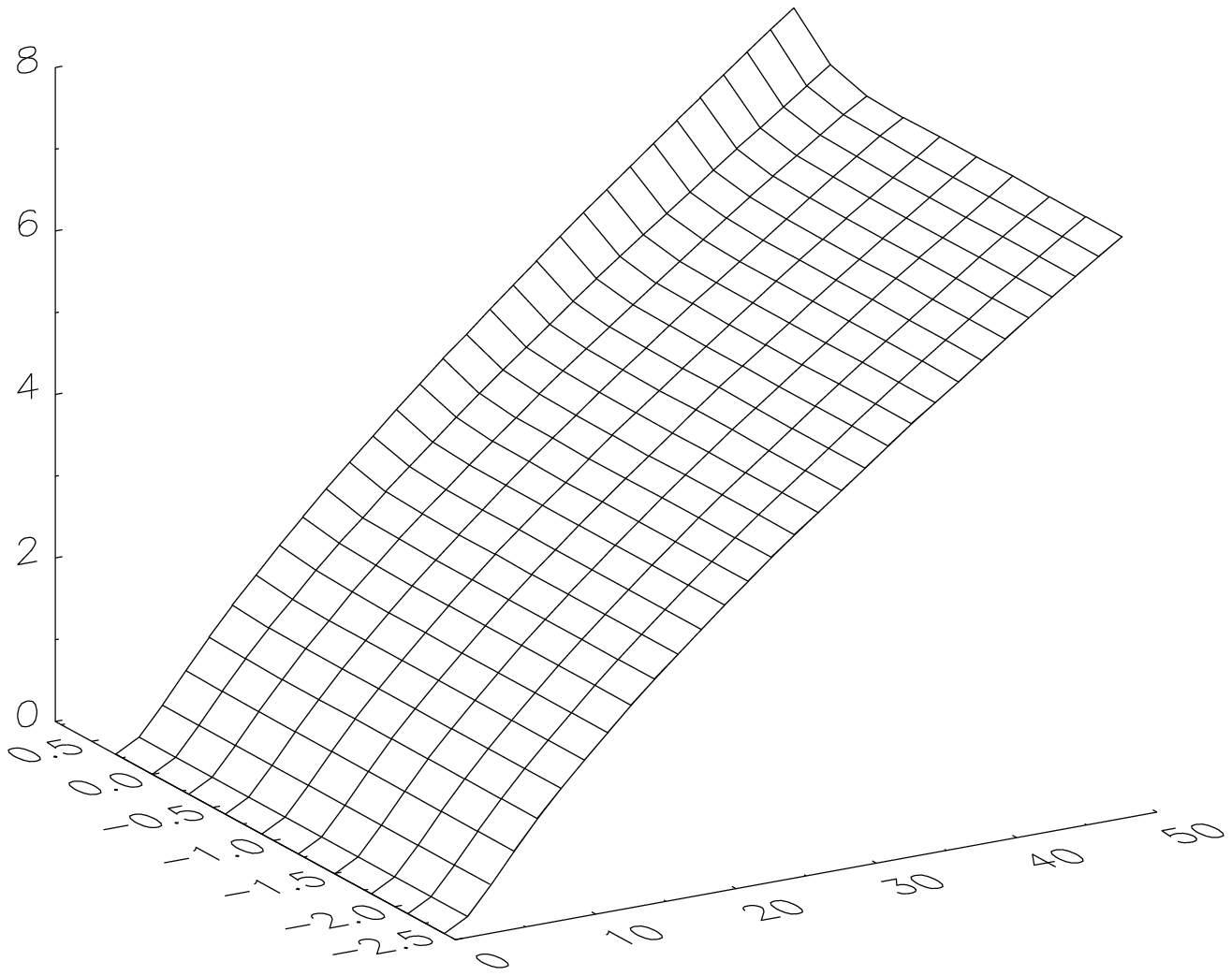 scaled 1000}}
\setbox\xlabela=\hbox{\MyBig $l$}
\setbox\ylabela=\hbox{\MyBig $\log \eps$}
\setbox\zlabela=\hbox{\MyBig $EL(l,\eps)$}
\setbox\zlabela=\hbox{\rotl\zlabela}
\setbox\xlabelb=\hbox{\MyBig $l$}
\setbox\ylabelb=\hbox{\MyBig $\log \eps$}
\setbox\zlabelb=\hbox{\MyBig $ER_x(l,\eps)$}
\setbox\zlabelb=\hbox{\rotl\zlabelb}
\setbox\boxa=%
\hbox to 18.0cm{% Specifies the width of the box
\vtop to 22.0cm{% Specifies the depth of the box, the box has zero height.
\at( 0.0,10.0){\box\imagea}
\at(13.5, 9.2){\box\xlabela}
\at( 3.0, 8.5){\box\ylabela}
\at( 0.5, 4.0){\box\zlabela}
\at( 0.0,21.5){\box\imageb}
\at(13.5,20.7){\box\xlabelb}
\at( 3.0,20.0){\box\ylabelb}
\at( 0.5,15.5){\box\zlabelb}
\at( 1.0,22.0){\box\caption}
\vfill}\hfill}
%
% Now place the boxes on the page.
%
\centerline{\box\boxa}\vfill\eject
%-----------------------------------------------------------------------------

\bye

%% file: gr-qc-figs.tex
%
%   These macros allow you to rotate or flip a \TeX\ box.  Very useful for
%   sideways tables or upsidedown answers.
%
%   To use, create a box containing the information you want to rotate.
%   (An hbox or vbox will do.)  Now call \rotr\boxnum to rotate the
%   material and create a new box with the appropriate (flipped) dimensions.
%   \rotr rotates right, \rotl rotates left, \rotu turns upside down, and
%   \rotf flips.  These boxes may contain other rotated boxes.
%
\newdimen\rotdimen
\def\vspec#1{\special{ps:#1}}%  passes #1 verbatim to the output
\def\rotstart#1{\vspec{gsave currentpoint currentpoint translate
   #1 neg exch neg exch translate}}% #1 can be any origin-fixing transformation
\def\rotfinish{\vspec{currentpoint grestore moveto}}% gets back in synch
%
%   First, the rotation right. The reference point of the rotated box
%   is the lower right corner of the original box.
%
\def\rotr#1{\rotdimen=\ht#1\advance\rotdimen by\dp#1%
   \hbox to\rotdimen{\hskip\ht#1\vbox to\wd#1{\rotstart{90 rotate}%
   \box#1\vss}\hss}\rotfinish}
%
%   Next, the rotation left. The reference point of the rotated box
%   is the upper left corner of the original box.
%
\def\rotl#1{\rotdimen=\ht#1\advance\rotdimen by\dp#1%
   \hbox to\rotdimen{\vbox to\wd#1{\vskip\wd#1\rotstart{270 rotate}%
   \box#1\vss}\hss}\rotfinish}%
%
%   Upside down is simple. The reference point of the rotated box
%   is the upper right corner of the original box. (The box's height
%   should be the current font's xheight, \fontdimen5\font,
%   if you want that xheight to be at the baseline after rotation.)
%
\def\rotu#1{\rotdimen=\ht#1\advance\rotdimen by\dp#1%
   \hbox to\wd#1{\hskip\wd#1\vbox to\rotdimen{\vskip\rotdimen
   \rotstart{-1 dup scale}\box#1\vss}\hss}\rotfinish}%
%
%   And flipped end for end is pretty ysae too. We retain the baseline.
%
\def\rotf#1{\hbox to\wd#1{\hskip\wd#1\rotstart{-1 1 scale}%
   \box#1\hss}\rotfinish}%
%
%==============================================================================
%
  %%
 %%%%%%%%%%%%%%%%%%%%%%%%%%%%%%%%%%%%%%%%%%%%%%%%%%%%%%%%%%%%%
  %%
 %%%%%   BoxedEPS.tex FOR FIGURE INSERTS OF EPSF NORM  %%%%%
 %%%%%   (EPSF = Encapsulated PostScript File)
  %%
 %%%%%%%%%%%%%%%%%%%%%%%%%%%%%%%%%%%%%%%%%%%%%%%%%%%%%%%%%%%%%
  %%  
 %%%  AUTHOR: Laurent Siebenmann
  %%    lcs@matups.matups.fr
  %%  
 %%%  VERSIONS: Feb 1991 -- 17 Sept, 1991
  %%  
 %%%  SOMMAIRE: BoxedEPS.tex d\'efinit des macro-commandes
  %%    qui permettent d'int\'egrer dans un document TeX des 
  %%    objets graphiques d\'ecrits par fichier de norme EPSF,
  %%    tout en accordant a chacun le statut d'une bo\^ite TeX ayant 
  %%    les bonnes dimensions.  La (seule!) contribution unique 
  %%    de ce fichier est de faire cela d'une fa{\c}con universelle.
  %%    C'est a dire de fa{\c}con \`a pouvoir commod\'ement 
  %%    servir avec tout pilote d'imprimante de norme 
  %%    PostScript --- malgr\'e l'absence d'une norme 
  %%    pour \special. 
  %%  
 %%%  POSTINGS: anonymous ftp 
  %%  ---  ftp 28.146.7.200 (alias shape.mps.ohio-state.edu); login:
  %%  anonymous; password: <anything>; directory pub/osutex
  %%  
  %%  ---  ftp 130.84.128.100 (alias rsovax.circe.fr); 
  %%  login: anonymous; password: <anything>; directory 
  %%  [anonymous.siebenmann]
  %%  
 %%%% DOCUMENTATION:
  %%  --- see BoxedEPS.doc
  %%  
 %%%% ACTIVATION:
  %%    by a driver-by-driver protocol
  %%    see \SetTexturesEPSFSpecial 
  %%    and its companions below.
  %%  

 \ifx\MYUNDEFINED\BoxedEPSF
   \let\temp\relax
 \else
   \message{}
   \message{ !!! BoxedEPS %
         or BoxedArt macros already defined !!!}
   \let\temp 
 \fi
  \temp
 
 \chardef\CatAt\the\catcode`\@
 \catcode`\@=11
 \chardef\C@tColon\the\catcode`\:
 \chardef\C@tSemicolon\the\catcode`\;
 \chardef\C@tQmark\the\catcode`\?
 \chardef\C@tEmark\the\catcode`\!

 \def\PunctOther@{\catcode`\:=12
   \catcode`\;=12 \catcode`\?=12 \catcode`\!=12}
 \PunctOther@

 %%temporarily suppress Plain's logging of allocations
 \let\wlog@ld\wlog 
 \def\wlog#1{\relax} 

 %% New for TOOLS
 \newif\ifIN@
 \newdimen\XShift@ \newdimen\YShift@ 
 \newtoks\Realtoks
 
 %%% New for Boxed EPSF
  %
 \newdimen\Wd@ \newdimen\Ht@
 \newdimen\Wd@@ \newdimen\Ht@@
 \newdimen\TT@
 \newdimen\LT@
 \newdimen\BT@
 \newdimen\RT@
 \newdimen\XSlide@ \newdimen\YSlide@ 
 \newdimen\TheScale  %% secretly scale in mils: 1pt= 1mil 
 \newdimen\FigScale  %% secretly scale in mils: 1pt= 1mil 
 \newdimen\ForcedDim@@

 \newtoks\EPSFDirectorytoks@
 \newtoks\EPSFNametoks@
 \newtoks\BdBoxtoks@
 \newtoks\LLXtoks@  %% useful info for Oz
 \newtoks\LLYtoks@

 \newif\ifNotIn@
 \newif\ifForcedDim@
 \newif\ifForceOn@
 \newif\ifForcedHeight@
 \newif\ifPSOrigin

 \newread\EPSFile@ 
 
 %%%% MESSAGES (separate macro needed for Europe)
  %%  
  \def\ms@g{\immediate\write16}

 %%%% WORD-PROCESSING MACROS
  %%
  %%% \IN@0#1@#2@ : Is 1st exp of #1 in 1st exp of #2 ??
   %% Answer in \ifIN@
 \newif\ifIN@\def\IN@{\expandafter\INN@\expandafter}
  \long\def\INN@0#1@#2@{\long\def\NI@##1#1##2##3\ENDNI@
    {\ifx\m@rker##2\IN@false\else\IN@true\fi}%
     \expandafter\NI@#2@@#1\m@rker\ENDNI@}
  \def\m@rker{\m@@rker}

  %%%  \SPLIT@0#1@#2@  :  Split 1st exp of #2 at 1st exp of #1
   %%  \Initialtoks@ , \Terminaltoks@ will contain pieces
  \newtoks\Initialtoks@  \newtoks\Terminaltoks@
  \def\SPLIT@{\expandafter\SPLITT@\expandafter}
  \def\SPLITT@0#1@#2@{\def\TTILPS@##1#1##2@{%
     \Initialtoks@{##1}\Terminaltoks@{##2}}\expandafter\TTILPS@#2@}

 %%%% MACROS TO TRIM  \ForeTrim@0#1@ and \Trim@0#1@  
   %% result appears in \Trimtoks@
   %% LIMITATION: assume no multiple spaces to trim

  \newtoks\Trimtoks@

  %%% \ForeTrim@0#1@ trims initial space of first erpansion of #1
   %% #1 of form \the\toks0 or \mymacro
 \def\ForeTrim@{\expandafter\ForeTrim@@\expandafter}
 \def\ForePrim@0 #1@{\Trimtoks@{#1}}
 \def\ForeTrim@@0#1@{\IN@0\m@rker. @\m@rker.#1@%
     \ifIN@\ForePrim@0#1@%
     \else\Trimtoks@\expandafter{#1}\fi}
   %%\m@rker expands here to \m@@rker since spot initial,
   %% so no confusuion with \m@rker

  %%% \Trim@0#1@ trims init and terminal spaces 
   %% Same syntax.
   %% Warns if internal spaces found.
   %% 
  \def\Trim@0#1@{%
      \ForeTrim@0#1@%
      \IN@0 @\the\Trimtoks@ @%
        \ifIN@ 
             \SPLIT@0 @\the\Trimtoks@ @\Trimtoks@\Initialtoks@
             \IN@0\the\Terminaltoks@ @ @%
                 \ifIN@
                 \else \Trimtoks@ {FigNameWithSpace}%
                 \fi
        \fi
      }

  %%%% MATH MACROS (provisional)
    %% use dimen registers for reals; unit 1pt
    %% (numerical dimension arguments OK unless contrary noted)

  %%%% One needs the point token seq (pt with cat 12) USES dimen 0
   \newtoks\pt@ks
   \def \getpt@ks 0.0#1@{\pt@ks{#1}}
   \dimen0=0pt\expandafter\getpt@ks\the\dimen0@

   %%% Convert dimen to "decimal multiplier"% USES dimens 0,2
  \newtoks\Realtoks% the output!
  \def\Real#1{%
    \dimen2=#1%
      \SPLIT@0\the\pt@ks @\the\dimen2@%%  lop off the points
       \Realtoks=\Initialtoks@%\showthe\Realtoks
            }

   %%% Multiplication 
      % USES dimens 0,2,4,6; preserves args; output \Product
   \newdimen\Product
   \def\Mult#1#2{%
     \dimen4=#1\relax
     \dimen6=#2%
     \Real{\dimen4}%
     \Product=\the\Realtoks\dimen6%
        }

   %%% Inverse 
     % USES dimens 0; preserves arg; output \Inverse
 \newdimen\Inverse
 \newdimen\hmxdim@ \hmxdim@=8192pt%halfmaxdimen
 \def\Invert#1{%
  \Inverse=\hmxdim@
  \dimen0=#1%
  \divide\Inverse \dimen0%
  \multiply\Inverse 8}

 %%% \Rescale#1#2#3  % USES dimens 0,2,4,6
  %%  alters dimen register #1 by ratio #2/#3 
  %%  where #2,#3 can be raw dimensions OR dimen registers
   \def\Rescale#1#2#3{% Adequate accuracy. Can improve. 
              \divide #1 by 100\relax
              \dimen2=#3\divide\dimen2 by 100 \Invert{\dimen2}% 
              \Mult{#1}{#2}%
              \Mult\Product\Inverse 
              #1=\Product}

 %%% \Scale#1 scales dimen register #1 
   %  by dimen register real \TheScale; USES dimens 0
  \def\Scale#1{\dimen0=\TheScale %
      \divide #1 by  1280 %% 1280*5120*10=1000*2^16 
      \divide \dimen0 by 5120 % 
      \multiply#1 by \dimen0 
      \divide#1 by 10   %% max size of #1 about 32000/10 pt
     }
 
 %%% SCRUNCHING BOXES AND SHIFTING CONTENTS
  %% TeX has to do this in general
  %% since some drivers do not let 
  %% one do it readily using Postscript

 \newbox\scrunchbox

 %%% \Scrunched#1 puts #1 in an hbox
  %%    then in effect zeros the dimensions of this box
 \def\Scrunched#1{{\setbox\scrunchbox\hbox{#1}%
   \wd\scrunchbox=0pt
   \ht\scrunchbox=0pt
   \dp\scrunchbox=0pt
   \box\scrunchbox}}

  %%% \Shifted@#1 puts #1 in \hbox 
   %% then locates basepoint to bottom left corner
   %% then translates ink only by \XShift@,\YShift@
   %% with Postscript convention
   %% For simplicity use only on scrunched boxes
  %\newdimen\XShift@ 
  %\newdimen\YShift@ 
 \def\Shifted@#1{%
   \vbox {\kern-\YShift@
       \hbox {\kern\XShift@\hbox{#1}\kern-\XShift@}%
           \kern\YShift@}}

  %%% \cBoxedEPSF#1 the main macro
   %%  component macros are explained in order below

 \def\cBoxedEPSF#1{{}\leavevmode %{} fixes box mirage for \Mas
   \ReadNameAndScale@{#1}%
   \SetEPSFSpec@
   \ReadEPSFile@ \ReadBdB@x  
   %% Calculations
     \TrimFigDims@ 
     \CalculateFigScale@  
     \ScaleFigDims@
     \SetInkShift@
   \hbox{$\mathsurround=0pt\relax
         \vcenter{\hbox{%
             \FrameSpider{\hskip-.4pt\vrule}%
             \vbox to \Ht@{\offinterlineskip\parindent=\z@%
                \FrameSpider{\vskip-.4pt\hrule}\vfil 
                \hbox to \Wd@{\hfil}%
                \vfil
                \InkShift@{\EPSFSpecial{\EPSFSpec@}{\FigSc@leReal}}%
             \FrameSpider{\hrule\vskip-.4pt}}%
         \FrameSpider{\vrule\hskip-.4pt}}}%
     $}%
    \CleanRegisters@ 
    \ms@g{ *** Box composed for the % 
         EPSF file \the\EPSFNametoks@}%
    }
 
 \def\tBoxedEPSF#1{\setbox4\hbox{\cBoxedEPSF{#1}}%
     \setbox4\hbox{\raise -\ht4 \hbox{\box4}}%
     \box4
      }

 \def\bBoxedEPSF#1{\setbox4\hbox{\cBoxedEPSF{#1}}%
     \setbox4\hbox{\raise \dp4 \hbox{\box4}}%
     \box4
      }

  \let\BoxedEPSF\cBoxedEPSF% default setting

  %% Some compatibility with BoxedArt.tex
   %

  %% Some compatibility with Sweet-teX
   %
  \def\gLinefigure[#1scaled#2]_#3{%
        \BoxedEPSF{#3 scaled #2}}
    
  %% Some compatibility with Rokicki's dvips
   %

  \def\EPSFxsize{\afterassignment\ForceW@\ForcedDim@@}
      \def\ForceW@{\ForcedDim@true\ForcedHeight@false}
  
  \def\EPSFysize{\afterassignment\ForceH@\ForcedDim@@}
      \def\ForceH@{\ForcedDim@true\ForcedHeight@true}

 %%% \ReadNameAndScale@#1
  %
 \def\ReadNameAndScale@#1{\IN@0 scaled@#1@% DOUBLE BARRELED
   \ifIN@\ReadNameAndScale@@0#1@%
   \else \ReadNameAndScale@@0#1 scaled\DefaultMilScale @
   \fi}
  
 \def\ReadNameAndScale@@0#1scaled#2@{% HELPER MACRO
    \let\OldBackslash@\\%
    \def\\{\OtherB@ckslash}%
    \edef\temp@{#1}%
    \Trim@0\temp@ @%
    \EPSFNametoks@\expandafter{\the\Trimtoks@ }%
    \FigScale=#2 pt%
    \let\\\OldBackslash@
    }
 
 \def\SetDefaultEPSFScale#1{%
      \global\def\DefaultMilScale{#1}}

 \SetDefaultEPSFScale{1000}

 %%% \ReadEPSFile@
  %
 \def \SetBogusBbox@{%
     \global\BdBoxtoks@{ BoundingBox:0 0 100 100 }%
     \global\def\BdBoxLine@{ BoundingBox:0 0 100 100 }%
     \ms@g{ !!! Will use placeholder !!!}%
     }

 \def\ReadEPSFile@{%\show\EPSFSpec@%
     \openin\EPSFile@\EPSFSpec@
     \relax  %necessary to prevent precocious expansion of \ifeof
  \ifeof\EPSFile@
     \ms@g{}%
     \ms@g{ !!! EPS FILE \the\EPSFDirectorytoks@
       \the\EPSFNametoks@\ WAS NOT FOUND !!!}
     \SetBogusBbox@
  \else%\fi
   \begingroup%%
   \catcode`\%=12\catcode`\:=12\catcode`\!=12
   \catcode`\G=14\catcode`\\=14\relax% 14 is comment
   \global\read\EPSFile@ to \BdBoxLine@%\show\BdBoxLine@
   \IN@0!PS@\BdBoxLine@ @%
   \ifIN@
     \NotIn@true %!PS OK so BdBox search!!
     \loop   
       \ifeof\EPSFile@\NotIn@false 
         \ms@g{}%
         \ms@g{ !!! BoundingBox NOT FOUND IN %
            \the\EPSFDirectorytoks@\the\EPSFNametoks@\ !!! }%
         \SetBogusBbox@
       \else\global\read\EPSFile@ to \BdBoxLine@
       %\show\BdBoxLine@
       \fi
       \global\BdBoxtoks@\expandafter{\BdBoxLine@}%
       \IN@0BoundingBox:@\the\BdBoxtoks@ @%
       \ifIN@\NotIn@false\fi%
     \ifNotIn@\repeat
   \else
         \ms@g{}%
         \ms@g{ !!! \the\EPSFNametoks@\ not PS!\  !!!}%
         \SetBogusBbox@
   \fi
  \endgroup\relax
  \fi
  \closein\EPSFile@ 
   }

  %%% \ReadBdB@x
   % Rmk For simplicity 0 not used in syntax 
   %  of \ReadBdB@x@,  \ReadBdB@x@@ 
  \def\ReadBdB@x{% PART 0
   \expandafter\ReadBdB@x@\the\BdBoxtoks@ @}
  
  \def\ReadBdB@x@#1BoundingBox:#2@{% PART 1
    \ForeTrim@0#2@%
    \IN@0atend@\the\Trimtoks@ @
       \ifIN@\Trimtoks@={0 0 100 100 }
         \ms@g{}%
         \ms@g{ !!! BoundingBox not found in %
         \the\EPSFDirectorytoks@\the\EPSFNametoks@\space !!!}%
         \ms@g{ !!! It must not be at end of EPSF !!!}%
         \ms@g{ !!! Will use placeholder !!!}%
       \fi%% cf \SetBogusBbox@
    \expandafter\ReadBdB@x@@\the\Trimtoks@ @%
   }
    
  \def\ReadBdB@x@@#1 #2 #3 #4@{% PART 2
      \Wd@=#3bp\advance\Wd@ by -#1bp%
      \Ht@=#4bp\advance\Ht@ by-#2bp%
       \Wd@@=\Wd@ \Ht@@=\Ht@ %% useful info for Clark
       \LLXtoks@={#1}\LLYtoks@={#2}%% useful info for Oz
      \ifPSOrigin\XShift@=-#1bp\YShift@=-#2bp\fi 
     }

  %%% \SetEPSFDirectory 
   %
   \def\G@bbl@#1{}
   \bgroup
     \global\edef\OtherB@ckslash{\expandafter\G@bbl@\string\\}
   \egroup

  \def\SetEPSFDirectory{%  Part 1
           \bgroup\PunctOther@\relax
           \let\\\OtherB@ckslash
           \SetEPSFDirectory@}

 \def\SetEPSFDirectory@#1{% Part 2
    \edef\temp@{#1}%
    \Trim@0\temp@ @%  result in \Trimtoks@
    \global\toks1\expandafter{\the\Trimtoks@ }\relax
    \egroup
    \EPSFDirectorytoks@=\toks1
    }

  %%% \SetEPSFSpec@
 \def\SetEPSFSpec@{%
     \bgroup
     \let\\=\OtherB@ckslash
     \global\edef\EPSFSpec@{%
        \the\EPSFDirectorytoks@\the\EPSFNametoks@}%
     \global\edef\EPSFSpec@{\EPSFSpec@}%
     \egroup}

 %%% \TrimFigDims@ 
  % 
 \def\TrimTop#1{\advance\TT@ by #1}
 \def\TrimLeft#1{\advance\LT@ by #1}
 \def\TrimBottom#1{\advance\BT@ by #1}
 \def\TrimRight#1{\advance\RT@ by #1}

 \def\TrimFigDims@{%
    \advance\Wd@ by -\LT@ 
    \advance\Wd@ by -\RT@ \RT@=\z@
    \advance\Ht@ by -\TT@ \TT@=\z@
    \advance\Ht@ by -\BT@ 
    }

 %%% \CalculateFigScale@
  %
  \def\ForceWidth#1{\ForcedDim@true
       \ForcedDim@@#1\ForcedHeight@false}
  
  \def\ForceHeight#1{\ForcedDim@true
       \ForcedDim@@=#1\ForcedHeight@true}

  \def\ForceOn{\ForceOn@true}
  \def\ForceOff{\ForceOn@false\ForcedDim@false}
  
  \def\epsfxsize{\afterassignment\ForceW@\ForcedDim@@}
      \def\ForceW@{\ForcedDim@true\ForcedHeight@false}
  
  \def\epsfysize{\afterassignment\ForceH@\ForcedDim@@}
      \def\ForceH@{\ForcedDim@true\ForcedHeight@true}
  
  \def\CalculateFigScale@{%
            %Have default \FigScale or read \FigScale
     \ifForcedDim@\FigScale=1000pt% %% start afresh
           \ifForcedHeight@
                \Rescale\FigScale\ForcedDim@@\Ht@
           \else
                \Rescale\FigScale\ForcedDim@@\Wd@
           \fi
     \fi
     \Real{\FigScale}%
     \edef\FigSc@leReal{\the\Realtoks}%
     }
   
  \def\ScaleFigDims@{\TheScale=\FigScale
      \ifForcedDim@
           \ifForcedHeight@ \Ht@=\ForcedDim@@  \Scale\Wd@
           \else \Wd@=\ForcedDim@@ \Scale\Ht@
           \fi
      \else \Scale\Wd@\Scale\Ht@        
      \fi
      \ifForceOn@\relax\else\global\ForcedDim@false\fi
      \Scale\LT@\Scale\BT@  %%%\Scale\Wd@\Scale\Ht@
      \Scale\XShift@\Scale\YShift@
      }
      
  %%% \ShowReservedBoxes
   %%  shows (prints) corrected scaled and positioned
   %%  bounding boxes; for diagnostics
  %%% \HideReservedBoxes makes them invisible again
   %%

 \let\HideDisplacementBoxes\HideReservedBoxes  %% some synonyms
 \let\ShowDisplacementBoxes\ShowReservedBoxes

  \ShowDisplacementBoxes
 
  %%% \hSlide#1, \vSlide#1
   %%
 \def\hSlide#1{\advance\XSlide@ by #1}
 \def\vSlide#1{\advance\YSlide@ by #1}
 
  %%% \SetInkShift@, \InkShift@#1
   %%
  \def\SetInkShift@{%
            \advance\XShift@ by -\LT@
            \advance\XShift@ by \XSlide@
            \advance\YShift@ by -\BT@
            \advance\YShift@ by -\YSlide@
             }
  \def\InkShift@#1{\Shifted@{\Scrunched{#1}}}
 
  %%% \CleanRegisters@
   %
  \def\CleanRegisters@{%
      \globaldefs=1\relax
        \XShift@=\z@\YShift@=\z@\XSlide@=\z@\YSlide@=\z@
        \TT@=\z@\LT@=\z@\BT@=\z@\RT@=\z@
      \globaldefs=0\relax}

 %%% Special syntax for several drivers. The macros 
  %% \SetTexturesEPSFSpecial  %% Textures 
  %% \SetUnixCoopEPSFSpecial %% dvi2ps early unix 
  %% \SetBetcholsheimEPSFSpecial %% dvi2ps by S.P.Betcholsheim
  %% \SetLisEPSFSpecial %% dvi2ps by Tony Lis
  %% \SetRokickiEPSFSpecial  %% dvips by Tom Rokicki
  %% \SetOzTeXEPSFSpecial  %% OzTeX by Andrew Trevorrow
  %% \SetPSprintEPSFSpecial %% PSprint by Andrew Trevorrow
  %% \SetArborEPSFSpecial  %% ArborTeX DVILASER/PS
  %% \SetClarkEPSFSpecial %% dvitops by James Clark
  %% \SetDVIPSoneEPSFSpecial %% DVIPSONE of Y&Y 
  %% \SetBeebeEPSFSpecial %% DVIALW by N. Beebe
  %% \SetStandardEPSFSpecial %% Nonexistant: Placebo below
  %% These macros adapt to various drivers by (re-)defining
  %% the macro \EPSFSpecial#1#2, where
  %% #1 = EPS file pathname (use \\ for the letter backslash)
  %% #2 = scale in mils 
    %% Be wary of using strange characters in pathnames!
 
 %% Textures, Blue Sky Research, Barry Smith
 \def\SetTexturesEPSFSpecial{\PSOriginfalse%\PSOrigintrue
  \gdef\EPSFSpecial##1##2{\relax
    \edef\specialthis{##2}%
    \SPLIT@0.@\specialthis.@\relax
    %\showthe\Initialtoks@
    \special{illustration ##1 scaled
                        \the\Initialtoks@}}}
 
  %% Unix : dvi2ps by:  Mark Senn, Stephan  Bechtolsheim,  
   % Bob  Brown, Richard, Furuta, James Schaad, 
   % Robert  Wells, Norm Hutchinson, Neal Holtz.
   % Introduced by B. Horn <bkph@ai.mit.edu>
  \def\SetUnixCoopEPSFSpecial{\PSOrigintrue % Please test!
   \gdef\EPSFSpecial##1##2{%
      \dimen4=##2pt% convert real to dimen
      \divide\dimen4 by 1000\relax
      \Real{\dimen4}%dimens 0,2 used here
      \edef\Aux@{\the\Realtoks}%  
      %%convert dimen to real
      \includegraphics{##1\space}}}

  %% dvi2ps by S.P. Bechtolsheim,
   % implantations? ; dates?; availability?
   % Introduced by B. Horn <bkph@ai.mit.edu>; please test!!
  \def\SetBechtolsheimRokickiEPSFSpecial{\PSOrigintrue 
   \gdef\EPSFSpecial##1##2{%
      \dimen4=##2pt% convert real to dimen
      \divide\dimen4 by 1000\relax
      \Real{\dimen4}% dimens 0,2 used here
      \edef\Aux@{\the\Realtoks}%  
      %%convert dimen to real
      \special{ps: psfiginit}%
      \special{ps: literal 1 1 0 0 1 1 startTexFig
           \the\mag\space 1000 div \Aux@\space mul 
           \the\mag\space 1000 div \Aux@\space mul scale}%
      \special{ps: include  ##1}%
      \special{ps: literal endTexFig}%
        }}

  %% dvi2ps by Tony Lis,
   % implantations? ; dates?; availability?
   % Introduced by B. Horn <bkph@ai.mit.edu>; please test!!
  \def\SetLisEPSFSpecial{\PSOrigintrue 
   \gdef\EPSFSpecial##1##2{%
      \dimen4=##2pt% convert real to dimen
      \divide\dimen4 by 1000\relax
      \Real{\dimen4}% dimens 0,2 used here
      \edef\Aux@{\the\Realtoks}%  
      %%convert dimen to real
      \special{pstext="1 1 0 0 1 1 startTexFig\space
           \the\mag\space 1000 div \Aux@\space mul 
           \the\mag\space 1000 div \Aux@\space mul scale}%
      \includegraphics{##1}%
      \special{pstext=endTexFig}%
        }}

  %% dvips by Tom Rokicki; driver in portable C 
   % This driver improves on dvi2ps; its Postscript
   % output is thee times as compact as that 
   % produced by dvi2ps
   % Introduced by W.D. Neumann <neumann@mps.ohio-state.edu>
  \def\SetRokickiEPSFSpecial{\PSOrigintrue 
   \gdef\EPSFSpecial##1##2{%
      \dimen4=##2pt% convert real to dimen
      \divide\dimen4 by 10\relax
      \Real{\dimen4}% dimens 0,2 used here
      \edef\Aux@{\the\Realtoks}%  
      %%convert dimen to real
      \includegraphics{##1}}}

  \def\SetInlineRokickiEPSFSpecial{\PSOrigintrue 
   \gdef\EPSFSpecial##1##2{%
      \dimen4=##2pt% convert real to dimen
      \divide\dimen4 by 1000\relax
      \Real{\dimen4}% dimens 0,2 used here
      \edef\Aux@{\the\Realtoks}%  
      %%convert dimen to real
      \special{ps::[begin] 1 1 0 0 1 1 startTexFig\space
           \the\mag\space 1000 div \Aux@\space mul 
           \the\mag\space 1000 div \Aux@\space mul scale}%
      \special{ps: plotfile ##1}%
      \special{ps::[end] endTexFig}%
        }}

  %% OzTeX, by AndrewTrevorrow, 
   %  complete public domain TeX for Macintosh
   %  Send 10 UNFORMATTED 800K disks 
   %  with return postage 
   %  Peter Abbott, Computing Service, 
   %  Aston University, Aston Triangle, Birmingham B4 7ET
  \def\SetOzTeXEPSFSpecial{\PSOriginfalse % artifice; see below
  \gdef\EPSFSpecial##1##2{%note order
     \special{##1\space 
       ##2 1000 div \the\mag\space 1000 div mul
       ##2 1000 div \the\mag\space 1000 div mul scale
       \the\LLXtoks@\space neg \the\LLYtoks@\space neg translate
             }}} 
 
 %% PSprint,  by AndrewTrevorrow for VaX VMS
  % diagnosed and tested 2-91 by Max Calviani 
  % <ISICA@ASTRPD.infn.it>

 %% DVILASER/PS driver originally written by David Fuchs
  % marketed and supported by ArborTeXt  535 W. William St.
  % Suite 300, Ann Arbor, MI 48103, U.S.A
  % (313) 996-3566 (313) 996-3573
  % help@arbortext.com, Andrew Dobrowolski
 \def\SetArborEPSFSpecial{\PSOriginfalse % check!
   \gdef\EPSFSpecial##1##2{%
     \edef\specialthis{##2}%
     \SPLIT@0.@\specialthis.@\relax % suppress decimals (nec!)
     \special{ps: epsfile ##1\space \the\Initialtoks@}}}

 %% dvitops, (c) James Clark <jjc@jclark.uucp>
  % public domain; distributed by UK TeX Archive
  % computers: unix, msdos, vms, primos and vm/cms,
  % (introduced by S. Ratz <spqr@uk.ac.southampton.ecs>)
 \def\SetClarkEPSFSpecial{\PSOriginfalse % please test!
   \gdef\EPSFSpecial##1##2{%
     \Rescale {\Wd@@}{##2pt}{1000pt}%
     \Rescale {\Ht@@}{##2pt}{1000pt}%
     \special{dvitops: import 
           ##1\space\the\Wd@@\space\the\Ht@@}}}

 %% DVIPSONE, for PC compatibles
  % Y&Y, 106 Indian Hill, Carlisle MA 01741, USA
  % (508) 371-3286
  % (introduced by B. Horn <bkph@ai.mit.edu>)

 %% DVIALW by N. Beebe, public domain, charge $100 
  % DVI Driver Distribution, Center for Scientific Computing,
  % Department of Mathematics, 220 South Physics Building,
  % University of Utah, Salt Lake City, UT 84112, USA
  % (introduced by B. Horn <bkph@ai.mit.edu>)

 \def\SetStandardEPSFSpecial{%
   \gdef\EPSFSpecial##1##2{%
     \ms@g{}
     \ms@g{%
       !!! Sorry! There is still no standard for \string%
       \special\ EPSF integration !!!}%
     \ms@g{%
      --- So you will have to identify your driver using a command}%
     \ms@g{%
      --- of the form \string\Set...EPSFSpecial, in order to get}%
     \ms@g{%
      --- your graphics to print.  See BoxedEPS.doc.}%
     \ms@g{}
     \KillEPSFSpecial
     }}

  \def\KillEPSFSpecial{\gdef\EPSFSpecial##1##2{}}

  \SetStandardEPSFSpecial %% currently gives warning
 
 \let\wlog\wlog@ld %%restore logging 

 \catcode`\:=\C@tColon
 \catcode`\;=\C@tSemicolon
 \catcode`\?=\C@tQmark
 \catcode`\!=\C@tEmark

 \catcode`\@=\CatAt

%%%%%%%%%%%% ASCII Character test
 %
 %       Upper case letters: ABCDEFGHIJKLMNOPQRSTUVWXYZ
 %       Lower case letters: abcdefghijklmnopqrstuvwxyz
 %                                   Digits: 0123456789
 % Square, curly, angle braces, parentheses: [] {} <> ()
 %           Backslash, slash, vertical bar: \ / |
 %                              Punctuation: . ? ! , : ;
 %          Underscore, hyphen, equals sign: _ - =
 %                Quotes--right left double: ' ` "
 %"at", "number" "dollar", "percent", "and": @ # $ % &
 %           "hat", "star", "plus", "tilde": ^ * + ~
 %
 %%%%%%%%%%%%%%%%%%%%%%%%
 %
 % Une seule erreur de transmission peut empoisoner un programme!
 %
 % A single transmission error can poison a whole program.
 %
 %%%%%%%%%%%%%%%%%%%%%%%%